\documentclass[11pt,twoside]{article}
\usepackage{colordvi}
\usepackage{latexsym} \usepackage{epsfig} \topmargin -1cm
\begin{document} \thispagestyle{empty}
 \begin{titlepage}
  \begin{center}
{ \Large \bf   BEYOND THE STANDARD MODEL$^*$\\[7mm] \large (~IN SEARCH OF SUPERSYMMETRY~)}
 \vspace{1cm}

{\large \bf D. I.~Kazakov } \\[5mm]

{\it BLTP, JINR, Dubna and ITEP, Moscow
 \\ e-mail: kazakovd@thsun1.jinr.ru}
  \end{center}
\vspace{1cm}

 \begin{center}
  \leavevmode
  \epsfxsize=7cm \epsfysize=7cm
 \epsffile{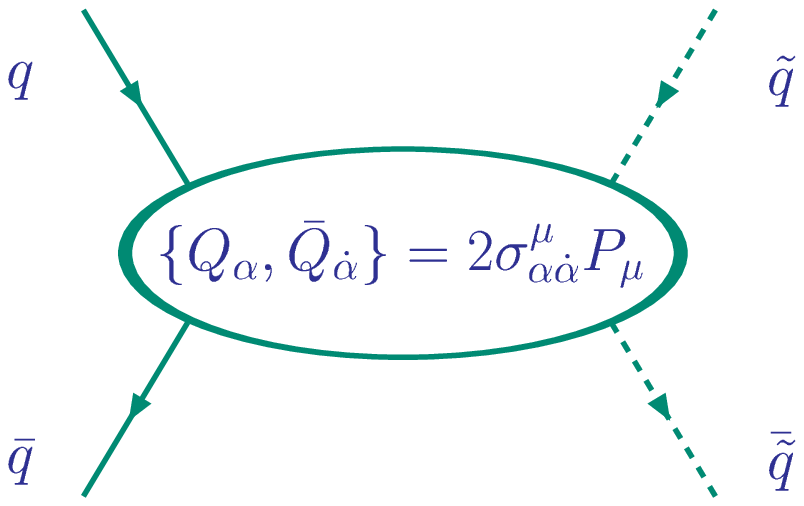}
 \end{center}

\vspace{1cm}

 \begin{center}
  Abstract
 \end{center}

The present lectures contain an introduction to low energy
supersymmetry, a new symmetry that relates bosons and fermions, in
particle physics.  The Standard Model of fundamental interactions
is briefly reviewed, and the motivation to introduce supersymmetry
is discussed. The main notions of supersymmetry  are introduced.
The supersymmetric extension of the Standard Model - the  Minimal
Supersymmetric Standard Model -  is considered in more detail.
Phenomenological features of the MSSM as well as possible
experimental signatures of SUSY are described. An intriguing
situation with the supersymmetric Higgs boson is discussed.
\vspace{3cm}

\noindent
------------------------------------------------------------

$^*$ Lectures given at the European School on High Energy
Physics, Aug.-Sept. 2000, Caramulo, Portugal

\pagebreak \thispagestyle{empty} \tableofcontents \vglue 0.34cm {\bf
References} \hfill {\bf 77}
 \end{titlepage}

\renewcommand{\thesubsection}{\thesection.\arabic{subsection}}
\renewcommand{\thesubsubsection}{\thesubsection.\arabic{subsubsection}}
\renewcommand{\theequation}{\thesection.\arabic{equation}}

\section{Introduction. The Standard Model and beyond}
 \setcounter{equation} 0

The Standard Model (SM) of fundamental interactions describes {\em
strong, weak} and {\em electromagnetic } interactions of
elementary particles  \cite{SM}. It is based on a {\em gauge
principle}, according to which all the forces of Nature are
mediated by an exchange of the gauge fields of the corresponding
local symmetry group. The symmetry group of the SM is
 \begin{equation}
SU_{colour}(3)\otimes SU_{left}(2)\otimes U_{hypercharge}(1),
 \label{a}
 \end{equation}
whereas the field content is the following:

\vspace{0.4cm} \noindent\underline{{\em Gauge sector} :\ \ Spin =
1} \par \vspace{0.4cm}

The gauge bosons are spin 1 vector particles belonging to the
adjoint representation of the group (\ref{a}). Their quantum
numbers with respect to $SU(3)\otimes SU(2)\otimes U(1)$ are
 \begin{equation}
 \begin{array}{ccccc}
\mbox{\em gluons}& G_\mu^a :& \ (\underline{8},\underline{1},0)& \
\ SU_c(3)& \ \ g_{s},\\ {\begin{array}{cc} \mbox{\em
intermediate}\\ \mbox{\em weak bosons} \end{array}}& W_\mu^i :& \
(\underline{1}, \underline{3},0)& \ \ SU_L(2)& \ \ g,\\ \mbox{\em
abelian boson}& B_\mu :& \ (\underline{1},\underline{1},0)& \ \
U_Y(1)&\ \ g',
 \end{array} \nonumber\end{equation}
where the coupling constants are usually denoted by $g_s$, $g$ and
$g'$, respectively.

\vspace{0.4cm} \noindent\underline{{\em Fermion sector} : \ \ Spin
= 1/2} \par \vspace{0.4cm}

The matter fields are fermions belonging to the fundamental
representation of the gauge group. These are believed to be quarks
and leptons of at least of three generations. The SM is left-right
asymmetric. Left-handed and right-handed fermions have different
quantum numbers
\begin{equation}
 \begin{array}{cccccccc}
\mbox{\em quarks} \ & & & & & & & \\
 Q_{\alpha L}^i = & \left(\begin{array}{c}U_\alpha^i\\D_\alpha^i
 \end{array} \right)_L & = & \left(\begin{array}{c}u^i\\d^i \end{array}\right)_L ,
& \left(\begin{array}{c}c^i\\s^i \end{array}\right)_L,&
\left(\begin{array}{c}t^i
\\b^i \end{array} \right)_L,& \ldots & (\underline{3},\underline{2},1/3)\\
  &  U_{\alpha R}^i & = & u_{iR},& c_{iR},& t_{iR},& \ldots &
(\underline{3}^*, \underline{1}, 4/3)\\ &  D_{\alpha R}^i & = &
d_{iR},&
  s_{iR},& b_{iR},& \ldots &(\underline{3}^*, \underline{1}, -2/3)\\
\mbox{\em leptons} \ &  L_{\alpha L} & = &
\left(\begin{array}{c}\nu_e\\ e \end{array} \right)_L, &
\left(\begin{array}{c}\nu_\mu\\ \mu  \end{array} \right)_L, &
\left(\begin {array}{c}\nu_\tau\\ \tau \end{array} \right)_L,&
\ldots & (\underline{1}, \underline{2},-1)\\ &  E_{\alpha R} & = &
e_R,&
  \mu_R,& \tau_R,& \ldots & (\underline{1},\underline{1}, -2) \end{array}
    \nonumber\end{equation}
$i = 1,2,3$ - colour, $\alpha = 1,2,3,\ldots$ - generation.

\vspace{0.4cm} \noindent\underline{{\em Higgs sector} : \ \ Spin =
0} \par \vspace{0.4cm}

 In the minimal version of the SM there is  one doublet of Higgs
 scalar fields
 \begin{equation}
H  =  \left(\begin{array}{c}H^{0}\\ H^- \end{array} \right) \ \ \
\ (\underline{1}, \underline{2},-1),
 \end{equation}
which is introduced in  order to give masses to quarks, leptons
and intermediate weak bosons via spontaneous  breaking of
electroweak symmetry.

In the framework of Quantum Field Theory  the SM is described by
the following Lagrangian:
 \begin{equation}
{\cal L} ={\cal L}_{gauge} + {\cal L}_{Yukawa} + {\cal L}_{Higgs},
\label{SM}
 \end{equation}
 \begin{eqnarray}
{\cal L}_{gauge} & = & -\frac{1}{4} G_{\mu\nu}^aG_{\mu\nu}^a -
\frac{1}{4} W_{\mu\nu}^iW_{\mu\nu}^i -\frac{1}{4}
B_{\mu\nu}B_{\mu\nu}\\ & &  +
 i\overline{L}_{\alpha}\gamma^{\mu}D_{\mu}L_{\alpha} +
i\overline{Q}_{\alpha}\gamma^{\mu}D_{\mu}Q_{\alpha} +
i\overline{E}_{\alpha} \gamma^{\mu}D_{\mu}E_{\alpha} \nonumber \\
 & &  + i\overline{U}_{\alpha}\gamma^{\mu}D_{\mu}U_{
\alpha} + i\overline{D}_{\alpha}\gamma^{\mu}D_{\mu}D_{\alpha} +
(D_{\mu}H)^{\dagger}(D_{\mu}H),  \nonumber
 \end{eqnarray}
where
 \begin{eqnarray*}
G_{\mu\nu}^a & = &
\partial_{\mu}G_{\nu}^a-\partial_{\nu}G_{\mu}^a+g_{s}
f^{abc}G_{\mu}^bG_{\nu}^c,\\ W_{\mu\nu}^i & = &
\partial_{\mu}W_{\nu}^i-\partial_{\nu}W_{\mu}^i+g\epsilon
^{ijk}W_{\mu}^jW_{\nu}^k,\\ B_{\mu\nu} & = &
\partial_{\mu}B_{\nu}-\partial_{\nu}B_{\mu},\\ D_{\mu}L_{\alpha} &
= &
(\partial_{\mu}-i\frac{g}{2}\tau^iW_{\mu}^i+i\frac{g'}{2}B_{\mu})L_{\alpha},\\
D_{\mu}E_{\alpha} & = & (\partial_{\mu}+ig'B_{\mu})E_{\alpha},\\
D_{\mu}Q_{\alpha} & = &
(\partial_{\mu}-i\frac{g}{2}\tau^iW_{\mu}^i-i\frac{g'}{6}B_{\mu}-
i\frac{g_s}{2}\lambda^aG_{\mu}^a)Q_{\alpha},\\ D_{\mu}U_{\alpha} &
= &
(\partial_{\mu}-i\frac{2}{3}g'B_{\mu}-i\frac{g_s}{2}\lambda^aG_{\mu}^a)
U_{\alpha},\\ D_{\mu}D_{\alpha} & = &
(\partial_{\mu}+i\frac{1}{3}g'
B_{\mu}-i\frac{g_s}{2}\lambda^aG_{\mu}^a)D_{\alpha}.
 \end{eqnarray*}
 \begin{equation}
{\cal L}_{Yukawa} =
y_{\alpha\beta}^L\overline{L}_{\alpha}E_{\beta}H + y_{\alpha
\beta}^D\overline{Q}_{\alpha}D_{\beta}H +
y_{\alpha\beta}^U\overline{Q}_{\alpha} U_{\beta}\tilde{H} + h.c.,
\label{yukawa}
 \end{equation}
where $\tilde{H}=i\tau_2H^{\dagger}$.
 \begin{equation}
{\cal L}_{Higgs} = - V = m^2H^{\dagger}H -
\frac{\lambda}{2}(H^{\dagger}H)^2.\label{hig}
 \end{equation}
Here $\{y\}$ are the Yukawa and $\lambda$ is the Higgs coupling
constants, both di\-men\-si\-on\-less, and $m$ is the only
dimensional mass parameter\footnote{We use the usual for particle
physics units $c=\hbar=1$}.

The Lagrangian of the SM contains the following set of free
parameters:
 \begin{itemize}
     \item 3 gauge couplings $g_s, g, g' $;
  \item 3 Yukawa matrices $y_{\alpha\beta}^L, y_{\alpha\beta}^D, y_{\alpha\beta}^U
$;
     \item Higgs coupling  constant $\lambda$;
     \item Higgs mass parameter $m^2$;
     \item number of matter fields (generations).
 \end{itemize}

 \begin{figure}[tb]
 \begin{center}
 \leavevmode
  \epsfxsize=11cm
 \epsffile{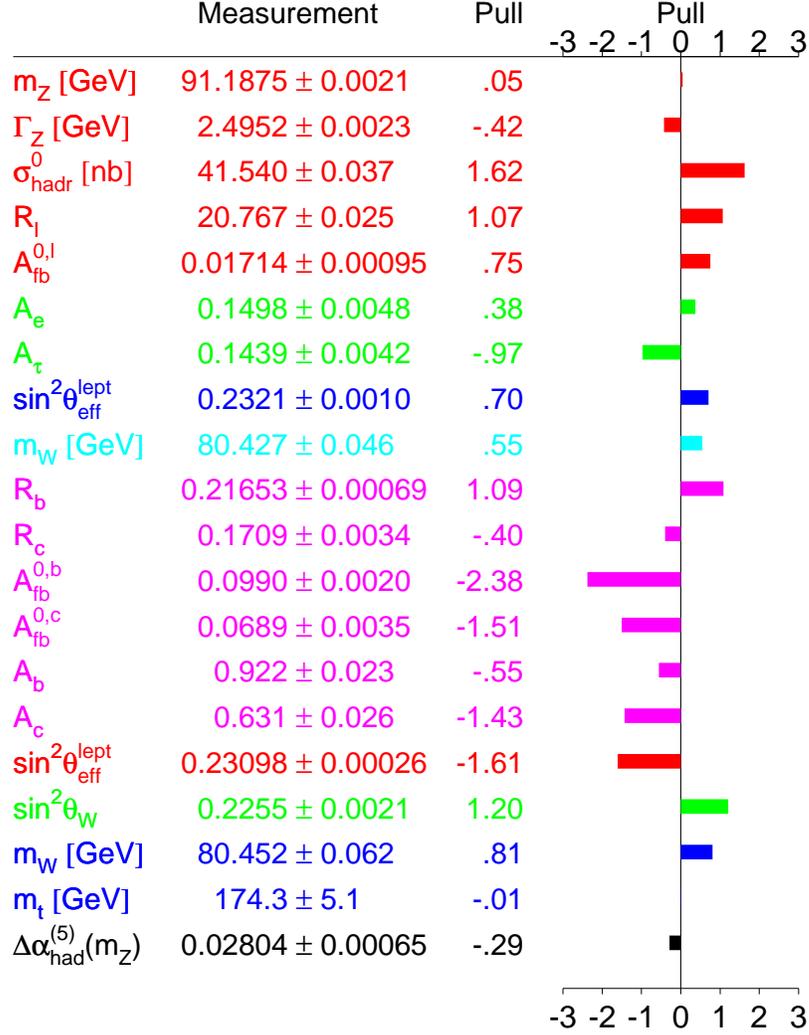}\vspace{-1cm}
 \caption{Global Fit of the Standard Model} \label{smfit}
 \end{center}
 \end{figure}
 \begin{figure}[tb]
 \begin{center}
 \leavevmode
  \epsfxsize=10cm
 \epsffile{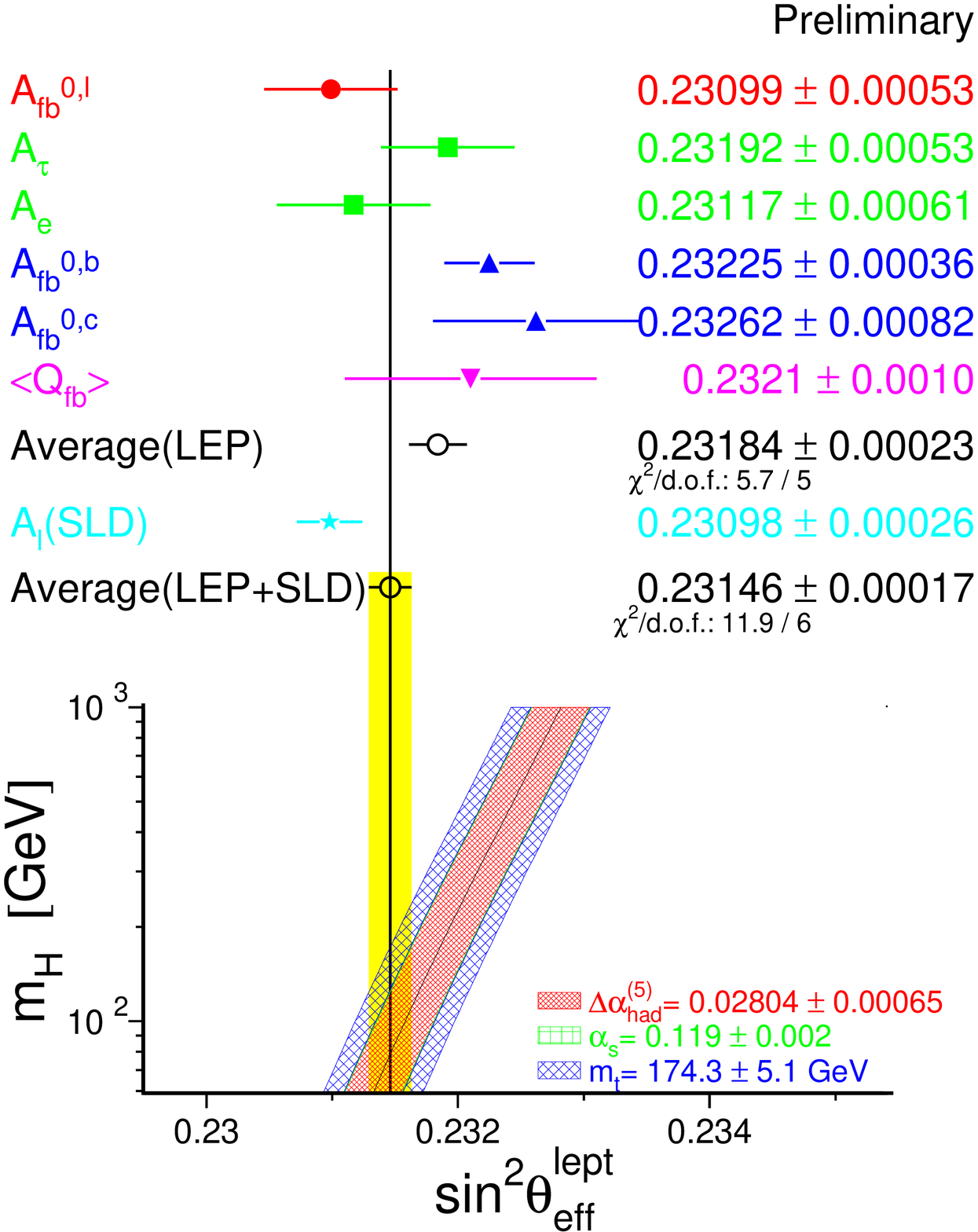}
 \caption{Weak mixing angle and the Higgs boson mass} \label{mixang}
 \end{center}
 \end{figure}
All the particles obtain their masses due to spontaneous breaking
of $SU_{left}(2)$ symmetry group via a non-zero vacuum expectation
value (v.e.v.) of the Higgs field
 \begin{equation}
<H> = \left(\begin{array}{c}v\\ 0\end{array}\right),\ \ \ \
v=m/\sqrt{\lambda}. \label{vac}
 \end{equation}
As a result,  the gauge group of the SM is spontaneously  broken
down to $$SU_c(3)\otimes SU_L(2)\otimes U_Y(1) \Rightarrow
SU_c(3)\otimes U_{EM}(1).$$ The physical weak intermediate bosons
are  linear combinations of the gauge ones
 \begin{equation} W_{\mu}^{\pm} =  \frac{W_{\mu}^1\mp iW_{\mu}^2}{\sqrt{2}},\
\ \ \ Z_{\mu} = -\sin{\theta_W}B_{\mu} + \cos{\theta_W}W_{\mu}^3
 \end{equation}
with masses
 \begin{equation}
m_W=\frac{1}{\sqrt{2}}gv,\ \ \ \  m_Z=m_W/\cos{\theta_W},\ \ \ \
\tan{\theta_W}=g'/g, \label{Z}
 \end{equation}
while the photon field
 \begin{equation}
\gamma_{\mu} = \cos{\theta_W}B_{\mu} + \sin{\theta_W}W_{\mu}^3
 \end{equation}
remains massless.

The matter fields acquire masses proportional to the corresponding
Yukawa couplings:
 \begin{equation}
M_{\alpha\beta}^u = y_{\alpha\beta}^uv,\ M_{\alpha\beta}^d =
y_{\alpha\beta}^dv,\ M_{\alpha\beta}^l = y_{\alpha\beta}^lv, \ m_H
= \sqrt{2}m. \label{mass}
 \end{equation}
Explicit mass terms in the Lagrangian are forbidden because they
are not $SU_{left}(2)$ symmetric and would destroy the
renormalizability of the Standard Model.

The SM has been constructed as a result of numerous  efforts both
theoretical and experimental.  At present, the SM is extraordinary
successful, the achieved accuracy of its predictions corresponds
to  experimental data within 5 \% \cite{SM,test-sm}. The combined
results of the Global SM fit are shown in Fig.\ref{smfit}
\cite{test-sm}. All the particles, except for  Higgs boson, have
been discovered experimentally. And the mass of the Higgs boson is
severely constrained from precision electroweak data (see
Fig.\ref{mixang}~\cite{test-sm}).

However, the SM has its natural drawbacks and unsolved problems.
Among them are
\begin{itemize}
\item inconsistency of the SM as a QFT (Landau pole),
\item large number of free parameters,
\item formal unification of strong and electroweak interactions,
\item still unclear mechanism of EW symmetry breaking: The Higgs  boson has
not yet been observed and it is not clear whether it is
fundamental or composite,
\item the problem of CP-violation is not well understood including
CP-violation in a strong interaction,
\item flavour mixing and the number of generations are arbitrary,
\item the origin of the mass spectrum is unclear.
\end{itemize}

The answer to these problems lies beyond the SM. There are two
possible ways of going beyond the SM
\begin{itemize}
\item[$\Rightarrow$] To consider the {\it same} fundamental fields
with {\it new} interactions. This way leads us to supersymmetry,
Grand Unification, String Theory, etc. It seems to be favoured  by
modern experimental data.
\item[$\Rightarrow$] To consider {\it new} fundamental fields with
{\it new} interactions. This way leads us to compositeness,
fermion-antifermion  condensates, Technicolour, extended
Technicolour, preons, etc. It is not favoured by data at the
moment.
\end{itemize}

There are also possible exotic ways out of the SM: gravity at TeV
energies, large extra dimensions, brane world, etc. We do not
consider them here. In what follows we go along the lines of the
first possibility and describe supersymmetry as a nearest option
for the new physics on TeV scale.

\section{What is supersymmetry?  Motivation in particle physics}
 \setcounter{equation} 0

Supersymmetry or fermion-boson symmetry has not yet been observed
in Nature. This is a purely theoretical invention~\cite{super}.
Its validity in particle physics follows from the common belief in
unification. Over 30 years thousands of papers have been written
on supersymmetry. For reviews see, e.g.
Refs.\cite{Rev}-\cite{Weinberg}.

\subsection{Unification with gravity}

 The {\em general idea} is a unification of all forces
of Nature. It defines the {\em strategy} : increasing unification
towards smaller distances up to $l_{Pl} \sim 10^{-33} $ cm
including quantum gravity. However, the graviton has spin 2, while
the other gauge bosons (photon, gluons, $W$ and $Z$ weak bosons)
have spin 1. Therefore, they correspond to different
representations of the Poincar\'e algebra. Attempts to unify all
four forces within the same algebra face a problem. Due to no-go
theorems~\cite{theorem}, unification of spin 2 and spin 1 gauge
fields within a unique algebra is forbidden. The only exception
from this theorem is supersymmetry algebra. The {\em uniqueness}
of SUSY is due to a strict mathematical statement that algebra of
SUSY is the {\em only} graded (i.e. containing anticommutators as
well as commutators) Lie algebra possible within relativistic
field theory~\cite{theorem}.

If $Q$ is a generator of SUSY algebra, then
 $$ Q| boson> = |fermion> \ \ \ {\rm and} \ \ \
 Q| fermion> = | boson> .$$
 Hence, starting with the graviton  state of spin 2 and acting by SUSY
generators we get the following chain of states:
 $$spin \ 2\ \
\rightarrow \ \ spin \ 3/2 \ \ \rightarrow \ \ spin \ 1 \ \
\rightarrow \ \ spin \ 1/2 \ \ \rightarrow \ \ spin \ 0 .$$
 Thus, a partial unification of matter (fermions) with forces (bosons)
naturally arises from an attempt to unify gravity with other
interactions.

SUSY algebra appears as a generalization of Poincar\'e algebra
(see next section) and links together various representations with
different spins. The key relation is given by  the anticommutator
$$\{Q_\alpha, \bar{Q}_{\dot \alpha}\}=2\sigma_{\alpha,\dot
\alpha}^\mu P_\mu .$$
 Taking infinitesimal transformations
$\delta_\epsilon = \epsilon^\alpha Q_\alpha, \ \bar{\delta}_{\bar
\epsilon} = \bar{Q}_{\dot \alpha}{\bar \epsilon}^{\dot \alpha},$
one gets
 \begin{equation}
\{\delta_\epsilon,\bar{\delta}_{\bar \epsilon} \}
 =2(\epsilon \sigma^\mu \bar \epsilon )P_\mu ,
 \label{com}
 \end{equation}
where $\epsilon$ is a transformation parameter. Choosing
$\epsilon$ to be local, i.e. a function of a space-time point
$\epsilon = \epsilon(x)$, one finds from eq.(\ref{com}) that an
anticommutator of two SUSY transformations is a local coordinate
translation. And a theory which is invariant under the general
coordinate transformation is General Relativity. Thus, making SUSY
local, one obtains General Relativity, or a theory of gravity, or
supergravity~\cite{SUGRA}.

Theoretical attractiveness of SUSY field theories is explained by
remarkable properties of SUSY models. This is first of all
cancellation of ultraviolet divergencies in rigid SUSY theories
which is the origin of
 \begin{itemize}
  \item possible solution of the hierarchy problem in GUTs;
  \item  vanishing of the cosmological constant;
  \item  integrability, allowing for an exact non-perturbative solution.
 \end{itemize}
 It is believed that along these lines one can also obtain
the unification of all forces of Nature including quantum
(super)gravity.

What is essential, the standard concepts of QFT allow SUSY without
any further assumptions, it is straightforward to construct the
supersymmetric generalization of the SM. Moreover, it can be
checked experimentally! In recent years, supersymmetry became a
subject of intensive experimental tests. Its predictions can be
verified at modern and future colliders.

\subsection{Unification  of  gauge couplings}

Since the main motivation for SUSY is related with the unification theory,
let us briefly recall the main ideas of the Grand Unification \cite{GUT}.

The philosophy of Grand Unification is based on a {\em
hypothesis}: Gauge symmetry increases with energy. Having in mind
unification of
 all forces of Nature on a common basis and neglecting gravity
for the time being due to its weakness, the idea of GUTs is the
following:

All known interactions are different branches of a unique
interaction associated with a simple gauge group. The unification
 (or splitting) occurs at high energy
$$\begin{array}{ccccl}
 & \mbox{Low energy} & & \Rightarrow & \mbox{High energy} \\ \\
SU_c(3) \otimes & SU_L(2) \otimes & U_Y(1) & \Rightarrow &
 G_{GUT} \ \ (or \ G^n + \mbox{discrete  symmetry}) \\
\mbox{gluons} & W , Z & \mbox{photon} & \Rightarrow & \mbox{ gauge bosons}
\\ \mbox{quarks} & \mbox{leptons} & & \Rightarrow & \mbox{fermions} \\ g_3
& g_2 & g_1 & \Rightarrow & g_{GUT} \end{array} $$

At first sight this is impossible due to a big difference in the
values of the couplings of strong, weak and electromagnetic
interactions. However, this is not so. The crucial point here is the
running coupling constants.
It is a generic property of quantum field theory which has an analogy
in classical physics.

\begin{figure}[t]\vspace{-1cm}
 \begin{center}
 \leavevmode
  \epsfxsize=14cm
 \epsffile{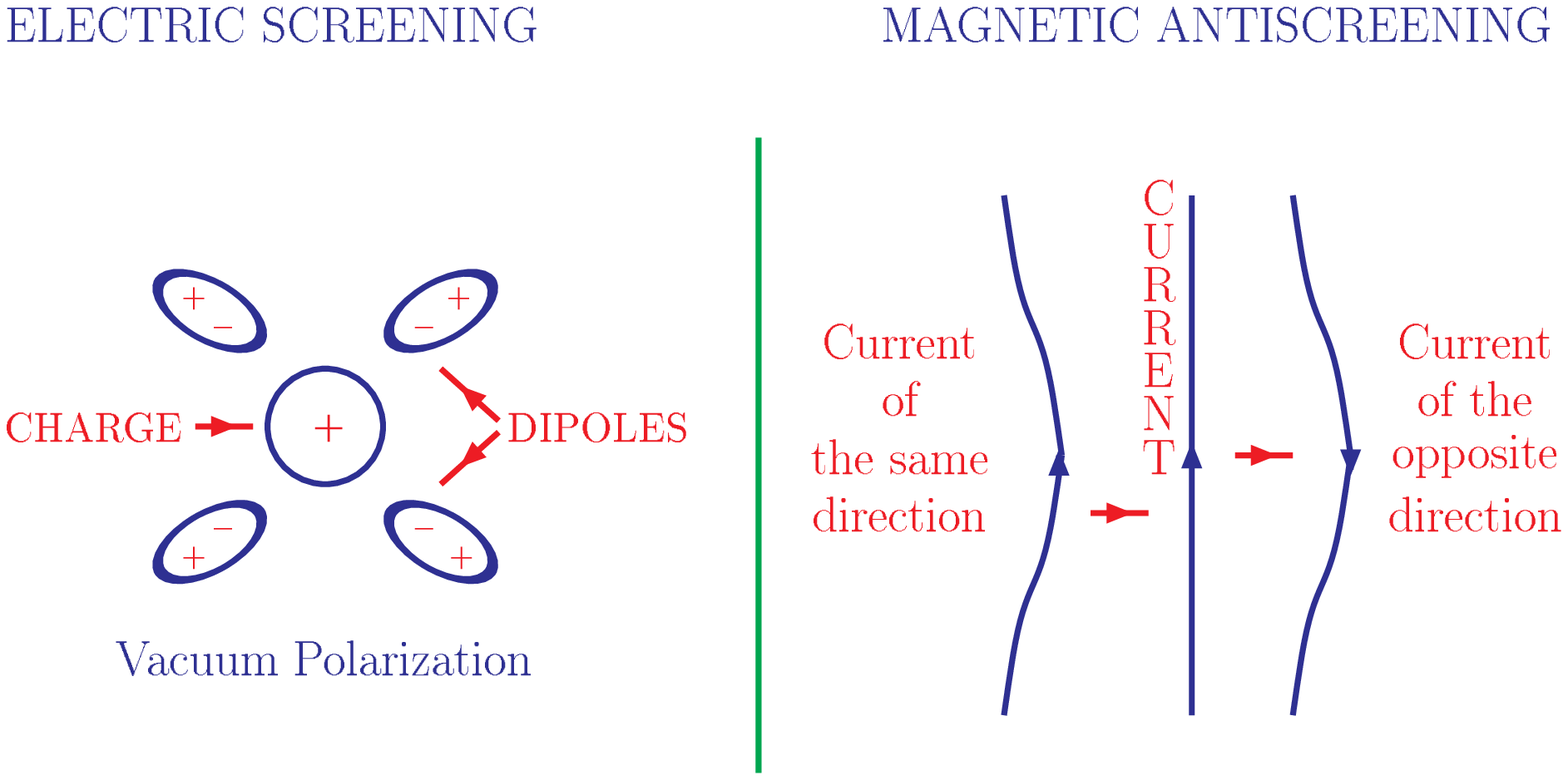}
 \end{center}\vspace{-1cm}
 \caption{Electric screening and magnetic antiscreening}
 \label{fig:run}
 \end{figure}

Indeed, consider electric and magnetic phenomena. Let us take some
dielectric medium and put a sample electric charge in it. What
happens is that  the medium is polarized. It contains  electric
dipoles which are arranged in such a way as  to screen the charge
(see Fig.\ref{fig:run}). It is a consequence of the Coulomb law:
attraction of the opposite charges and repulsion of the same ones.
This is the origin of electric screening.

The opposite situation occurs  in a magnetic medium. According to
the Biot-Savart law, electric currents of the same direction are
attracted to each other, while those of the opposite one are
repulsed (see Fig.\ref{fig:run}). This leads to antiscreening of
electric currents in a magnetic medium.

In QFT, the role of the medium is played by the vacuum. Vacuum is
polarized due to the presence of virtual pairs of particles in it.
The matter fields and transverse quanta of  vector fields in this
case behave like dipoles in a dielectric medium and cause
screening, while the longitudinal quanta of vector fields behave
like currents and cause antiscreening. These two effects compete
each other (see eq.(\ref{regeq1}) below).

\begin{figure}[ht]
 \begin{center}
 \leavevmode
  \epsfxsize=9cm
 \epsffile{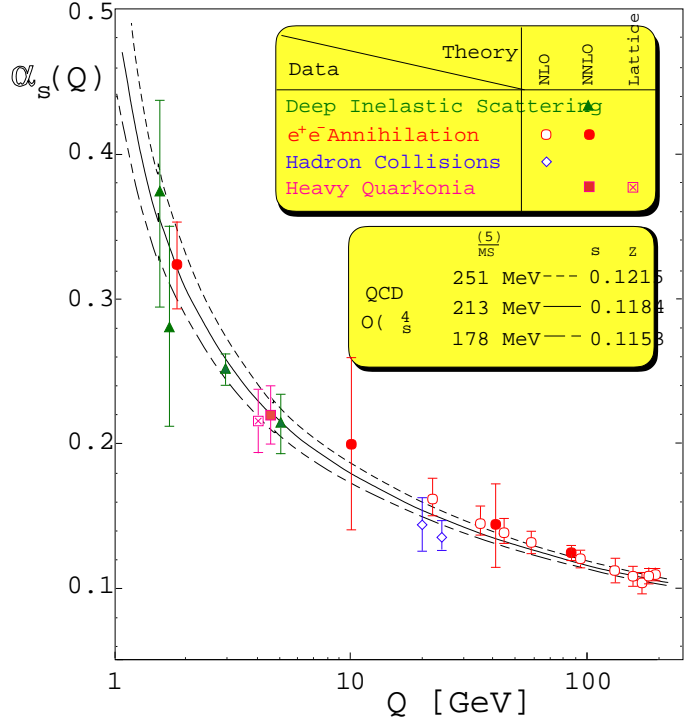}
 \end{center}
 \caption{Summary of running of the strong coupling $\alpha_s$~\cite{bethke}}
 \label{fig:alpha}
 \end{figure}

Thus, the couplings become the functions of a distance or an
energy scale
 $$\alpha_i =
\alpha_i(\frac{Q^2}{\Lambda^2}) = \alpha_i(\mbox{distance}),\ \ \ \ \ \
\alpha_i \equiv g_i^2/4\pi. $$
This dependence is described by the renormalization group equations and is
confirmed experimentally (see Fig.\ref{fig:alpha}).

In the SM the strong and weak couplings associated with
non-Abelian gauge groups decrease with energy, while the
electromagnetic one associated with the Abelian group on the
contrary increases. Thus, it becomes possible that at some energy
scale they become equal. According to the GUT idea, this equality
is not occasional but is a manifestation of a unique origin of
these three interactions. As a result of spontaneous symmetry
breaking, the unifying group is broken and the unique interaction
is splitted into three branches which we call strong, weak
 and electromagnetic interactions. This happens at a very
high energy of an order of $10^{15\div 16}$ GeV. Of course, this
energy is out of the range of accelerators; however, some
 crucial predictions follow from the very fact of unification.

After the precise measurement of the $SU(3)\times SU(2) \times
U(1)$ coupling constants, it has become possible to check the
unification numerically.

The three coupling constants to be compared are
 \begin{eqnarray}
\alpha_1&=&(5/3)g^{\prime2}/(4\pi)=5\alpha/(3\cos^2\theta_W),
\nonumber \\ \alpha_2&=& g^2/(4\pi)=\alpha/\sin^2\theta_W, \\
\alpha_3&=& g_s^2/(4\pi) \nonumber
 \end{eqnarray}
where $g',~g$ and $g_s$ are the usual $U(1)$, $SU(2)$ and $SU(3)$
coupling constants and $\alpha$ is the fine structure constant.
The factor of 5/3 in the definition of $\alpha_1$ has been
included for  proper normalization of the generators.

The couplings, when defined as renormalized values including loop
corrections require the specification of a renormalization
prescription for which the modified minimal subtraction
($\overline{MS}$) scheme~\cite{msbar} is used.

In this scheme, the world averaged values of the coup\-lings at
the Z$^0$ energy are obtained from a fit to the LEP and Tevatron
data~\cite{fine},\cite{test-sm},\cite{bethke}:
 \begin{eqnarray}
  \label{worave}
  \alpha^{-1}(M_Z)             & = & 128.978\pm 0.027   \nonumber\\
  \sin^2\theta_{\overline{MS}} & = & 0.23146\pm 0.00017\\
  \alpha_s                     & = & 0.1184\pm 0.0031, \nonumber
 \end{eqnarray}
that gives
 \begin{equation} \alpha_1(M_Z)=0.017 ,\ \ \alpha_2(M_Z)=0.034, \
\ \alpha_3(M_Z)=0.118\pm 0.003.
 \end{equation}
 Assuming that the SM is valid up to the unification
scale, one can then use the known RG equations for the three
couplings. They are the following:
 \begin{equation}
\frac{d\tilde{\alpha}_i}{dt} =  b_i\tilde{\alpha}_i^2, \ \ \ \
\tilde{\alpha}_i=\frac{\alpha_i}{4\pi}, \ \ \ \ \
t=log(\frac{Q^2}{\mu^2}), \label{alpha}
 \end{equation}
where for the SM the coefficients $b_i$ are
 \begin{equation}
b_i=\left( \begin{array}{r} b_1 \\ b_2 \\b_3 \end{array} \right)
   =
\left( \begin{array}{r}           0    \\
                             - 22 / 3  \\
                                -11    \end{array} \right) +N_{Fam}
\left( \begin{array}{r}         4 / 3  \\
                                4 / 3  \\
                                4 / 3  \end{array} \right) + N_{Higgs}
\left( \begin{array}{r}         1 / 10 \\
                                1 / 6  \\
                                  0    \end{array} \right) .
 \label{regeq1}
 \end{equation}
Here  $N_{Fam}$ is the number of generations of matter multiplets
and $N_{Higgs}$ is the number of Higgs doublets. We use
$N_{Fam}=3$ and $N_{Higgs}=1$ for the minimal SM, which gives
$b_i=(41/10, -19/6, -7)$.

Notice a positive contribution (screening) from the matter
multiplets and negative one (antiscreening) from the gauge fields.
For the Abelian group $U(1)$ this contribution is absent due to
the absence of a self-interaction of Abelian gauge fields.

\begin{figure}[t]
 \begin{center}
  \leavevmode
  \epsfxsize=15cm
 \epsffile{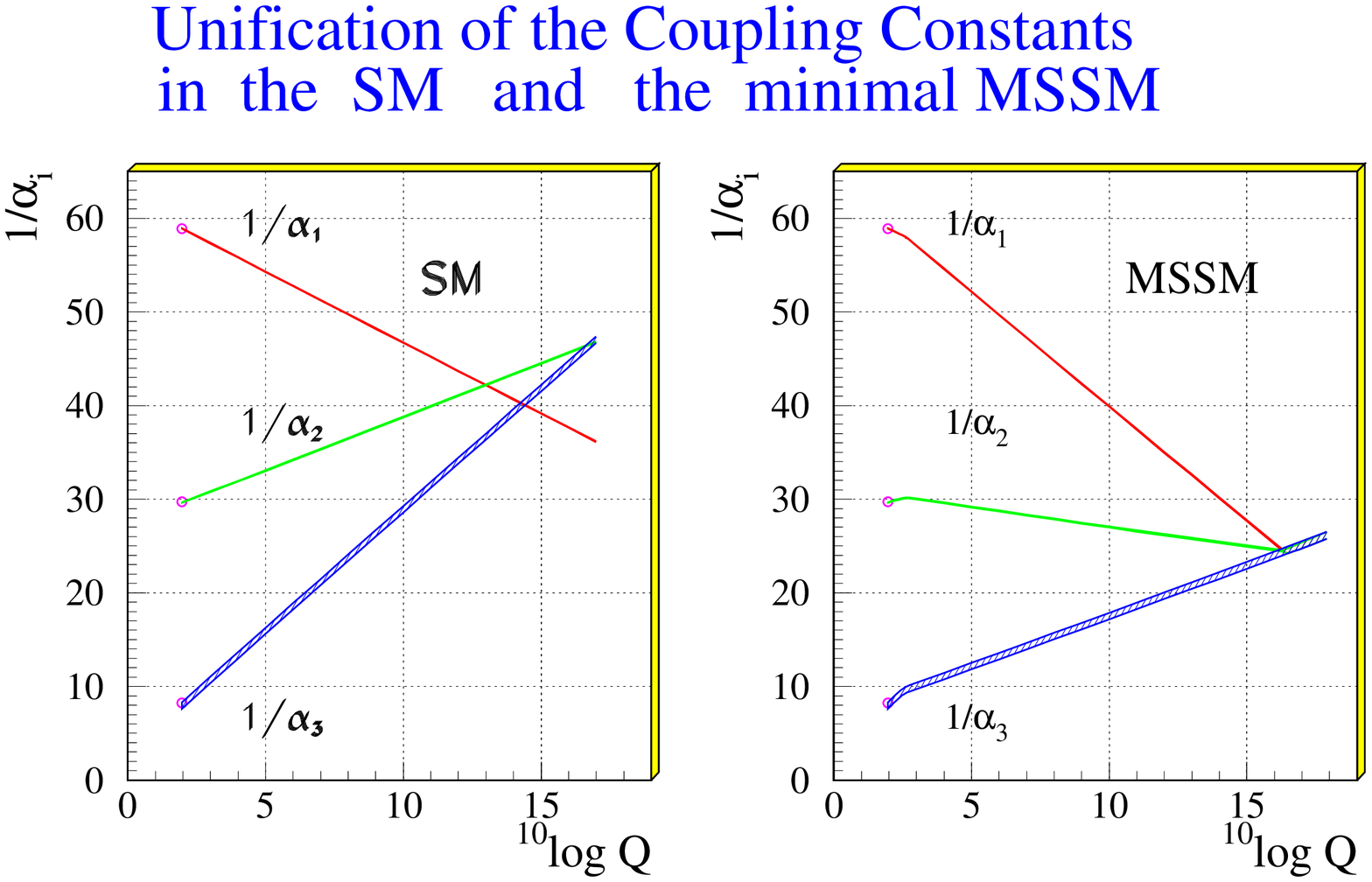}
 \end{center}\vspace{-1cm}
 \caption{Evolution of the inverse of the three coupling constants
in the Standard Model (left) and in the supersymmetric extension
of the SM (MSSM) (right). Only in the latter case unification is
obtained. The SUSY particles are assumed to contribute only above
the effective SUSY scale $M_{SUSY}$ of about  1 TeV, which causes
a change in the slope in the evolution of couplings. The thickness
of the lines represents the error in the coupling constants
\cite{ABF}.}\label{unif}
 \end{figure}

The solution to eq.(\ref{alpha}) is very simple
 \begin{equation}
\frac{1}{\tilde{\alpha}_i(Q^2)} = \frac{1}{\tilde{\alpha}_i(\mu^2)}-
 b_i log(\frac{Q^2}{\mu^2}). \label{alphasol}
 \end{equation}
The result is demonstrated in  Fig.\ref{unif} showing the
evolution of the inverse of the couplings as a function of the
logarithm of energy. In this presentation, the evolution becomes a
straight line in first order. The second order corrections are
small and do not cause any visible deviation from a straight line.
Fig.\ref{unif} clearly demonstrates that within the SM the
coupling constant unification at a single point is impossible. It
is excluded by more than 8 standard deviations. This result means
that the unification can only be obtained if new physics enters
between the electroweak and the Planck scales!

Since we do not know what kind of  new physics it may be, there is
a lot of arbitrariness. In this situation, some guiding idea is
needed. It is very tempting to try to check whether unification is
possible within a supersymmetric generalization of the SM. In the
SUSY case, the slopes of the RG evolution curves are modified. The
coefficients $b_i$ in eq.(\ref{alpha}) now are
 \begin{equation}
b_i=\left( \begin{array}{r} b_1 \\ b_2 \\b_3 \end{array} \right)
   =
\left( \begin{array}{r}          0     \\
                                -6     \\
                                -9      \end{array} \right) +N_{Fam}
\left( \begin{array}{r}          2     \\
                                 2     \\
                                 2      \end{array} \right) +N_{Higgs}
\left( \begin{array}{r}          3/10  \\
                                 1/2   \\
                                 0      \end{array} \right)     ,
 \label{susy1}
 \end{equation}
where we use  $N_{Fam}=3$ and $N_{Higgs}=2$ in the minimal SUSY
model which gives $b_i=(33/5, 1, -3)$.

\begin{figure}[htb]\vspace{-0.3cm}
 \begin{center}
 \epsfysize=6cm
 \epsfxsize=12cm
\epsffile{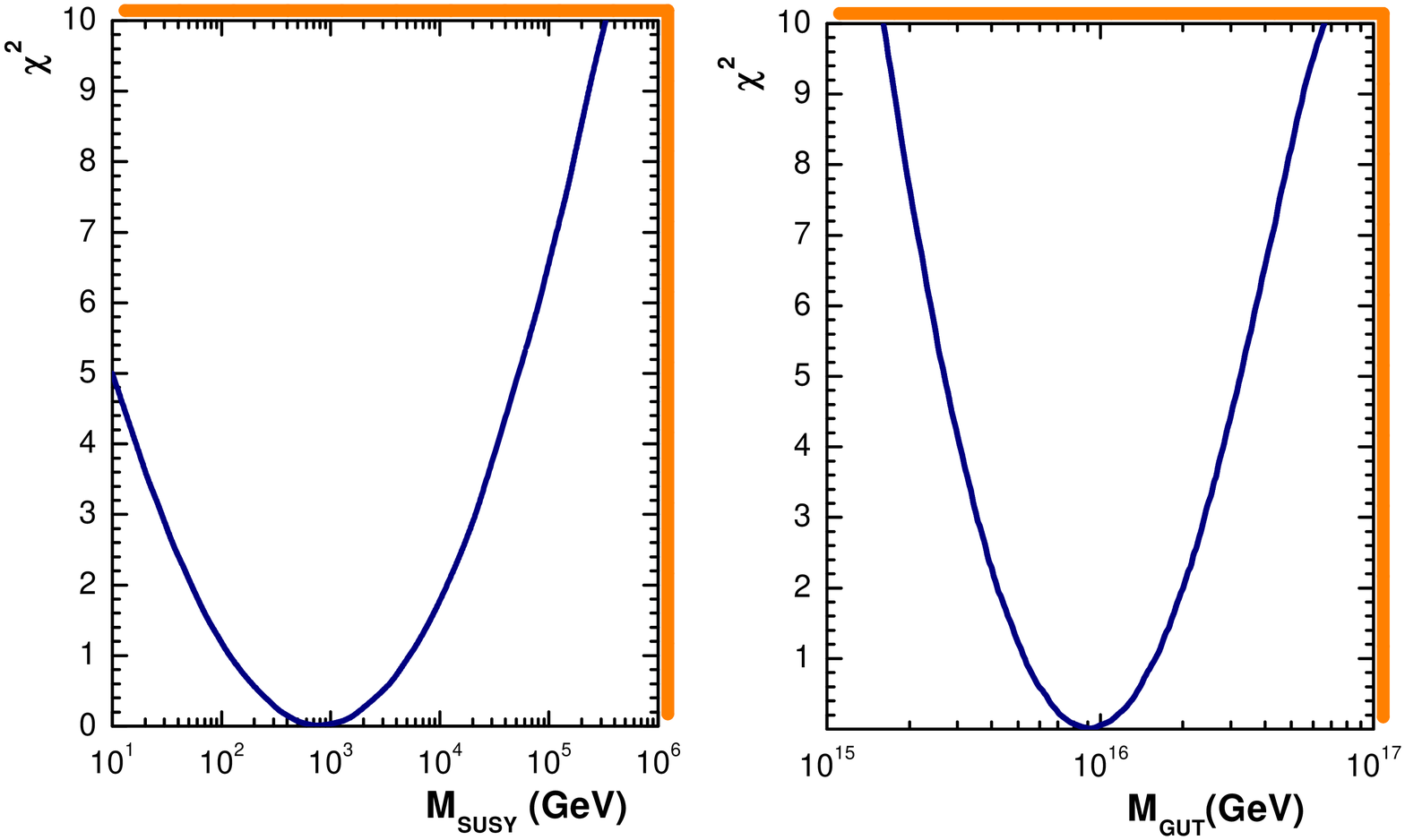} \vspace{-0.5cm}
 \caption{The $\chi^2$ distributions of $M_{SUSY}$ and
$M_{GUT}$}\label{chisusy}
 \end{center}
 \end{figure}

It turns out that within the SUSY model a perfect unification can
be obtained if the SUSY  masses are of an order of 1 TeV. This is
shown in Fig.\ref{chisusy}; the SUSY particles are assumed to
effectively contribute  to the running of the coupling constants
only for energies above the typical SUSY mass scale, which causes
the change in the slope of the lines near 1 TeV. From the fit
requiring unification one finds for the break point $M_{SUSY}$ and
the unification point $M_{GUT}$~\cite{ABF}
 \begin{eqnarray}
M_{SUSY} &= & 10^{3.4\pm 0.9\pm 0.4} \ GeV , \nonumber\\ M_{GUT}
&= & 10^{15.8\pm 0.3\pm 0.1} \ GeV , \label{MSUSY}\\
\alpha^{-1}_{GUT} &= & 26.3 \pm 1.9 \pm 1.0 , \nonumber
 \end{eqnarray}
where $\alpha_{GUT}=g_5^2/4\pi$. The first error originates from
the uncertainty in the coupling constant, while the second one is
due to the uncertainty in the mass splittings between the SUSY
particles.
%
The $\chi^2$ distributions of $M_{SUSY}$ and $M_{GUT}$ are shown
in Fig.\ref{chisusy}~\cite{ABF}, where
 \begin{equation}
\chi^2=\sum_{i=1}^{3}\frac{(\alpha^{-1}_i-\alpha^{-1}_{GUT}
)^2}{\sigma_i^2}.
 \end{equation}

For SUSY models, the dimensional reduction $\overline{DR}$ scheme
is a more appropriate renormalization scheme~\cite{akt}.  In this
scheme, all thresholds are treated by simple step approximations,
and unification occurs if all three $\alpha$'s meet exactly at one
point. This crossing point corresponds to the mass of the heavy
gauge bosons. The $\overline{MS}$ and $\overline{DR}$ couplings
differ by a small offset
 \begin{equation}
{{1\over\alpha_i^{\overline{DR}}}=
{1\over\alpha_i^{\overline{MS}}}-{C_i\over\strut12\pi},}
 \end{equation}
where  $C_i$ are the quadratic Casimir operators of the group
($C_i=N$ for SU($N$) and 0 for U(1) so $\alpha_1$ remains the
same).

This observation was considered as the first "evidence" for
supersymmetry, especially since $M_{SUSY}$  was found in the range
preferred by the fine-tuning arguments.

It should be noted that the unification of the three curves at a
single point is not that trivial as it may seem from the existence
of  three free parameters ($ M_{SUSY}, M_{GUT}$ and
$\alpha_{GUT}$). Out of more than a thousand models tried, only a
handful yielded unification. The reason is simple: Introducing new
particles one influences all three curves simultaneously, thus
giving rise to strong correlations between the slopes of the three
lines. For example, adding new generations and/or new Higgs
doublets never yields unification! Nevertheless, unification does
not prove supersymmetry. The real proof would be the observation
of the sparticles.

\subsection{Solution of the hierarchy problem}

The appearance of two different scales $V \gg v$ in a GUT theory,
namely, $M_W$ and $M_{GUT}$, leads to a very  serious problem
which is called the {\em hierarchy problem}. There are two aspects
of this problem.

The first one is the very existence of the hierarchy. To get the
desired spontaneous symmetry breaking pattern, one needs
 \begin{equation}
 \begin{array}{ccccc}
m_H & \sim & v & \sim & 10^2 \ \ \mbox{GeV} \\ m_{\Sigma} & \sim &
V & \sim & 10^{16} \ \ \mbox{GeV}
 \end{array} \ \ \ \ \ \
\frac{m_H}{m_{\Sigma}} \ \sim \ 10^{-14} \ \ \ll 1 , \label{hier}
 \end{equation}
where $H$ and $\Sigma$ are the Higgs fields responsible for the
spontaneous breaking of the $SU(2)$ and the GUT groups,
respectively.

The question arises of how to get so small number in a natural
way. One needs some kind of fine tuning in a theory, and we don't
know if
 there anything behind it.

\begin{figure}[ht]\vspace{-1cm}
 \begin{center}
 \leavevmode
  \epsfxsize=10cm
 \epsffile{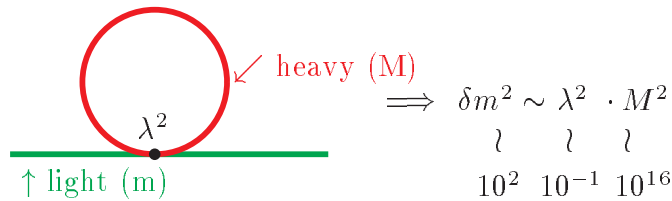}
 \end{center}\vspace{-1cm}
 \caption{Radiative correction to the light Higgs boson mass}\label{fig:hierar}
 \end{figure}

The second aspect of the hierarchy problem is connected with the
preservation of a given hierarchy. Even if we choose the hierarchy
like eq.(\ref{hier}) the radiative corrections will destroy it! To
see this, consider the radiative correction to the light Higgs
mass. It is given by the  Feynman diagram shown in
Fig.\ref{fig:hierar}
\noindent and is proportional to the mass squared of the heavy
particle. This correction obviously spoils the hierarchy if it is
not cancelled. This very accurate cancellation with a precision
$\sim 10^{-14}$ needs a fine tuning of the coupling constants.

\begin{figure}[ht]
 \begin{center}
 \leavevmode
  \epsfxsize=10cm
 \epsffile{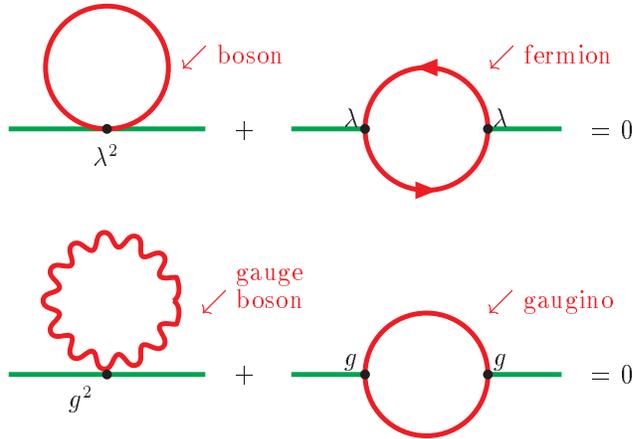}
 \end{center}\vspace{-1cm}
 \caption{Cancellation of quadratic terms (divergencies)}\label{fig:cancel}
 \end{figure}

The only known way of achieving this kind of cancellation of
quadratic terms (also known as the cancellation of the quadratic
divergencies) is supersymmetry. Moreover, SUSY automatically
cancels quadratic corrections in all orders of PT. This is due to
the contributions of superpartners of  ordinary particles. The
contribution from boson loops cancels those from the fermion ones
because of an additional factor (-1) coming from Fermi statistics,
as shown in Fig.\ref{fig:cancel}.
One can see here  two types of contribution. The first line is the
contribution of the heavy Higgs boson and its superpartner. The
strength of interaction is given by the Yukawa coupling $\lambda$.
The second line represents the gauge interaction proportional to
the gauge coupling constant $g$ with the contribution from the
heavy gauge boson and heavy gaugino.

In both the cases the cancellation of quadratic terms takes place.
This cancellation is true in the case of unbroken supersymmetry
due to the following sum rule relating the masses of superpartners
\begin{equation}
 \sum_{bosons} m^2 = \sum_{fermions} m^2
\end{equation}
and is violated when SUSY is broken. Then, the cancellation is
true up to the SUSY breaking scale, $M_{SUSY}$, since
\begin{equation}
 \sum_{bosons} m^2 - \sum_{fermions} m^2= M_{SUSY}^2,
\end{equation}
which  should not be very large ($\leq$ 1 TeV) to make the
fine-tuning natural. Indeed, let us take the Higgs boson mass.
Requiring for consistency of perturbation theory that the
radiative corrections to the Higgs  boson mass do not  exceed the
mass itself gives
 \begin{equation}
\delta M_h^2 \sim g^2 M^2_{SUSY} \sim M_h^2.\label{del}
 \end{equation}
 So, if $M_h \sim 10^2$ GeV and $g \sim 10^{-1}$, one
needs $M_{SUSY} \sim 10^3$ GeV in order that the relation
(\ref{del}) is valid. Thus, we again get the same rough estimate
of $M_{SUSY} \sim $ 1 TeV as from the gauge coupling unification
above. Two requirements match together.

 That is why it is usually said that supersymmetry solves the
hierarchy problem. Moreover, sometimes it is said that: "There is
no GUT without SUSY". However, this is only the second aspect of
the problem, the preservation of the hierarchy.  The origin of the
hierarchy is  the other part of the problem. We show below how SUSY can
 explain this part as well.

\subsection{Beyond GUTs: superstring}

Another motivation for supersymmetry follows from even more
radical changes of basic ideas related to the ultimate goal of
construction of consistent unified theory of everything. At the
moment the only viable conception is the superstring
theory~\cite{string}, which pretends to be a self-consistent
quantum field theory in a non-perturbative sense allowing exact
non-perturbative solutions in the quantum case. In the superstring
theory, strings are considered as fundamental objects, closed or
open, and are nonlocal in nature. Ordinary particles are
considered as string excitation modes. String interactions, which
are local, generate  proper interactions of usual particles,
including gravitational ones.

To be consistent, the string theory should be conformal invariant
in D-dimensional target space and have a stable
vacuum~\cite{Polyakov}. The first requirement is valid in
classical theory but may be violated by quantum anomalies.
Cancellation of quantum anomalies takes place when space-time
dimension of a target space equals  a critical one. For  a bosonic
string  the critical dimension is $D=26$, and for a fermionic one
it is $D=10$.

The second requirement is that the massless  string excitations
(the particles of the SM)  are stable. This assumes the absence of
tachyons, the states with imaginary mass, which can be guaranteed
only in supersymmetric string theories!

Thus, the superstring theory proves to be the only known
consistent quantum theory. This serves as  justification of
research in spite of absence of even a shred of experimental
evidence. However, many ingredients of this theory are still
unclear.

\section{Basics of supersymmetry}
 \setcounter{equation} 0

Supersymmetry trans\-for\-ma\-tions differ from ordinary global
transformations as far as they convert bosons into fermions and
vice versa. Indeed, if we symbolically write SUSY transformation
as
 $$ \delta B = \varepsilon \cdot f,$$
 where $B$ and $f$ are boson and fermion fields, respectively, and $\varepsilon $ is an
infinitesimal trans\-for\-mation parameter, then from the usual
(anti)commutation relations for (fermions) bosons
 $$ \{f,f\} = 0, \ \ \ \ [B,B] = 0  $$
we immediately find
 $$\{ \varepsilon , \varepsilon \} = 0. $$
This means that all the generators of SUSY must be {\em fermionic}, i.e. they
must change the spin by a half-odd amount and change the statistics.

\subsection{Algebra of SUSY}

Combined with the usual Poincar\'e and internal symmetry algebra
the Super-Poincar\'e Lie algebra contains additional SUSY
generators $Q_{\alpha}^i$ and $\bar Q_{\dot
\alpha}^i$~\cite{super}
 \begin{equation}  \begin{array}{l}
{[} P_{\mu},P_{\nu}{]}  =   0,  \\ {[} P_{\mu},M_{\rho\sigma}{]} =
i(g_{\mu \rho}P_{\sigma} -
 g_{\mu \sigma}P_{\rho}),  \\
{[} M_{\mu \nu} , M_{\rho \sigma} {]}   =    i(g_{\nu \rho}M_{\mu
\sigma} - g_{\nu \sigma}M_{\mu \rho} - g_{\mu \rho}M_{\nu \sigma}
+ g_{ \mu \sigma}M_{\nu \rho}), \\
 {[} B_r , B_s {]}    =    iC_{rs}^t B_{t}, \\
 {[} B_r , P_{\mu} {]}    =    {[} B_r ,M_{\mu \sigma} {]} = 0,  \\
 {[} Q_{\alpha}^i , P_{\mu} {]} =  {[} \bar Q_{\dot \alpha}^i , P_{\mu} {]} = 0, \\
 {[} Q_{\alpha}^i , M_{\mu \nu} {]}  =   \frac{1}{2} (\sigma_{\mu \nu})_{\alpha}
^{\beta}Q_{\beta}^i , \ \ \ \ {[}\bar Q_{\dot \alpha}^i ,M_{\mu
\nu}{]} = - \frac{1}{2} \bar Q_{\dot \beta}^i (\bar \sigma_{\mu
\nu})_{\dot \alpha} ^{\dot \beta} , \\
 {[} Q_{\alpha}^i , B_r {]}  =  (b_r)_{j}^i Q_{\alpha}^j , \ \ \ {[}
 \bar Q_{\dot \alpha}^i , B_r {]} = - \bar Q_{\dot \alpha}^j (b_r)_j^i , \\
 \{ Q_{\alpha}^i , \bar Q_{\dot \beta}^j \}
  =   2 \delta^{ij} (\sigma ^{\mu})_{\alpha \dot \beta }P_{\mu} , \\
 \{ Q_{\alpha}^i , Q_{\beta}^j \}  =   2 \epsilon_{\alpha \beta}Z^{ij} ,
\ \ \ Z_{ij} = a_{ij}^r b_r , \ \ \ \ Z^{ij} = Z_{ij}^+ , \\
 \{ \bar Q_{\dot \alpha}^i , \bar Q_{\dot \beta}^j \}  =  - 2 \epsilon
_{\dot \alpha \dot \beta}Z^{ij} , \ \ \ {[}Z_{ij} , anything {]} =
0 , \\
 \alpha , \dot \alpha  =  1,2 \ \ \ \ i,j = 1,2, \ldots , N .
 \end{array}  \label{group}
 \end{equation}

Here $P_{\mu}$ and $M_{\mu \nu}$  are four-momentum and angular
momentum operators, respectively, $B_r$ are the internal symmetry
generators, $Q^i$ and $\bar Q^i$ are the spinorial SUSY generators
and $Z_{ij}$ are the so-called central charges; $\alpha , \dot
\alpha, \beta , \dot \beta $ are the spinorial indices. In the
simplest case one has one spinor generator $Q_\alpha$ (and the
conjugated one $\bar Q_{\dot{\alpha}}$) that corresponds to an
ordinary or N=1 supersymmetry. When $N>1$ one has an extended
supersymmetry.

A natural question arises: how many SUSY generators are possible,
i.e. what is the value of $N$? To answer this question, consider
massless states~\cite{WessB}. Let us start with the ground state
labeled by energy and helicity, i.e. projection of a spin on the
direction of momenta, and let it be annihilated by $Q_i$
 $${\rm Vacuum} =|E,\lambda >, \ \ \ \ \ \ \  Q_i|E,\lambda >=0. $$
Then  one and more  particle states can be constructed with the
help of a creation operators as $$\begin{array}{lll}
\mbox{\underline{State}} & \mbox{\underline{Expression}}  & \# \
\mbox{ \underline{of States}} \\  \\ \mbox{vacuum} & |E,\lambda >
& \ \ \ \  1 \\
 1-\mbox{particle  state} \ \ \ & \bar Q_i|E,\lambda>= |E,\lambda +1/2>_i
& \left(\begin{array}{c} N\\ 1 \end{array}\right)=N \\
 2-\mbox{particle  state} & \bar Q_i\bar Q_j|E,\lambda>= |E,\lambda +1>_{ij}
& \left(\begin{array}{c} N\\ 2 \end{array}\right)= \frac{N(N-1)}{2} \\ ... &
... & ... \\
 N-\mbox{particle  state} & \bar Q_1\bar Q_2 ... \bar Q_N|E,\lambda>= |E,\lambda +N/2>
& \left(\begin{array}{c} N\\ N \end{array}\right)= 1\\ \\ \mbox{Total \#
of States} &\displaystyle \sum_{k=0}^{N}\left(\begin{array}{c} N\\ k
\end{array}\right) =2^N=2^{N-1}\ \mbox{bosons} + 2^{N-1}\
\mbox{fermions},& \end{array} $$
 where the energy $E$ is not changed, since according to (\ref{group})
 the operators $ \bar Q_i$ commute with the Hamiltonian.

Thus, one has a sequence of bosonic and fermionic states and the
total number of bosons equals  that of fermions. This is a generic
property of any supersymmetric theory. However, in CPT invariant
theories the number of states is doubled, since CPT transformation
changes the sign of helicity. Hence, in CPT invariant theories,
one has to add the states with opposite helicity to the above
mentioned ones.

Consider some examples. Let us take $N=1$ and $\lambda= 0$. Then
one has the following set of states:

 $$\begin{array}{lccccccc} &
\mbox{helicity}&0&1/2& & \mbox{helicity}&0&-1/2 \\
 N=1 \ \ \  \lambda =0 \ \ \  & & & & \stackrel{CPT}{\Longrightarrow} &&  &\\
 & \# \ \mbox{of states}& 1& 1&&\# \ \mbox{of states} &1& 1
\end{array}$$
Hence, a complete $N=1$ multiplet is
 $$ \begin{array}{llccc} N=1 \ \ &
\mbox{helicity}&-1/2&0&1/2  \\ &
 \# \ \mbox{of states}&\ 1& 2&1
\end{array}$$ which contains one complex scalar and
one spinor with two helicity states.

This is an example of the so-called self-conjugated multiplet.
There are also self-conjugated multiplets with $N>1$ corresponding
to extended supersymmetry. Two particular examples are the $N=4$
super Yang-Mills multiplet and the $N=8$ supergravity multiplet
$$\begin{array}{llcccccccccc} N=4 & \mbox{SUSY YM} &
\mbox{helicity}&& &-1&-1/2&0&1/2&1 &&  \\
 & \lambda=-1 &\# \ \mbox{of states}&&&\ \ 1&\ 4&6 & 4 & 1 &&\\ &&&&&&&&& \\
N=8 & \mbox{SUGRA} &\mbox{helicity}&-2&-3/2&-1&-1/2&0&\ 1/2&\ 1&\ 3/2&\ 2
\\ & \lambda=-2 &\# \ \mbox{of states}&\ \ 1&\ 8&28 &\ \ 56 & 70 &\ 56&\
28& \ \ 8&\ 1 \end{array} $$ One can see that the multiplets of
extended supersymmetry are very rich and contain a vast number of
particles.

 The constraint on the number of SUSY generators comes from a requirement of consistency of
the corresponding QFT.  The number of supersymmetries and the maximal spin of
the particle in the multiplet are related by
 $$N \leq 4S,$$
where $S$ is the maximal spin. Since the theories with spin
greater than 1 are non-renormalizable and the theories with spin
greater than 5/2 have no consistent coupling to gravity, this
imposes a constraint on the number of SUSY generators
$$\begin{array}{cl} N \leq 4 & \ \ \ \ \mbox{for renormalizable
theories (YM),} \\ N \leq 8 & \ \ \ \ \mbox{for (super)gravity}.
\end{array}$$
 In what follows, we shall consider simple supersymmetry, or $N=1$
supersymmetry, contrary to extended supersymmetries with $N > 1$.
In this case, one has two types of supermultiplets: the so-called
chiral multiplet with $\lambda =0$, which contains two physical
states $(\phi,\psi)$ with spin 0 and 1/2, respectively, and the
vector multiplet with $\lambda =1/2$, which also contains two
physical states $(\lambda, A_\mu)$ with spin 1/2 and 1,
respectively.

\subsection{Superspace and superfields}

An elegant formulation of supersymmetry transformations and
invariants can be achieved in the framework of superspace
\cite{sspace}.
 Superspace differs from the ordinary Euclidean (Minkowski)
space by adding of two new coordinates, $\theta_{\alpha}$ and
$\bar \theta_{\dot \alpha}$, which are Grassmannian,
 i.e. anti\-com\-muting, variables
 $$\{ \theta_{\alpha}, \theta_{\beta} \} = 0 , \ \ \{\bar
\theta_{\dot \alpha}, \bar \theta_{\dot \beta} \} = 0, \ \
\theta_{\alpha}^2 = 0,\ \ \bar \theta_{\dot \alpha}^2=0, \ \
\alpha,\beta, \dot\alpha, \dot\beta =1,2.$$
 Thus, we go from space to superspace
$$\begin{array}{cc} Space & \ \Rightarrow \ \ Superspace \\
x_{\mu} & \ \ \ \ \ \ \ x_{\mu}, \theta_{\alpha} , \bar \theta_
{\dot \alpha} \end{array}$$
 A SUSY group element can be constructed in
superspace in the same way as an ordinary translation in the usual
space
\begin{equation}
 G(x,\theta ,\bar \theta ) = e^{\displaystyle
i(-x^{\mu}P_{\mu} + \theta Q + \bar \theta \bar Q)}.\label{st}
\end{equation}
 It leads to a supertranslation in superspace
 \begin{equation}
 \begin{array}{ccl}
x_{\mu} & \rightarrow & x_{\mu} + i\theta \sigma_{\mu} \bar
\varepsilon
 - i\varepsilon \sigma_{\mu} \bar \theta, \\
\theta & \rightarrow & \theta + \varepsilon ,\label{sutr} \\ \bar
\theta & \rightarrow & \bar \theta + \bar \varepsilon ,
\end{array}
 \end{equation}
where $\varepsilon $ and $\bar \varepsilon $ are Grassmannian
transformation parameters. From eq.(\ref{sutr}) one can easily
obtain the representation for the supercharges (\ref{group})
acting on the superspace
 \begin{equation}
Q_\alpha =\frac{\partial }{\partial \theta_\alpha
}-i\sigma^\mu_{\alpha \dot \alpha}\bar{\theta}^{\dot \alpha
}\partial_\mu , \ \ \ \bar{Q}_{\dot \alpha} =-\frac{\partial
}{\partial \bar{\theta}_{\dot \alpha}
}+i\theta_\alpha\sigma^\mu_{\alpha \dot \alpha}\partial_\mu .
\label{q}
 \end{equation}
Taking the  Grassmannian transformation parameters  to be local,
or space-time dependent, one gets a local translation. As has
already been mentioned, this leads to a theory of (super) gravity.

To define the fields  on a superspace, consider  representations
of the Super-Poincar\'e group (\ref{group}) \cite{WessB}. The
simplest one is a scalar superfield $F(x,\theta , \bar \theta )$
which is SUSY invariant. Its Taylor expansion in $\theta$ and
$\bar \theta $ has only several terms due to the nilpotent
character of Grassmannian parameters. However, this superfield  is
a reducible representation of SUSY. To get an irreducible one, we
define a {\em chiral} superfield which obeys the equation
 \begin{equation}
\bar D F = 0 , \ \ \ \ \mbox{where} \ \bar D = -\frac{\partial}{
\partial \overline{ \theta}} - i \theta \sigma^{\mu}
\partial_{\mu} \label{k}
 \end{equation}
 is a superspace covariant derivative.

For the chiral superfield Grassmannian Taylor expansion looks like
($y =x + i\theta \sigma \bar \theta $)
 \begin{eqnarray}
\Phi (y, \theta ) & = & A(y) + \sqrt{2} \theta \psi (y) + \theta
 \theta F(y) \nonumber \\
  & = & A(x) + i\theta \sigma^{\mu} \bar \theta \partial_{\mu}A(x)
 + \frac{1}{4} \theta \theta \bar \theta \bar \theta \Box A(x) \nonumber\\
  & + & \sqrt{2} \theta \psi (x) - \frac{i}{\sqrt{2}} \theta
\theta \partial_{\mu} \psi (x) \sigma^{\mu} \bar \theta + \theta
 \theta F(x).\label{field}
 \end{eqnarray}
The coefficients are ordinary functions of $x$ being the usual
fields. They are called the {\em components} of a superfield. In
eq.(\ref{field}) one has 2 bosonic (complex scalar field $A$) and
2 fermionic (Weyl spinor field $\psi$) degrees of freedom. The
component fields $A$ and $\psi$ are called the {\em
superpartners}. The field $F$ is  an {\em auxiliary} field, it has
the ``wrong'' dimension and has no physical meaning. It is needed
to close the algebra (\ref{group}). One can get rid of the
auxiliary fields with the help of equations of motion.

Thus, a superfield contains an equal number of bosonic and
fermionic degrees of freedom. Under SUSY transformation they
convert  into one another
 \begin{eqnarray}
\delta_\varepsilon A &=& \sqrt 2 \varepsilon \psi, \nonumber \\
\delta_\varepsilon \psi &=& i \sqrt 2 \sigma^\mu \bar \varepsilon
\partial_\mu A + \sqrt 2 \varepsilon F, \\
\delta_\varepsilon F &=&i \sqrt 2  \bar \varepsilon\sigma^\mu\partial_\mu\psi.
\nonumber
 \end{eqnarray}
Notice that the variation of the $F$-component is a total derivative, i.e.
it vanishes when integrated over the space-time.

One can also construct an antichiral superfield $\Phi^+$ obeying
the equation
$$D\Phi^+ = 0, \ \ \ \mbox{with}  \ \
 D = \frac{\partial}{\partial \theta} + i \sigma^{\mu}\bar \theta
\partial_{\mu}. $$
The product of chiral (antichiral)  superfields $\Phi^2 , \Phi^3
$, etc is also a chiral (antichiral) superfield, while the product
of chiral and antichiral ones $\Phi^+ \Phi$ is a general
superfield.

For  any arbitrary function of chiral superfields one has
 \begin{eqnarray}
{\cal W}(\Phi_i)&=&{\cal W}(A_i+\sqrt{2}\theta \psi_i+\theta
\theta F) \nonumber \\ &=&{\cal W}(A_i)+\frac{\partial {\cal
W}}{\partial A_i}\sqrt{2}\theta \psi_i +\theta \theta
\left(\frac{\partial {\cal W}}{\partial A_i}F_i -
\frac{1}{2}\frac{\partial^2 {\cal W}}{\partial A_i\partial
A_j}\psi_i\psi_j \right). \label{sp}
 \end{eqnarray}
The ${\cal W}$ is usually referred to as a superpotential which
replaces the usual potential for the scalar fields.

To construct the gauge invariant interactions, one needs a real
vector superfield $V = V^+$. It is not chiral but rather a general
superfield with the following Grassmannian expansion:
 \begin{eqnarray}
V(x, \theta, \bar \theta) & = & C(x) + i\theta \chi (x) -i\bar
\theta \bar \chi (x)  \nonumber \\
 & + & \frac{i}{2} \theta \theta [M(x) + iN(x)] - \frac{i}{2} \bar
 \theta \bar \theta [M(x) - iN(x)]  \nonumber \\
 & - & \theta \sigma^{\mu} \bar \theta v_{\mu}(x) + i \theta \theta
\bar \theta [\lambda (x) + \frac{i}{2}\bar \sigma^{\mu} \partial
_{\mu} \chi (x)]  \nonumber \\
 & - & i\bar \theta \bar \theta \theta [\lambda + \frac{i}{2}
\sigma^{\mu} \partial_{\mu} \bar \chi (x)] + \frac{1}{2} \theta
\theta \bar \theta \bar \theta [D(x) + \frac{1}{2}\Box C(x)].
 \label{p}
 \end{eqnarray}
The physical degrees of freedom corresponding to a real vector
superfield $V$ are the vector gauge field $v_{\mu}$ and the
Majorana spinor field $\lambda$. All other components are
unphysical and can be eliminated. Indeed, under the Abelian
(super)gauge transformation the superfield $V$ is transformed as
$$V\ \ \ \rightarrow \ \ \ V + \Phi + \Phi^+ , $$
 where $\Phi$ and $\Phi^+$ are some chiral superfields. In components it looks like
 \begin{eqnarray}
C \ \ \ & \rightarrow & \ \ \ C + A + A^* ,\nonumber \\ \chi \ \ \
& \rightarrow & \ \ \ \chi -i\sqrt{2} \psi, \nonumber \\ M + iN \
\ \ & \rightarrow & \ \ \ M + iN - 2iF, \nonumber \\ v_{\mu} \ \ \
& \rightarrow & \ \ \ v_{\mu} -i\partial_{\mu} (A - A^*),
\label{n}  \\ \lambda \ \ \ & \rightarrow & \lambda , \nonumber \\
D \ \ \ & \rightarrow & D , \nonumber
 \end{eqnarray} and
corresponds to ordinary gauge transformations for physical
components. According to eq.(\ref{n}), one can choose a gauge (the
Wess-Zumino gauge~\cite{WZgauge}) where $C = \chi = M = N =0 $,
leaving one with only physical degrees of freedom except for the
auxiliary field $D$. In this gauge
 \begin{eqnarray}
V & = & - \theta \sigma^{\mu} \bar \theta v_{\mu}(x) + i \theta
\theta \bar \theta \bar \lambda (x) -i\bar \theta \bar \theta
\theta \lambda (x) + \frac{1}{2} \theta \theta \bar \theta \bar
\theta D(x) , \nonumber\\ V^2 & = & - \frac{1}{2} \theta \theta
\bar \theta \bar \theta v_{\mu}(x)v^{\mu}(x) , \nonumber\\ V^3 & =
& 0, \ \ \ etc.
 \end{eqnarray}
 One can define also a field
strength tensor (as analog of $F_{\mu \nu}$ in gauge theories)
 \begin{eqnarray}
W_{\alpha} & = & - \frac{1}{4} \bar D^2 e^V D_{\alpha} e^{-V} , \nonumber\\
\bar W_{\dot \alpha} & = & - \frac{1}{4} D^2 e^V \bar D_{\alpha} e^{-V},
\label{str}
 \end{eqnarray}
 which is a polynomial in the
Wess-Zumino gauge. (Here $Ds$ are the supercovariant derivatives.)

The strength tensor is a chiral superfield
$$ \bar D_{\dot
\beta}W_{\alpha} = 0, \ \ \ \ D_{\beta}\bar W_{\dot \alpha} = 0 .
$$  In the Wess-Zumino gauge it is a polynomial over component
fields:
 \begin{equation}
W_\alpha = T^a\left(-i\lambda^a_\alpha +\theta_\alpha D^a-
\frac{i}{2}(\sigma^\mu \bar{\sigma}^\nu \theta)_\alpha F^a_{\mu
\nu }+ \theta^2\sigma^\mu D_\mu \bar{\lambda}^a \right) ,
 \end{equation}
where
 $$F^a_{\mu \nu }=\partial_\mu v^a_\nu - \partial_\nu v^a_\mu
+f^{abc} v^b_\mu v^c_\nu , \ \ \ D_\mu \bar{\lambda }^a=\partial
\bar{\lambda }^a +f^{abc}v^b_\mu \bar{\lambda }^c.$$
 In Abelian case eqs.(\ref{str}) are simplified and take form
$$W_\alpha= - \frac{1}{4} \bar
D^2 D_{\alpha}V , \ \ \ \bar W_{\dot \alpha}= - \frac{1}{4} D^2 \bar
D_{\alpha}V .$$

\subsection{Construction of SUSY Lagrangians}

Let us start with the Lagrangian which has no local gauge
invariance. In the superfield notation SUSY invariant Lagrangians
are the polynomials of superfields. Having in mind that for
component fields one should have  ordinary terms and the above
mentioned property of SUSY invariance of the highest dimension
components of a superfield, the general SUSY invariant Lagrangian
has the form
 \begin{equation}
{\cal L}  = \Phi_i^+ \Phi_i |_{\theta \theta \bar \theta \bar
\theta} + [(\lambda_i \Phi_i + \frac{1}{2}m_{ij}\Phi_i \Phi_j +
\frac{1}{3} g_{ijk} \Phi_i \Phi_j \Phi_k ) |_{\theta \theta } +
h.c. ].  \label{ll}
 \end{equation}
Hereafter the vertical line means the corresponding term of a
Taylor expansion.

 The first term is a kinetic term. It contains both the chiral and
antichiral superfields $\Phi_i$ and $\Phi_i^+$, respectively, and
is a function of Grassmannian parameters $\theta$ and
$\bar\theta$. Being expanded over $\theta$ and $\bar\theta$ it
leads to the usual kinetic terms for the
corresponding component fields.

The terms in the bracket form the superpotential. It is composed
of the chiral fields only (plus the hermitian conjugated
counterpart composed of antichiral superfields) and  is a chiral
superfield. Since the  products of a chiral superfield and
antichiral one produce a general superfield, they are not allowed
in a superpotential. The last coefficient of its expansion over
the parameter $\theta$ is supersymmetrically invariant and gives
the usual potential after getting rid of the auxiliary fields, as
it will be clear later.

The Lagrangian (\ref{ll}) can be written in a much more elegant
way in superspace. The same way as an ordinary action is an
integral over space-time of Lagrangian density, in supersymmetric
case the action is an integral over the superspace. The space-time
Lagrangian density then is~\cite{WessB, West, sspace}
 \begin{eqnarray}
{\cal L} & = &\int d^2\theta d^2\bar \theta ~\Phi_i^+ \Phi_i+ \int
d^2\theta
 ~[\lambda_i \Phi_i + \frac{1}{2}m_{ij}\Phi_i \Phi_j + \frac{1}{3} y_{ijk}
\Phi_i \Phi_j \Phi_k] + h.c. \label{l}
 \end{eqnarray}
where the first part is a kinetic term and the second one is a
superpotential  ${\cal W}$. Here instead of taking the proper
components we use integration over the superspace according to the
rules of Grassmannian integration~\cite{ber}
 $$\int \ d\theta_\alpha =
0 , \ \ \ \ \int \theta _\alpha\ d\theta _\beta=
\delta_{\alpha\beta}. $$
 Performing explicit integration over the Grassmannian parameters,
we get from eq.(\ref{l})
 \begin{eqnarray}
{\cal L} & = & i\partial_{\mu}\bar \psi_i \bar \sigma^{\mu}\psi_i
+ A_i^{\ast} \Box A_i + F_i^{\ast}F_i \label{20} \\
 & + & [\lambda_i F_i + m_{ij}(A_iF_j - \frac{1}{2}\psi_i \psi_j )
+ y_{ijk}(A_iA_jF_k - \psi_i \psi_j A_k ) + h.c. ] . \nonumber
 \end{eqnarray}
The last two terms 
are the interaction ones. To obtain a familiar form of the
 Lagrangian, we have to solve the constraints
 \begin{eqnarray}
\frac{\partial {\cal L}}{\partial F_k^*} & = & F_k + \lambda_k^* +
m_{ik}^*A_i^* + y_{ijk}^* A_i^*A_j^* = 0, \\ \frac{\partial {\cal
L}}{\partial F_k} & = & F_k^* + \lambda_k + m_{ik}A_i + y_{ijk}
A_iA_j = 0.
\end{eqnarray}
Expressing the auxiliary fields $F$ and $F^*$  from these
equations, one finally gets
 \begin{eqnarray} {\cal L} & = & i\partial_{\mu}\bar \psi_i \bar
\sigma^{\mu}\psi_i + A_i^* \Box A_i - \frac{1}{2}m_{ij}\psi_i
\psi_j - \frac{1}{2}m_{ij}^* \bar \psi_i \bar \psi_j \nonumber \\
& & - y_{ijk}\psi_i \psi_j A_k - y_{ijk}^* \bar \psi_i \bar \psi_j
A_k^* - V(A_i,A_j),  \label{m}
 \end{eqnarray}
where the scalar potential $V = F_k^* F_k $. We will return to the
discussion of the form of the scalar potential in SUSY theories
later.

Consider now the gauge invariant SUSY Lagrangians. They should
contain gauge invariant interaction of the matter fields with the
gauge ones and the kinetic term and the self-interaction of the
gauge fields.

Let us start with the gauge field kinetic terms. In the
Wess-Zumino gauge one has
 \begin{equation} W^{\alpha}W_{\alpha} |_{\theta \theta}= -2i\lambda
\sigma^{\mu}D_{\mu}\bar \lambda -
 \frac{1}{2}F_{\mu \nu}F^{\mu \nu}+\frac{1}{2}D^2
+i \frac{1}{4}F^{\mu \nu}F^{\rho \sigma}\epsilon_{\mu \nu \rho
\sigma },
 \end{equation}
where $D_\mu = \partial_\mu +ig[v_\mu, ]$ is the usual covariant
derivative and the last, the so-called topological $\theta$
term,\footnote{ Terminology comes from the $\theta$ term of
 QCD~\cite{theta} and has nothing to do with the Grassmannian
parameter $\theta$.} is the total derivative.

The gauge invariant Lagrangian now has a familiar form
 \begin{eqnarray}
{\cal L} & = & \frac{1}{4}\int d^2\theta ~W^{\alpha}W_{\alpha}
 + \frac{1}{4} \int d^2 \bar \theta ~\bar W^{\dot \alpha}\bar W_{\dot \alpha}
  \nonumber \\
 & = & \frac{1}{2}D^2 - \frac{1}{4}F_{\mu \nu}F^{\mu \nu} -
 i \lambda \sigma^{\mu}D_{\mu}\bar \lambda. \label{29}
 \end{eqnarray}
To obtain a gauge-invariant interaction with matter chiral
superfields, consider their gauge transformation (Abelian)
 $$\Phi\ \ \rightarrow \ \ e^{-ig\Lambda} \Phi , \ \ \ \Phi^+ \ \
\rightarrow \ \ \Phi^+ e^{ig\Lambda^+} , \ \ \ V \ \ \rightarrow \
\ V + i(\Lambda - \Lambda^+) , $$ where $\Lambda $ is a gauge
parameter (chiral superfield).

It is clear now how to construct both the SUSY and gauge invariant
kinetic term  (compare with the covariant derivative in a usual
gauge theory)
 \begin{equation}
\Phi_i^+  \Phi_i |_{\theta \theta \bar \theta \bar \theta}
\Rightarrow \Phi_i^+ e^{gV} \Phi_i |_{\theta \theta \bar \theta
\bar \theta}
 \end{equation}
A complete SUSY and gauge invariant Lagrangian then looks like
 \begin{eqnarray}
{\cal L}_{inv} & = & \frac{1}{4}\int d^2\theta
~W^{\alpha}W_{\alpha}  + \frac{1}{4}\int d^2 \bar\theta ~\bar
W^{\dot \alpha} \bar W_{\dot \alpha}
 + \int d^2\theta  d^2\bar \theta ~\Phi_i^+ e^{gV} \Phi_i \label{30} \\
 & + & \int d^2\theta ~(\frac{1}{2}m_{ij}\Phi_i \Phi_j + \frac{1}{3} y_{ijk}
\Phi_i \Phi_j \Phi_k )  + h.c.  \nonumber
 \end{eqnarray}
In particular, the SUSY generalization of QED looks as follows:
 \begin{eqnarray}
{\cal L}_{SUSY \ QED} & = & \frac{1}{4}\int d^2\theta
~W^{\alpha}W_{\alpha}  + \frac{1}{4}\int d^2\bar \theta ~\bar
W^{\dot \alpha} \bar W_{\dot \alpha}  \nonumber\\
 & + & \int d^4\theta ~(\Phi_+^+ e^{gV}\Phi_+ + \Phi_-^+ e^{-gV}
 \Phi_-) \\
  &+& \int d^2\theta ~m~\Phi_+ \Phi_-  + \int d^2\bar \theta ~m~\Phi_+^+ \Phi_-^+ ,
  \nonumber
 \end{eqnarray}
where two superfields $\Phi_+$ and $\Phi_-$ have been introduced
in order to have both left- and right-handed fermions.

The non-Abelian generalization is straightforward
 \begin{eqnarray}
{\cal L}_{SUSY \ YM} & = & \frac{1}{4}\int d^2 \theta
~Tr(W^{\alpha}W_{\alpha})   + \frac{1}{4}\int d^2 \bar{\theta}
~Tr(\bar{W}^{\alpha}\bar{W}_{\alpha})  \label{nonab}\\ &+& \int
d^2 \theta d^2 \bar \theta ~\bar \Phi_{ia}(e^{gV})_b^a\Phi_i^b
+\int d^2 \theta ~{\cal W}(\Phi_i)   +\int d^2 \bar{\theta}
~\bar{{\cal W}}(\bar{\Phi}_i)  , \nonumber
 \end{eqnarray}
where ${\cal W}$ is a  superpotential, which should be invariant
under the group of symmetry of a particular model.

In terms of component fields the above Lagrangian takes the form
 \begin{eqnarray}
{\cal L}_{SUSY \ YM} & = & -\frac{1}{4}F^a_{\mu \nu }F^{a\mu \nu
}-i\lambda^a\sigma^\mu D_\mu \bar{\lambda}^a+\frac{1}{2}D^aD^a
\nonumber\\ &+&(\partial_\mu A_i -igv^a_\mu T^aA_i)^\dagger
(\partial_\mu A_i -igv^{a\mu}T^aA_i)
-i\bar{\psi}_i\bar{\sigma}^\mu (\partial_\mu \psi_i
-igv^{a\mu}T^a\psi_i) \nonumber \\ &-& D^aA^\dagger_i
T^aA_i-i\sqrt{2}A^\dagger_iT^a\lambda^a\psi_i +
i\sqrt{2}\bar{\psi}_iT^aA_i\bar{\lambda}^a+F^\dagger_iF_i
\nonumber \\ &+& \frac{\partial {\cal W}}{\partial A_i} F_i+
\frac{\partial \bar{{\cal W}}}{\partial A_i^\dagger}F^\dagger_i
-\frac{1}{2}\frac{\partial^2 {\cal W}}{\partial A_ \partial
A_j}\psi_i\psi_j -\frac{1}{2}\frac{\partial^2 \bar{{\cal
W}}}{\partial A_i^\dagger \partial
A_j^\dagger}\bar{\psi}_i\bar{\psi}_j  .\label{sulag}
 \end{eqnarray}
Integrating out the auxiliary fields $D^a$ and $F_i$, one
reproduces the usual Lagrangian.

\subsection{The scalar potential}

Contrary to the SM, where the scalar potential is arbitrary and is
defined only by the requirement of the gauge invariance, in
supersymmetric theories it is completely  defined by the
superpotential. It consists of the contributions from the
$D$-terms and $F$-terms. The kinetic energy of the gauge fields
(recall eq.(\ref{29}) yields the $1/2 D^aD^a$ term, and the
matter-gauge interaction (recall eq.(\ref{30}) yields the
$gD^aT^a_{ij}A^*_iA_j$ one. Together they give
 \begin{equation}
{\cal L}_D=\frac{1}{2}D^aD^a + gD^aT^a_{ij}A^*_iA_j. \label{d}
 \end{equation}
The equation of motion reads
 \begin{equation}
D^a=-gT^a_{ij}A^*_iA_j. \label{sol}
 \end{equation}
Substituting it back into eq.(\ref{d}) yields the $D$-term part of
the potential
 \begin{equation} {\cal L}_D=-\frac{1}{2}D^aD^a \ \ \ \
\Longrightarrow V_D=\frac{1}{2}D^aD^a,
 \end{equation}
where $D$ is given by eq.(\ref{sol}).

The $F$-term contribution can be derived from the matter field
self-in\-ter\-action eq.(\ref{20}). For a general type
superpotential $W$ one has
 \begin{equation}
{\cal L}_F=F^*_iF_i+(\frac{\partial W}{\partial A_i}F_i + h.c.).
 \end{equation}
Using the equations of motion for the auxiliary field $F_i$
 \begin{equation}
F^*_i=-\frac{\partial W}{\partial A_i} \label{solf}
 \end{equation}
yields
 \begin{equation} {\cal L}_F=-F^*_iF_i \ \ \ \
\Longrightarrow V_F= F^*_iF_i ,
 \end{equation}
where $F$ is given by eq.(\ref{solf}). The full potential is the
sum of the two contributions
 \begin{equation}
V=V_D+V_F.
 \end{equation}

Thus, the form of the Lagrangian  is practically fixed by symmetry
requirements. The only freedom is the field content, the value of
the gauge coupling $g$, Yukawa couplings $y_{ijk}$ and the masses.
Because of the renormalizability constraint $V \leq A^4 $ the
superpotential should be limited by ${\cal W} \leq \Phi^3 $ as in
eq.(\ref{l}). All members of a supermultiplet have the same
masses, i.e. bosons and fermions are degenerate in masses. This
property of SUSY theories contradicts the phenomenology and
requires supersymmetry breaking.

 \subsection{Spontaneous breaking of SUSY}

Since supersymmetric algebra leads to mass degeneracy in a
supermultiplet, it should be broken  to explain the absence of
superpartners at modern energies. There are several ways of
supersymmetry breaking. It can be broken either explicitly or
spontaneously. Performing SUSY breaking one has to be careful not
to spoil the cancellation of quadratic divergencies which allows
one to solve the hierarchy problem. This is achieved by
spontaneous breaking of SUSY.

Apart from non-supersymmetric theories in SUSY models the energy
is always nonnegative definite. Indeed, according to quantum
mechanics $$ E = <0| \ H \ |0> $$ and due to SUSY algebra
eq.(\ref{group})
 $$\{Q_{\alpha}, \bar Q_{\dot \beta} \} =
2(\sigma^{\mu}) _{\alpha \dot \beta}P_{\mu} ,$$ taking into
account that $tr(\sigma^{\mu}P_{\mu}) = 2P_0 ,$
 one gets
$$E = \frac{1}{4} \sum_{\alpha = 1,2}<0| \{Q_{\alpha}, \bar
Q_{\alpha} \} |0> = \frac{1}{4} \sum_{\alpha} |Q_{\alpha}
  |0> |^2 \geq 0 .$$
Hence $$ E = <0| \ H \ |0> \neq 0 \ \ \ \ if \ and \ only \ if \ \
\ Q_{\alpha}|0> \neq 0 .$$

Therefore, supersymmetry is spontaneously broken, i.e. vacuum is
not invariant $(Q_{\alpha} |0> \neq 0 )$, {\em if and only if} the
minimum of the potential is positive $(i.e.\ E > 0)$ .

\begin{figure}[ht]
 \begin{center}
 \leavevmode
  \epsfxsize=14cm
 \epsffile{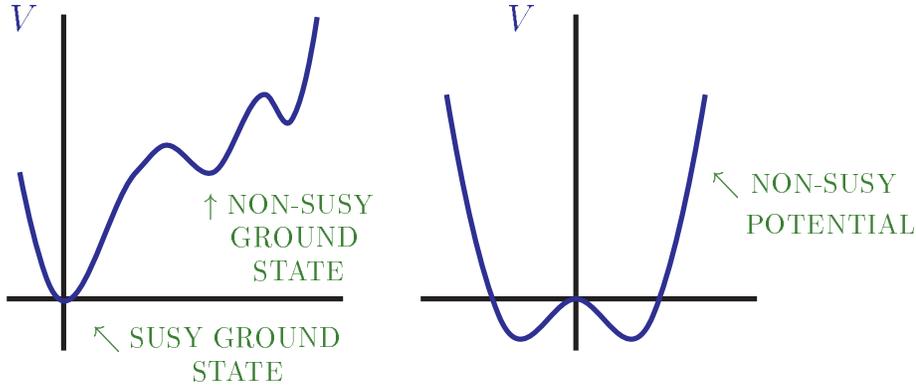}
 \end{center}\vspace{-1cm}
 \caption{Scalar potential in supersymmetric and non-supersymmetric theories}
 \label{poten}
 \end{figure}

The situation is illustrated in Fig.\ref{poten}. The SUSY ground
state has $E=0$, while a non-SUSY one has $E>0$. On the right-hand
side a non-SUSY potential is shown. It does not appear  even in
spontaneously broken SUSY theories. However, just this type of the
potential is used for spontaneous breaking of the gauge invariance
via  the Higgs mechanism.  This property has crucial consequences
for the spontaneous breaking of the gauge invariance. Indeed, as
will be seen later, in the MSSM spontaneous breaking of $SU(2)$
invariance takes place only after SUSY is broken.

\begin{figure}[ht]
 \begin{center}
 \leavevmode
  \epsfxsize=14cm
 \epsffile{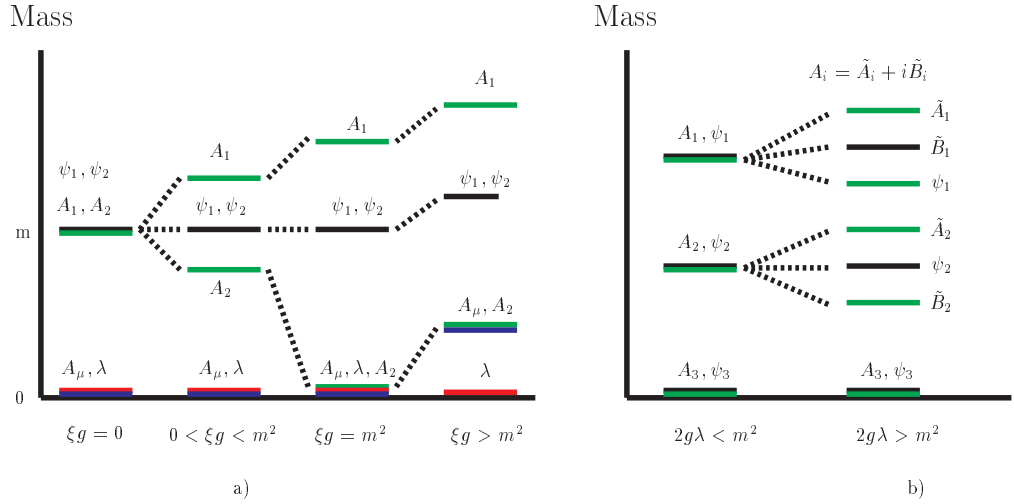}
 \end{center}\vspace{-1cm}
 \caption{Spectrum of spontaneously broken SUSY theories}\label{FI}
 \end{figure}

Spontaneous breaking of supersymmetry is achieved in the same way
as the electroweak symmetry  breaking. One introduces the field
whose vacuum expectation value is nonzero and breaks the symmetry.
However, due to a special character of SUSY, this should be a
superfield whose auxiliary $F$ and $D$ components acquire nonzero
v.e.v.'s. Thus,  among possible spontaneous SUSY breaking
mechanisms one distinguishes the $F$ and $D$ ones.

i) Fayet-Iliopoulos ($D$-term) mechanism \cite{Fayet}. \\
 In this case the, the linear $D$-term is added to the Lagrangian
 \begin{equation} \Delta {\cal
L} = \xi V\vert_{ \theta \theta \bar \theta \bar \theta} = \xi\int
d^4\theta\ V.
\end{equation}
 It is gauge and SUSY invariant by itself; however, it may
lead to spontaneous breaking of both of them depending on the value of
$\xi$. We show in Fig.\ref{FI}a the sample spectrum for two chiral matter
multiplets.
The drawback of this mechanism is the necessity of $U(1)$ gauge
invariance. It can be used in SUSY generalizations of the SM but
not in GUTs.

The mass spectrum also causes some troubles since the
following sum rule is always valid
 \begin{equation}
\sum_{boson \ states} m^2_i =
\sum_{fermion \ states} m^2_i , \label{sumrule}
 \end{equation}
which is bad for  phenomenology.

ii) O'Raifeartaigh ($F$-term) mechanism \cite{O'R}. \\
 In this case,
several chiral fields are needed and the superpotential should be
chosen in a way that trivial zero v.e.v.s for the auxiliary
$F$-fields be absent.  For instance, choosing the superpotential
to be
 $${\cal W}(\Phi)=
\lambda \Phi_3 +m\Phi_1\Phi_2 +g \Phi_3\Phi_1^2,$$
 one gets the equations for the auxiliary fields
 \begin{eqnarray*}
F^*_1&=&mA_2+2gA_1A_3, \\
F^*_2 &=& mA_1, \\
F^*_3 &=& \lambda +gA^2_1,
 \end{eqnarray*}
which have no solutions with $<F_i> =0$ and SUSY is spontaneously broken.
The sample spectrum is shown in Fig.\ref{FI}b.

The drawbacks of this mechanism is a lot of arbitrariness in the
choice of potential. The sum rule (\ref{sumrule}) is also valid
here.

Unfortunately, none of these mechanisms explicitly works in SUSY
generalizations of the SM. None of the fields of the SM can
develop nonzero v.e.v.s for their $F$ or $D$ components without
breaking  $SU(3)$ or $U(1)$ gauge invariance since they are not
singlets with respect to these groups. This requires the presence
of extra sources of spontaneous SUSY breaking, which we consider
below. They are based, however, on the same  $F$ and $D$
mechanisms.

\section{SUSY generalization of the Standard Model. The MSSM}
 \setcounter{equation} 0

As has been already mentioned, in SUSY theories the number of
bosonic degrees of freedom equals that of fermionic. At the same
time,  in the SM one has 28 bosonic and 90 fermionic degrees of
freedom (with massless neutrino, otherwise 96). So the SM is to a
great extent non-supersymmetric. Trying to add some new particles
to supersymmetrize the SM, one should take into account the
following observations:
 \begin{enumerate}
\item There are no fermions with quantum numbers of the
gauge bosons;
\item
Higgs fields have nonzero v.e.v.s; hence they cannot be
 superpartners of quarks and leptons since
this would induce  spontaneous violation of baryon and lepton
numbers;
\item
One needs at least two complex chiral Higgs multiplets to give
masses to Up and Down quarks.
 \end{enumerate}

The latter is due to the form of a superpotential and chirality of
matter superfields. Indeed, the superpotential should be invariant
under the $SU(3)\times SU(2)\times U(1)$ gauge group. If one looks
at the Yukawa interaction in the Standard Model,
eq.(\ref{yukawa}), one finds that it is indeed $U(1)$ invariant
since the sum of hypercharges in each vertex equals zero. In the
last term this is achieved by taking  the conjugated Higgs doublet
$\tilde{H}=i\tau_2H^\dagger$ instead of $H$. However, in SUSY $H$
is a chiral superfield and hence a superpotential, which is
constructed out of  chiral fields, can contain only $H$ but not
$\tilde H$ which is an antichiral superfield.

Another reason for the second  Higgs doublet is related to chiral
anomalies. It is known that chiral anomalies spoil the gauge
invariance and, hence, the renormalizability of the theory. They
are canceled in the SM between quarks and leptons in each
generation.

Indeed, chiral (or triangle anomaly) is proportional to the trace
of three hypercharges. In the SM one has
$$\begin{array}{ccrcccccccc} Tr Y^3 &=& 3 & \left(
\frac{1}{27}\right. &+\frac{1}{27}&-\frac{64}{27}&\left.
+\frac{8}{27} \right) & -1& -1& + 8& = 0. \\ & & \uparrow &
\uparrow & \uparrow &\uparrow &\uparrow &\uparrow & \uparrow &
\uparrow & \\ && colour &u_L & d_L & u_R & d_R & \nu_L & e_L & e_R
& \end{array} $$ However, if one introduces a chiral Higgs
superfield, it contains higgsinos, which are chiral fermions, and
contain anomalies. To cancel them one has to add the second Higgs
doublet with the opposite hypercharge.

Therefore, the Higgs sector in SUSY models is inevitably enlarged,
it contains an even number of doublets.

\vglue 0.4cm {\em Conclusion}: In SUSY models supersymmetry
associates {\em known} bosons with {\em new} fermi\-ons and {\em
known} fermi\-ons with {\em new} bosons.
 \begin{figure}[t]
\begin{center}\vspace{-1cm}
 \leavevmode
  \epsfxsize=12cm
 \epsffile{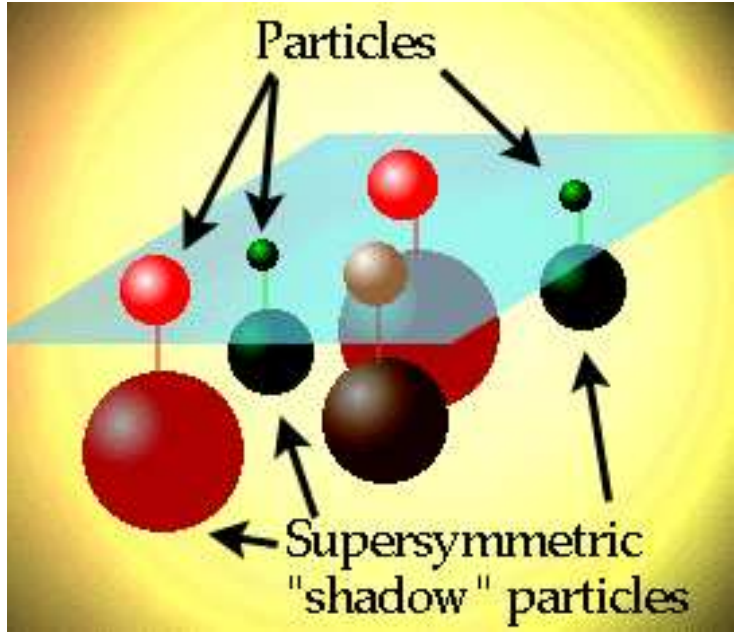}
\end{center}\vspace{-1cm} \caption{The shadow world of SUSY particles
\cite{shadow}}\label{fig:shadow}
 \end{figure}

\subsection{The field content}

Consider the particle content of the Minimal Supersymmetric
Standard Model \cite{MSSM}. According to the previous discussion,
in the minimal version  we double the number of particles
(introducing a superpartner to each particle) and add another
Higgs doublet (with its superpartner). The particle content of the
MSSM then appears as~\cite{haber}

\begin{center}
{\bf Particle Content of the MSSM} \\
\bigskip
\nopagebreak[4]
\renewcommand{\tabcolsep}{0.1cm}
\begin{tabular}{lllccc}
Superfield & \ \ \ \ \ \ \ Bosons & \ \ \ \ \ \ \ Fermions &
$SU_c(3)$& $SU_L(2)$ & $U_Y(1)$ \\ \hline \hline Gauge  &&&&& \\
${\bf G^a}$   & gluon \ \ \ \ \ \ \ \ \ \ \ \ \ \ \  $g^a$ &
gluino$ \ \ \ \ \ \ \ \ \ \ \ \ \tilde{g}^a$ & 8 & 0 & 0 \\ ${\bf
V^k}$ & Weak \ \ \ $W^k$ \ $(W^\pm, Z)$ & wino, zino \
$\tilde{w}^k$ \ $(\tilde{w}^\pm, \tilde{z})$ & 1 & 3& 0 \\ ${\bf
V'}$   & Hypercharge  \ \ \ $B\ (\gamma)$ & bino \ \ \ \ \ \ \ \ \
\ \ $\tilde{b}(\tilde{\gamma })$ & 1 & 1& 0 \\ \hline Matter &&&&
\\ $\begin{array}{c} {\bf L_i} \\ {\bf E_i}\end{array}$ & sleptons
\ $\left\{
\begin{array}{l} \tilde{L}_i=(\tilde{\nu},\tilde e)_L \\ \tilde{E}_i =\tilde
e_R \end{array} \right. $ & leptons \ $\left\{ \begin{array}{l}
L_i=(\nu,e)_L
\\ E_i=e_R \end{array} \right.$ & $\begin{array}{l} 1 \\ 1 \end{array} $  &
$\begin{array}{l} 2 \\ 1 \end{array} $ & $\begin{array}{r} -1 \\ 2
\end{array} $ \\ $\begin{array}{c} {\bf Q_i} \\ {\bf U_i} \\ {\bf D_i}
\end{array}$ & squarks \ $\left\{ \begin{array}{l}
\tilde{Q}_i=(\tilde{u},\tilde d)_L \\ \tilde{U}_i =\tilde u_R \\
\tilde{D}_i =\tilde d_R\end{array}  \right. $ & quarks \ $\left\{
\begin{array}{l} Q_i=(u,d)_L \\ U_i=u_R^c \\ D_i=d_R^c \end{array}
\right.$ & $\begin{array}{l} 3
\\ 3^* \\ 3^* \end{array} $  & $\begin{array}{l} 2 \\ 1 \\ 1 \end{array} $ &
$\begin{array}{r} 1/3 \\ -4/3 \\ 2/3 \end{array} $ \\ \hline Higgs
&&&& \\ $\begin{array}{c} {\bf H_1} \\ {\bf H_2}\end{array}$ &
Higgses \ $\left\{
\begin{array}{l} H_1 \\ H_2 \end{array}  \right. $ & higgsinos \ $\left\{
 \begin{array}{l} \tilde{H}_1 \\ \tilde{H}_2 \end{array} \right.$ &
$\begin{array}{l} 1 \\ 1 \end{array} $  & $\begin{array}{l} 2 \\ 2
\end{array} $ &
$\begin{array}{r} -1 \\ 1
\end{array} $
 \\ \hline \hline
\end{tabular}
\end{center}
\vglue .5cm

\noindent where $a=1,2,...,8$ and $k=1,2,3$ are the $SU(3)$ and
$SU(2)$ indices, respectively, and $i=1,2,3$ is the generation
index. Hereafter, tilde denotes a superpartner of an ordinary
particle.

Thus, the characteristic feature of any supersymmetric
generalization of the SM is the presence of superpartners (see
Fig.\ref{fig:shadow}).  If supersymmetry is exact, superpartners
of ordinary particles should have the same masses and have to be
observed. The absence of them at modern energies is believed to be
explained by the fact that their masses are very heavy, that means
that supersymmetry should be broken. Hence,  if the energy of
accelerators is high enough, the superpartners will be created.

The presence of an extra Higgs doublet in SUSY model is a novel
feature of the theory. In the MSSM  one has two doublets with the
quantum numbers (1,2,-1) and (1,2,1), respectively:
 \begin{equation}
H_1=\left( \begin{array}{c} H^0_1 \\ H_1^- \end{array}  \right) =
\left( \begin{array}{c} v_1 +\frac{\displaystyle
S_1+iP_1}{\sqrt{2}} \\ H^-_1 \end{array}\right), \ \ \ H_2=\left(
\begin{array}{c} H^+_2 \\ H_2^0 \end{array}  \right) = \left(
\begin{array}{c} H^+_2 \\ v_2 +\frac{\displaystyle
S_2+iP_2}{\sqrt{2}}
\end{array} \right),
 \end{equation}
where  $v_i$ are the vacuum expectation values of the neutral
components.

Hence, one has 8=4+4=5+3 degrees of freedom. As in the case of the
SM, 3 degrees of freedom can be gauged away, and one is left with
5 physical states compared to 1 state in the SM.

Thus, in the MSSM, as actually in any of two Higgs doublet models,
 one has  five  physical Higgs bosons: two CP-even neutral,
one CP-odd neutral  and two charged. We consider the mass
eigenstates below.

\subsection{Lagrangian of the MSSM}

 The Lagrangian of the MSSM consists of two parts; the
first part is SUSY generalization of the Standard Model, while the
second one represents the SUSY breaking as mentioned above.
 \begin{equation}
 {\cal L}={\cal L}_{SUSY}+{\cal L}_{Breaking},
 \end{equation}
where
 \begin{equation}
 {\cal L}_{SUSY}={\cal L}_{Gauge}+{\cal L}_{Yukawa}
 \end{equation}
and
 \begin{eqnarray}
{\cal L}_{Gauge} &= &\sum_{SU(3),SU(2),U(1)}^{}\frac{1}{4}
\left(\int d^2\theta \ Tr W^\alpha W_\alpha + \int d^2\bar\theta \
Tr \bar W^{\dot \alpha}\bar W_{\dot \alpha} \right) \nonumber \\
&& +\sum_{Matter}^{}\int d^2\theta d^2 \bar\theta \
\Phi^\dagger_ie^{\displaystyle g_3\hat V_3 + g_2\hat V_2 + g_1\hat
V_1}\Phi_i ,  \\ {\cal L}_{Yukawa}&= &\int d^2\theta \ ({\cal
W}_R+{\cal W}_{NR}) + h.c .
 \end{eqnarray}
The index $R$ in a superpotential refers to the so-called
$R$-parity~\cite{r-parity} which adjusts a "$+$" charge to all the
ordinary particles and a "$-$" charge to their superpartners. The
first part of ${\cal W}$ is R-symmetric
 \begin{equation}
W_{R} = \epsilon_{ij}(y^U_{ab}Q_a^j U^c_bH_2^i +
y^D_{ab}Q_a^jD^c_bH_1^i
       +  y^L_{ab}L_a^jE^c_bH_1^i + \mu H_1^iH_2^j), \label{R}
 \end{equation}
where $i,j=1,2,3$ are the $SU(2)$ and $a,b=1,2,3$ are the
generation indices; colour indices are suppressed. This part of
the Lagrangian almost exactly repeats that of the SM except that
the fields are now the superfields rather than the ordinary fields
of the SM. The only difference is the last term which describes
the Higgs mixing. It is absent in the SM since there is only one
Higgs field there.

The second part is R-nonsymmetric
 \begin{eqnarray}
W_{NR} &=& \epsilon_{ij}(\lambda^L_{abd}L_a^i L_b^jE_d^c +
\lambda^{L\prime}_{abd}L_a^iQ_b^jD_d^c +\mu'_aL^i_aH_2^j)
\nonumber\\ & + &  \lambda^B_{abd}U_a^cD_b^cD_d^c. \label{NR}
 \end{eqnarray}
These terms are absent in the SM. The reason is very simple: one
can not replace the superfields in eq.(\ref{NR}) by the ordinary
fields like in eq.(\ref{R}) because of the Lorentz invariance.
These terms have a different property, they violate either lepton
(the first line in eq.(\ref{NR})) or baryon number (the second
line). Since both effects are not observed in Nature, these terms
must be suppressed or be excluded. One can avoid such terms if one
introduces  special symmetry called the
$R$-symmetry~\cite{r-symmetry}. This is the global $U(1)_R$
invariance
 \begin{equation}
U(1)_R: \ \ \theta \to e^{i\alpha} \theta ,\ \  \Phi \to
e^{in\alpha}\Phi , \label{RS}
 \end{equation}
i.e., the superfield has the quantum number $R=n$. To preserve
$U(1)_R$ invariance the superpotential $W$ must have $R=2$. Thus,
to get $W_{NR}=0$ one must choose $R=1$ for all the Higgs
superfields and $R=1/2$ for quark and lepton ones. However, this
property happens to be too restrictive. Indeed, the gaugino mass
term, which is Lorentz and gauge invariant and is introduced while
supersymmetry breaking, happens to be $R$-invariant only for
$\alpha =\pm \pi $. This reduces the $R$-symmetry to the discrete
group $Z_2$, called the $R$-parity~\cite{r-parity}. The $R$-parity
quantum number
 is given by
 \begin{equation}
R=(-1)^{3(B-L)+2S} \label{par}
 \end{equation}
for particles with spin $S$. Thus, all the ordinary particles have
the $R$-parity quantum number equal to $R=+1$, while all the
superpartners have $R$-parity quantum number equal to $R=-1$. The
$R$-parity obviously forbids the $W_{NR}$ terms. It is usually
assumed that they are absent in the MSSM, i.e. $R$-parity is
preserved. However, there is no physical principle behind it. It
may well be that these terms are present, though experimental
limits on the couplings are very severe~\cite{r-con}
$$\lambda^L_{abc}, \ \ \lambda^{L\prime}_{abc} < 10^{-4}, \ \ \ \
\ \lambda^B_{abc} < 10^{-9}.$$

\subsection{Properties of interactions}

If one assumes that the $R$-parity is preserved, then the
 interactions of superpartners are essentially the same as in the
SM, but two of three particles involved into an interaction at any
vertex are replaced by superpartners. The reason for it, as we
discussed earlier, is the $R$-parity. According to eq.(\ref{par}),
all the ordinary particles are R-even, while all the superpartners
are R-odd.

Conservation of the $R$-parity has two consequences
 \begin{itemize}
\item the superpartners are created in pairs;
\item  the lightest superparticle (LSP) is stable.
 \end{itemize}
Usually it is  photino $\tilde \gamma $, the superpartner of a
photon with some admixture of neutral higgsino.

\begin{figure}[ht]\vspace{-1cm}
\begin{center}
 \leavevmode
  \epsfxsize=14cm
 \epsffile{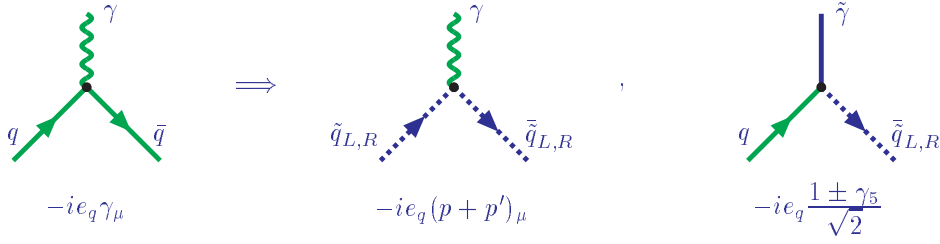}
 \end{center}
\vspace{-1cm}
 \caption{Gauge-matter interaction}\label{interact}
 \end{figure}

\begin{figure}[ht]
 \begin{center}
 \leavevmode
  \epsfxsize=14cm
 \epsffile{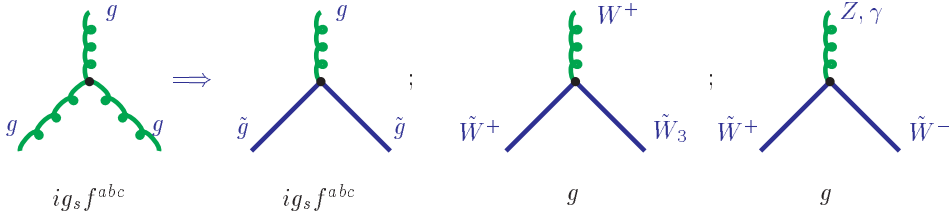}
 \end{center}
\vspace{-1.5cm}
 \caption{Gauge self-interaction}\label{gaugeint}
 \end{figure}

\begin{figure}[ht]\vspace{-0.5cm}
\begin{center}
 \leavevmode
  \epsfxsize=14cm
 \epsffile{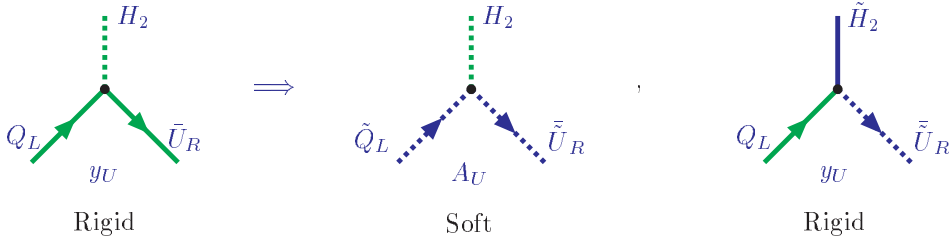}
 \end{center}
\vspace{-1.0cm}
 \caption{Yukawa-type interaction}\label{yukint}
 \end{figure}

Typical vertices are  shown in Figs.\ref{interact}-\ref{yukint}.
The tilde above a letter denotes the corresponding superpartner.
Note that the coupling is the same in all the vertices involving
superpartners.

In the case of $R$-parity violation one has additional vertices
with new types of interaction. As has been already mentioned, they
violate either the lepton or baryon number. The typical ones are
 \begin{eqnarray}
{\cal L}_{LLE}&=& \lambda' \left\{\tilde \nu_L e_Le^c_R - \tilde e_L
\nu_Le^c_R +\tilde{e}_R^* \nu_Le_R + \dots \right\}, \\ {\cal L}_{LQD}&=&
\lambda \left\{\tilde \nu_L d_L\bar d_R - \tilde e_L u_L¿\bar d_R +\tilde
d_L \nu_L\bar d_R - \tilde u_L e_L\bar d_R + \dots \right\}.
 \end{eqnarray}

\begin{figure}[ht]\vspace{-1cm}
\begin{center}
 \leavevmode
  \epsfxsize=12cm
 \epsffile{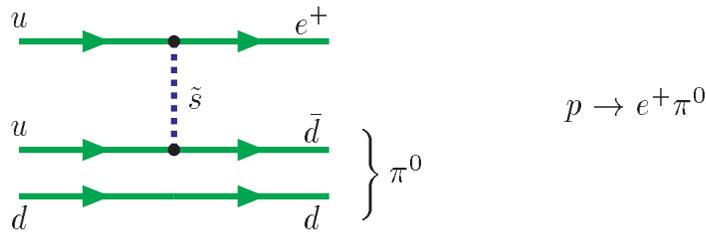}
 \end{center}\vspace{-1.5cm}
 \caption{Proton decay in R-parity violating models}\label{proton}
 \end{figure}

 There are also $UDD$ terms which violate the baryon number. These
terms together lead to a fast proton decay via the process shown
in Fig.\ref{proton}. To avoid it, one usually leaves either $L$ or
$B$ violating interactions.

The limits on $R$-parity violating couplings come from non-observation of
various processes, like proton decay, $\nu_\mu e$ scattering, etc and also
from the charged current universality: $\Gamma(\pi \to e\nu)/\Gamma(\pi \to
\mu\nu), \Gamma(\tau \to e\nu \bar \nu)/\Gamma(\tau \to \mu\nu\bar \nu)$,
etc.

\subsection{Creation and decay of superpartners}

\begin{figure}[ht]
 \begin{center}
 \leavevmode
  \epsfxsize=13cm
 \epsffile{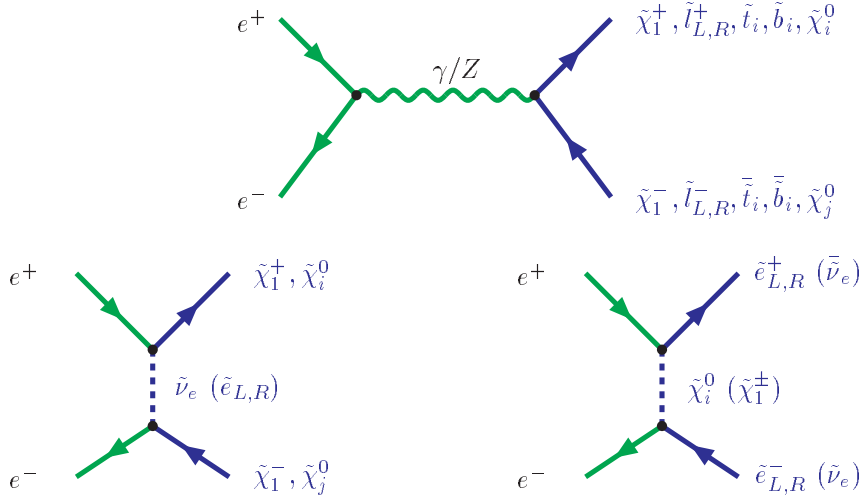}
 \end{center}\vspace{-1cm}
 \caption{Creation of superpartners}\label{creation}
 \end{figure}

The above-mentioned rule   together with the Feynman rules for the
SM enables us to draw diagrams describing creation of
superpartners.  One of the most promising processes is the
$e^+e^-$ annihilation (see Fig.\ref{creation}).
The usual kinematic restriction is given by the centre of mass
energy $$m^{max}_{sparticle} \leq \frac{\sqrt s}{2}.$$ Similar
processes take place at hadron colliders with electrons and
positrons being replaced by quarks and gluons.

Creation of superpartners can be  accompanied by creation of
ordinary particles as well. We consider various experimental
signatures for $e^+e^-$ and hadron colliders below. They crucially
depend on SUSY breaking pattern and on the mass spectrum of
superpartners.

\begin{figure}[htb]\vspace{-0.7cm}
 \begin{center}
 \leavevmode
  \epsfxsize=12cm
 \epsffile{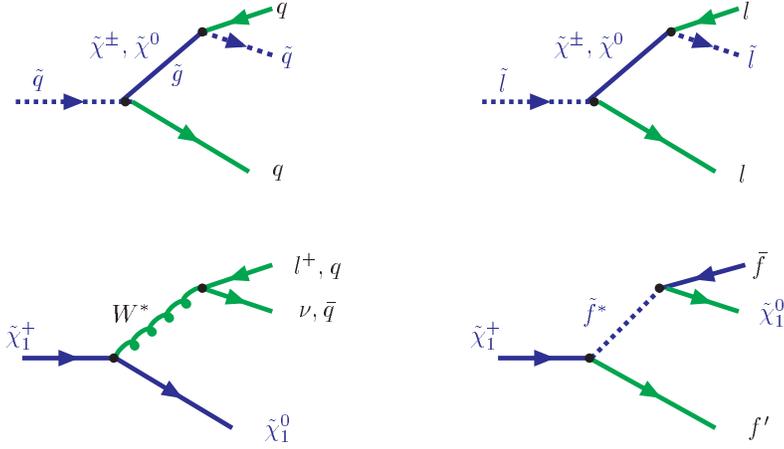}
 \end{center}\vspace{-1cm}
 \caption{Decay of superpartners}\label{decay}
 \end{figure}

The decay properties of superpartners also depend on their masses. For
the quark and lepton superpartners the main processes are shown in
Fig.\ref{decay}.

When the $R$-parity is conserved, new particles will eventually
end up giving neutralinos (the lightest superparticle) whose
interactions are comparable to those of neutrinos and they leave
undetected. Therefore, their signature would be missing energy and
transverse momentum. \vspace{0.5cm}

\noindent {\em Examples}. Consider some explicit examples of
superpartner decays.
 $$\begin{array}{lccll} \mbox {squarks} : &
\tilde{q}_{L,R} & \rightarrow & q + \tilde{\chi}^0_i & \mbox{(quark +
photino)} \\ &\tilde{q}_{L} & \rightarrow & q' + \tilde{\chi}^\pm_i &
\mbox{(quark + chargino)} \\ &  \tilde q & \rightarrow & q + \tilde g &
\mbox{(quark + gluino)} \ for \
 m_{\tilde{q}}>m_{\tilde{g}} \\
 &\tilde{t}_1 & \rightarrow & c + \tilde{\chi}^0_1 & \mbox{(main decay) \
 signal: 2 acollinear jets} + \Big/ \hspace{-0.3cm \Big/ \hspace{-0.3cm E_T}} \\
& \tilde{t}_1 & \rightarrow & b + \tilde{\chi}^+_1 & \hspace*{2.5cm}
 \mbox{signal: 2 b jets + 2 leptons} + \Big/ \hspace{-0.3cm E_T} \\
 & && \ \ \ \ \ \ \hookrightarrow \tilde{\chi}^0_1 f\bar f' &
  ( f\bar f'= l\bar \nu, q¿\bar q) \hspace*{2cm}  \mbox{(4 jets)}+
  \Big/ \hspace{-0.3cm E_T} \\
\end{array}$$  $$ \begin{array}{cccll} \mbox{sleptons} : & \tilde l &
\rightarrow & l + \tilde{\chi}^0_i & \mbox{(lepton + photino)} \\
&\tilde{l}_{L} & \rightarrow & \nu_l + \tilde{\chi}^\pm_i &
\mbox{(neutrino + chargino)} \\
 \mbox{gluino} : & \tilde g & \rightarrow &
q + \bar q + \tilde
 \gamma & \mbox{(quark + antiquark + photino)} \\
 & \tilde g & \rightarrow & g + \tilde \gamma & \mbox{(gluon + photino)} \\
\mbox{chargino} : & \tilde{\chi}^\pm_i & \rightarrow & e + \nu_e +
\tilde{\chi}^0_i & \mbox{(electron + neutrino + photino)}\\ &
\tilde{\chi}^\pm_i & \rightarrow & q + \bar q' + \tilde{\chi}^0_i &
\mbox{(quark + antiquark + photino)}\\ \mbox {neutralino} : &
\tilde{\chi}^0_2 & \rightarrow & \tilde{\chi}^0_1 + X & \ \end{array}$$
 In the last case there are many possible channels both visible and invisible.
$$\begin{array}{lll}& \mbox{\underline{Visible
Channels}}&\mbox{\underline{Final States}} \\ & & \\ \tilde{\chi}^0_2
&\rightarrow~~~ \tilde{\chi}^0_1 l^+l^- ~~~(l=e,\mu,\tau)~~~~ & \\ &
\rightarrow~~~ \tilde{\chi}^\pm_1 l^\mp \nu_l & l^+l^- + \Big/
\hspace{-0.3cm E_T} \\ &
 \hspace*{.9cm} \hookrightarrow \tilde{\chi}^0_1 l^\pm \nu_l & \\
& \rightarrow~~~ \tilde{\chi}^0_1 q\bar q & \mbox{2 jets} + \Big/
\hspace{-0.3cm E_T} \\ & \rightarrow~~~ \tilde{\chi}^0_ \gamma & \gamma +
\Big/ \hspace{-0.3cm E_T}\\&
\rightarrow~~~ \tilde{\chi}^\pm_1q\bar q' & \\ & \hspace*{.9cm}
\hookrightarrow \tilde{\chi}^0_1 l^\pm q\bar q' & \mbox{2 jets} + \Big/
\hspace{-0.3cm E_T} \\ & \rightarrow~~~ \tilde{\chi}^\pm_1 l^\mp \nu_l  &
\\ &
 \hspace*{.9cm} \hookrightarrow \tilde{\chi}^0_1 q\bar q' &
 l^\pm + \mbox{2 jets} + \Big/ \hspace{-0.3cm E_T}\\ &
\rightarrow~~~ \tilde{\chi}^\pm_1q\bar q' & \\ &
  \hspace*{.9cm} \hookrightarrow \tilde{\chi}^0_1 l^\pm \nu_l &
l^\pm + \mbox{2 jets} + \Big/ \hspace{-0.3cm E_T}
 \\ &  & \\ &
\mbox{\underline{Invisible Channel}}&\mbox{\underline{Final State}} \\ &&
\\ &  \rightarrow~~~ \tilde{\chi}^0_1 \nu_l \bar \nu_l &\ \ \ \ \ \ \
\Big/ \hspace{-0.3cm E_T} \end{array}$$ Thus, if supersymmetry exists in
Nature and if it is broken somewhere below 1 TeV, then it will be possible
to detect it in the nearest future.

\section{Breaking of SUSY in the MSSM}
 \setcounter{equation} 0

\begin{figure}[ht]
\begin{center}
\leavevmode
  \epsfxsize=14cm
 \epsffile{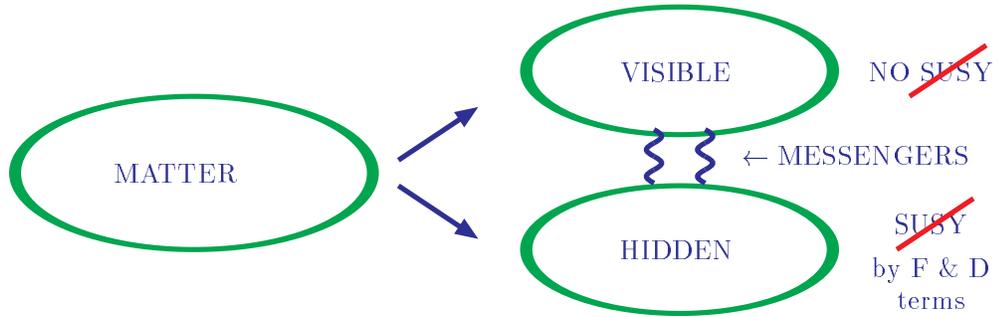}
\end{center}\vspace{-1cm}
\caption{Hidden Sector Scenario}\label{hidden}
\end{figure}

Since none of the fields of the MSSM can develop non-zero v.e.v.
to break SUSY without spoiling the gauge invariance, it is
supposed that spontaneous supersymmetry breaking takes place via
some other fields. The most common scenario for producing
low-energy supersymmetry breaking is called the {\em hidden
sector} one~\cite{hidden}. According to this scenario, there exist
two sectors: the usual matter belongs to the "visible" one, while
the second, "hidden" sector, contains fields which lead to
breaking of supersymmetry. These two sectors interact with each
other by exchange of some fields called {\em messengers}, which
mediate SUSY breaking from the hidden to the visible sector (see
Fig.\ref{hidden}).  There might be various types of messenger
fields: gravity, gauge, etc. Below we consider four possible
scenarios.

The hidden sector is the weakest part of the MSSM. It contains a lot of
ambiguities and leads to uncertainties of the MSSM predictions considered
below.

\subsection{The hidden sector: four scenarios}

So far there are known four main mechanisms to mediate SUSY breaking from a
hidden to a visible sector:\vspace{-0.2cm}
\begin{itemize}
\item Gravity mediation (SUGRA);\\[-0.7cm]
\item Gauge mediation;\\[-0.7cm]
\item Anomaly mediation;\\[-0.7cm]
\item Gaugino mediation.
\end{itemize}
Consider them in more detail. \vspace{0.2cm}

\noindent \underline{SUGRA}\vspace{0.2cm}

This mechanism is based on effective nonrenormalizable
interactions arising as a low-energy limit of supergravity
theories~\cite{gravmed}. In this case, two sectors interact with
each other via gravity. There are two types of scalar fields that
develop nonzero v.e.v.s, namely moduli fields $T$, which appear as
a result of compactification from higher dimensions, and the
dilaton field $S$, part of SUGRA  supermultiplet. These fields
obtain nonzero v.e.v.s for their $F$ components: $<F_T> \neq 0,
<F_S> \neq 0$, which leads to spontaneous SUSY breaking. Since in
SUGRA theory supersymmetry is local, spontaneous breaking leads to
Goldstone particle which is a Goldstone fermion in this case. With
the help of a super-Higgs effect this particle may be absorbed
into an additional component of a spin 3/2 particle, called {\em
gravitino}, which becomes massive.

SUSY breaking is then mediated to a visible sector via
gravitational interaction leading to the following SUSY breaking
scale:
 $$ M_{\ \ \ \ \Big/\hspace{-0.7cm}{SUSY}} \sim \frac{<F_T>}{M_{PL}}
  + \frac{<F_S>}{M_{PL}} ~\sim ~m_{3/2}, $$
where $m_{3/2}$ is the gravitino mass.

 The effective low-energy theory, which emerges, contains explicit soft
supersymmetry breaking terms
 \begin{eqnarray}
{\cal L}_{soft} &=& - \sum_i m_i^2 |A_i|^2 - \sum_i M_i(\lambda_i\lambda_i +
\bar \lambda_i \bar \lambda_i)
  -{\cal B}\ {\cal W}^{(2)}(A) - {\cal A}\ {\cal W}^{(3)}(A),
 \end{eqnarray}
where ${\cal W}^{(2)}$ and ${\cal W}^{(3)}$ are the quadratic and cubic
terms of a superpotential, respectively. The mass parameters are
 \begin{eqnarray*}
m_i^2 &\sim & \left(\frac{<F_S>}{M_{PL}}\right)^2 \sim m_{3/2}^2,
\ \ \ M_i \sim \frac{<F_S>}{M_{PL}} \sim m_{3/2},\\   {\cal B}
&\sim & \left(\frac{<F_T>}{M_{PL}}\right)^2 \sim m_{3/2}^2, \ \ \
{\cal A} \sim \frac{<F_{T,S}>}{M_{PL}} \sim m_{3/2}.
 \end{eqnarray*}
To have SUSY masses of an order of 1 TeV, one needs
$\sqrt{<F_{T,S}>} \sim 10 ^{11}$ GeV.

In spite of attractiveness of these mechanism in general, since we
know that gravity exists anyway, it is not truly substantiated due
to the lack of a consistent theory of quantum (super)gravity.
Among the problems of a supergravity mechanism  also are the large
freedom of parameters and the absence of automatic suppression of
flavour violation.\vspace{0.2cm}

 \noindent \underline{Gauge Mediation}\vspace{0.2cm}

In this version of a hidden sector scenario, the SUSY breaking
effects are mediated to the observable world not via gravity but
via gauge interactions \cite{gaugemed}. The messengers are the
gauge bosons and matter fields of the SM and of some GUT theory.
The hidden sector is necessary since the dynamical SUSY breaking
requires the fields with quantum numbers not compatible with the
SM. The advantage of this scenario is that one can construct a
renormalizable model with dynamic SUSY breaking, where in
principle all the parameters can be calculated.

\begin{figure}[ht]\vspace{-1cm}
\begin{center}
\leavevmode
  \epsfxsize=14cm
 \epsffile{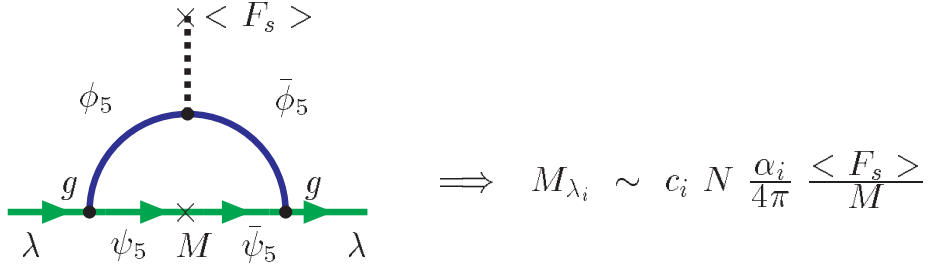}
\end{center}\vspace{-1cm}
\caption{Gaugino mass generation}\label{gaugino}
\end{figure}

\begin{figure}[ht]
\begin{center}
\leavevmode
  \epsfxsize=16cm
 \epsffile{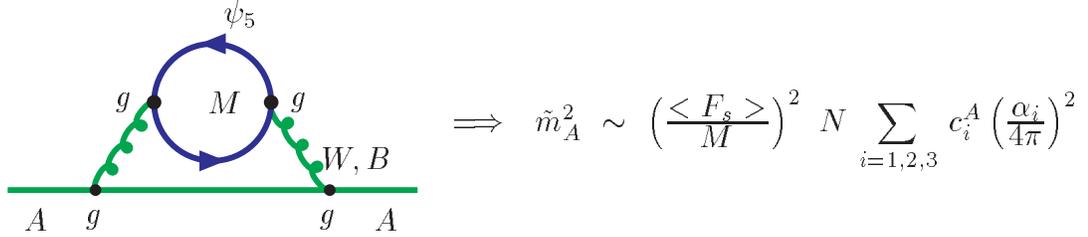}
\end{center}\vspace{-1cm}
\caption{Squark mass generation}\label{squark}
\end{figure}

Consider some simplest possibility where in a hidden sector one
has a singlet scalar superfield $S$ with nonzero v.e.v. $<F_S>\neq
0$. The messenger sector consists of some superfield $\Phi$, for
instance, $\bar 5$ of SU(5), that couples to $S$ and to the SM
fields with a superpotential
 \begin{equation}
{\cal W} \sim S \Phi^\dagger \Phi,  \ \ \ \  <S> =  M \neq 0.
 \end{equation}
Integrating out the messenger fields gives mass to gauginos at the one loop
level (see Fig.\ref{gaugino}) and to the scalar fields (squarks and
sleptons) at the two loop one (see Fig.\ref{squark}).
So, in gauge mediated scenario all the soft masses are correlated to the
gauge couplings and in this sense this scenario is more restrictive than the
SUGRA one. There is no problem with flavour violating processes as well,
since the soft terms automatically repeat the rigid sector.

It is remarkable that in this scenario the LSP happens to be the gravitino.
The mass of the gravitino is given by
 \begin{equation}
m_{\tilde G} \sim \frac{<F_S>}{M}\cdot \frac{M}{M_{PL}} \sim 10^{-14}
\frac{M}{[GeV]},
 \end{equation}
that leads to a very light gravitino field.

The problem of the gauge mediated SUSY breaking scenario emerges
in the Higgs sector since the Higgs mass mixing parameters, which
break an unwanted Peccei-Quin symmetry, cannot be generated by
gauge interactions only. In order to parameterize  some new
unknown interactions, two new inputs have to be introduced ($\mu$
and $B$ in SUGRA conventions). \vspace{0.2cm}

 \noindent \underline{Anomaly Mediation}\vspace{0.2cm}

An anomaly mediation mechanism assumes no SUSY breaking at the
tree level. SUSY breaking is generated due to conformal anomaly.
This mechanism refers to a hidden sector of a multidimensional
theory with the couplings being dynamic fields which may acquire
v.e.v.s. for their $F$ components \cite{anommed}. The external
field or scale dependence of the couplings emerges as a result of
conformal anomaly and that is why is proportional to the
corresponding $\beta$ functions. In the leading order one has
 \begin{eqnarray}
M_i(\Lambda) &\sim & b_i \alpha_i(\Lambda)\ \frac{<F_{T,S}>}{M_{PL}}\sim \
b_i \ \alpha_i \ m_{3/2}, \nonumber \\ m^2(\Lambda) &\sim& \ b_i^2\
\alpha_i^2(\Lambda)\  m^2_{3/2},
 \end{eqnarray}
where $b_i$ are the one-loop RG coefficients (see
eq.(\ref{susy1})).

This reminds supergravity mediation mechanism but with fixed coefficients.
It leads to two main differences: \\
 i) the inverted relation  between the gaugino masses at high energy scale
$$M_1:M_2:M_3 = b_1:b_2:b_3, $$
 ii) negative slepton mass squared (tachyons!) at the tree level.\\
This problem has to be cured. \vspace{0.2cm}

 \noindent \underline{Gaugino Mediation}\vspace{0.2cm}

\begin{figure}[ht]
\begin{center} \leavevmode
  \epsfxsize=11cm
 \epsffile{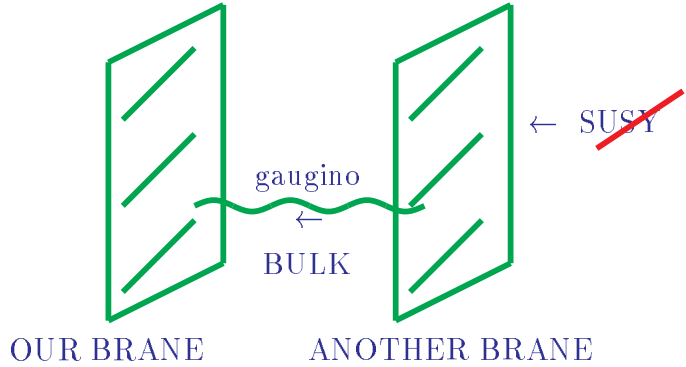}
\end{center}\vspace{-1cm}
\caption{Gaugino mediated SUSY breaking}\label{gaugmed}
\end{figure}

At last we would like to mention the gaugino mediation mechanism
of SUSY breaking \cite{gauginomed}. This is a less developed
scenario so far. It is based on a paradigm of a brane world.
According to this paradigm, there exists a multidimensional world
where our four dimensional space-time represents a brane of 4
dimensions. The fields of the SM live on the brane, while gravity
and some other fields can propagate in the bulk. There also exists
another brane where supersymmetry is broken.  SUSY breaking is
mediated to our brane via the fields propagating in the bulk. It
is assumed that the gaugino field plays an essential role in this
mechanism (see Fig.\ref{gaugmed})\vspace{-0.5cm}

All four mechanisms of soft SUSY breaking are different in details
but are  common in results. They generate gauge invariant soft
SUSY breaking operators of dimension $\leq 4$ of the form
 \begin{eqnarray}
{\cal L}_{soft} &=& - \sum_i m_i^2 |A_i|^2 - \sum_i M_i(\lambda_i\lambda_i +
\bar \lambda_i \bar \lambda_i) \nonumber \\ &&  -\sum_{ij}B_{ij}A_iA_j -
\sum_{ijk}A_{ijk}A_iA_jA_k + h.c., \label{br}
 \end{eqnarray}
where the bilinear and trilinear couplings $B_{ij}$ and $A_{ijk}$
are such that not to break the gauge invariance. These are the
only possible soft terms that do not break renormalizability of a
theory and preserve SUSY Ward identities for the rigid
terms~\cite{GG}.

\begin{figure}[ht]\vspace{-1cm} \begin{center}
\leavevmode
  \epsfxsize=12cm \epsfysize=11cm
 \epsffile{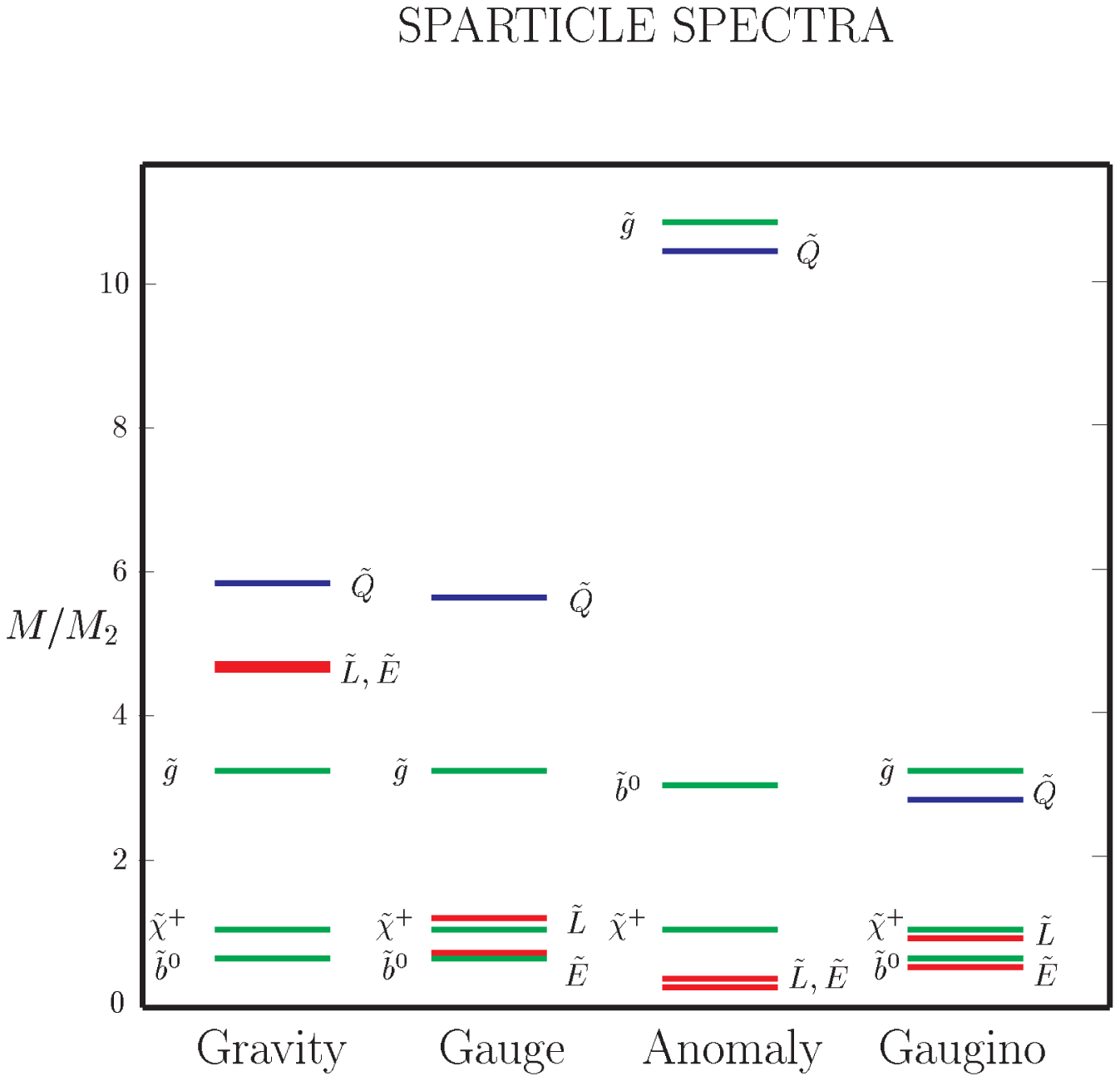}
\end{center}\vspace{-1cm} \caption{Superparticle spectra for various
mediation mechanisms}\label{spectra} \end{figure}

Predictions for the sparticle spectrum depend on the mechanism of
SUSY breaking. For comparison of four above-mentioned mechanisms
we show  in Fig.\ref{spectra} the sample spectra as the ratio to
the gaugino mass $M_2$~\cite{Peskin}.

In what follows, to calculate the mass spectrum of superpartners,
we need an explicit form of SUSY breaking terms. Applying
eq.(\ref{br}) to the MSSM and avoiding the $R$-parity violation
gives
\begin{eqnarray}
-{\cal L}_{Breaking} & = & \sum_{i}^{}m^2_{0i}|\varphi_i|^2+\left( \frac 12
\sum_{\alpha}^{}M_\alpha \tilde \lambda_\alpha\tilde \lambda_\alpha+
BH_1H_2\right. \label{soft}\\
 & + & \left. A^U_{ab}\tilde Q_a\tilde U^c_bH_2+A^D_{ab}\tilde Q_a
 \tilde D^c_bH_1+ A^L_{ab}\tilde L_a\tilde E^c_bH_1
  +h.c.\right) , \nonumber
\end{eqnarray}
where we have suppressed the $SU(2)$ indices.  Here $\varphi_i$
are all scalar fields, $\tilde \lambda_\alpha $  are the gaugino
fields, $\tilde Q, \tilde U, \tilde D$ and $\tilde L, \tilde E$
are the squark and slepton  fields, respectively, and $H_{1,2}$
are the SU(2) doublet Higgs fields.

Eq.(\ref{soft}) contains a vast number of free parameters which
spoils the prediction power of the model. To reduce their number,
we adopt the so-called universality hypothesis, i.e., we assume
the universality or equality of various soft parameters at a high
energy scale, namely, we put  all the spin 0 particle masses to be
equal to the universal value $m_0$, all the spin 1/2 particle
(gaugino) masses to be equal to $m_{1/2}$ and all the cubic and
quadratic terms, proportional to $A$ and $B$, to repeat the
structure of the Yukawa superpotential (\ref{R}). This is an
additional requirement motivated by the supergravity mechanism of
SUSY breaking. Universality is not a necessary requirement and one
may consider nonuniversal soft terms as well. However, it will not
change the qualitative picture presented below; so for simplicity,
in what follows we consider the universal boundary conditions. In
this case, eq.(\ref{soft}) takes the form
\begin{eqnarray}
-{\cal L}_{Breaking} & = & m_0^2\sum_{i}^{}|\varphi_i|^2+\left( \frac 12
m_{1/2}\sum_{\alpha}^{} \tilde \lambda_\alpha\tilde \lambda_\alpha\right.
\label{soft2}\\
 & + & \left. A[y^U_{ab}\tilde Q_a\tilde U^c_bH_2+y^D_{ab}\tilde Q_a
 \tilde D^c_bH_1+ y^L_{ab}\tilde L_a\tilde E^c_bH_1]
 +  B[\mu H_1H_2] +h.c.\right) , \nonumber
\end{eqnarray}

It should be noted that supergravity induced universality of the
soft terms is more likely to be valid at the Planck scale rather
than at the GUT one. This is because a natural scale for gravity
is $M_{Planck}$ while $M_{GUT}$ is the scale for  gauge
interactions. However, due to a small difference between these two
scales, it is usually ignored in the first approximation resulting
in minor uncertainties in the low-energy
predictions~\cite{polonsky}.

The soft terms explicitly break supersymmetry. As will be shown
later, they lead to the mass spectrum of superpartners different
from that of  ordinary particles. Remind that the masses of quarks
and leptons remain zero until $SU(2)$ invariance is spontaneously
broken.

\subsection{The soft terms and the mass formulas}

There are two main sources of the mass terms in the Lagrangian:
the $D$ terms and soft ones. With given values of
$m_0,m_{1/2},\mu,Y_t,Y_b,Y_\tau, A$, and $B$ one can construct the
mass matrices for all the particles. Knowing them at the GUT
scale, one can solve the corresponding RG equations, thus linking
the values at the GUT and electroweak scales. Substituting these
parameters into the mass matrices, one can predict the mass
spectrum of superpartners \cite{MSSM,spectrum,BEK}.

\subsubsection{Gaugino-higgsino mass terms}

The mass matrix for  gauginos, the superpartners of the gauge
bosons, and for higgsinos, the superpartners of the Higgs bosons,
is nondiagonal, thus leading to their mixing. The mass terms look
like
\begin{equation}
{\cal L}_{Gaugino-Higgsino}=
 -\frac{1}{2}M_3\bar{\lambda}_a\lambda_a
 -\frac{1}{2}\bar{\chi}M^{(0)}\chi -(\bar{\psi}M^{(c)}\psi + h.c.),
\end{equation}
where $\lambda_a , a=1,2,\ldots ,8,$ are the Majorana gluino
fields and
\begin{equation}
\chi = \left(\begin{array}{c}\tilde{B}^0 \\ \tilde{W}^3 \\
\tilde{H}^0_1 \\ \tilde{H}^0_2
\end{array}\right), \ \ \ \psi = \left( \begin{array}{c}
\tilde{W}^{+} \\ \tilde{H}^{+}
\end{array}\right)
\end{equation}
are, respectively, the Majorana neutralino and Dirac chargino
fields.

The neutralino mass matrix is
\begin{equation}
M^{(0)}=\left(
\begin{array}{cccc}
M_1 & 0 & -M_Z\cos\beta \sin_W & M_Z\sin\beta \sin_W \\ 0 & M_2 &
M_Z\cos\beta \cos_W   & -M_Z\sin\beta \cos_W  \\ -M_Z\cos\beta
\sin_W & M_Z\cos\beta \cos_W  & 0 & -\mu \\ M_Z\sin\beta \sin_W &
-M_Z\sin\beta \cos_W  & -\mu & 0
\end{array} \right),\label{neut}
\end{equation}
where $\tan\beta = v_2/v_1$ is the ratio of two Higgs v.e.v.s and
$\sin_W= \sin\theta_W$ is the usual sinus of the weak mixing
angle. The physical neutralino masses  $M_{\tilde{\chi}_i^0}$ are
obtained as eigenvalues of this matrix after diagonalization.

For charginos one has
\begin{equation}
M^{(c)}=\left(
\begin{array}{cc}
M_2 & \sqrt{2}M_W\sin\beta \\ \sqrt{2}M_W\cos\beta & \mu
\end{array} \right).\label{char}
\end{equation}
This matrix has two chargino eigenstates
$\tilde{\chi}_{1,2}^{\pm}$ with mass eigenvalues
\begin{equation}
M^2_{1,2}=\frac{1}{2}\left[M^2_2+\mu^2+2M^2_W \mp
\sqrt{(M^2_2-\mu^2)^2+4M^4_W\cos^22\beta
+4M^2_W(M^2_2+\mu^2+2M_2\mu \sin 2\beta )}\right].
\end{equation}

\subsubsection{Squark and slepton masses}

Non-negligible Yukawa couplings cause a mixing between the
electroweak eigenstates and the mass eigenstates of the third
generation particles.  The mixing matrices for
$\tilde{m}^{2}_t,\tilde{m}^{2}_b$ and $\tilde{m}^{2}_\tau$ are
\begin{equation} \label{stopmat}
\left(\begin{array}{cc} \tilde m_{tL}^2& m_t(A_t-\mu\cot \beta )
\\ m_t(A_t-\mu\cot \beta ) & \tilde m_{tR}^2 \end{array}  \right),
\nonumber
\end{equation}
\begin{equation} \label{sbotmat}
\left(\begin{array}{cc} \tilde  m_{bL}^2& m_b(A_b-\mu\tan \beta )
\\ m_b(A_b-\mu\tan \beta ) & \tilde  m_{bR}^2 \end{array}
\right), \nonumber
\end{equation}
\begin{equation} \label{staumat} \left(\begin{array}{cc}
\tilde  m_{\tau L}^2& m_{\tau}(A_{\tau}-\mu\tan \beta ) \\
m_{\tau}(A_{\tau}-\mu\tan \beta ) & \tilde m_{\tau R}^2
\end{array}  \right)               \nonumber
\end{equation}
with
\begin{eqnarray*}
  \tilde m_{tL}^2&=&\tilde{m}_Q^2+m_t^2+\frac{1}{6}(4M_W^2-M_Z^2)\cos
  2\beta ,\\
  \tilde m_{tR}^2&=&\tilde{m}_U^2+m_t^2-\frac{2}{3}(M_W^2-M_Z^2)\cos
  2\beta ,\\
  \tilde m_{bL}^2&=&\tilde{m}_Q^2+m_b^2-\frac{1}{6}(2M_W^2+M_Z^2)\cos
  2\beta ,\\
  \tilde m_{bR}^2&=&\tilde{m}_D^2+m_b^2+\frac{1}{3}(M_W^2-M_Z^2)\cos
  2\beta ,\\
 \tilde m_{\tau L}^2&=&\tilde{m}_L^2+m_{\tau}^2-\frac{1}{2}(2M_W^2-M_Z^2)\cos
2\beta ,\\ \tilde m_{\tau
R}^2&=&\tilde{m}_E^2+m_{\tau}^2+(M_W^2-M_Z^2)\cos
  2\beta
\end{eqnarray*}
and the  mass eigenstates are  the eigenvalues of these mass matrices. For
the light generations the mixing is negligible.

The first terms here ($\tilde{m}^2$) are the soft ones, which are
calculated using the RG equations starting from their values at
the GUT (Planck) scale. The second ones are the usual masses of
quarks and leptons and the last ones are the $D$ terms of the
potential.

\subsection{The Higgs potential}

As has already been mentioned, the Higgs potential in the MSSM is
totally defined by superpotential (and the soft terms). Due to the
structure of ${\cal W}$ the Higgs self-interaction is given by the
$D$-terms while the $F$-terms contribute only to the mass matrix.
The tree level potential is
\begin{eqnarray}
V_{tree}(H_1,H_2)&=&m^2_1|H_1|^2+m^2_2|H_2|^2-m^2_3(H_1H_2+h.c.)
\label{Higpot} \nonumber\\ &+&
\frac{g^2+g^{'2}}{8}(|H_1|^2-|H_2|^2)^2 +
\frac{g^2}{2}|H_1^+H_2|^2,
\end{eqnarray}
where $m_1^2=m^2_{H_1}+\mu^2, m_2^2=m^2_{H_2}+\mu^2$. At the GUT
scale $m_1^2=m^2_2=m_0^2+\mu^2_0, \ m^2_3=-B\mu_0$. Notice that
the Higgs self-interaction coupling in eq.(\ref{Higpot}) is  fixed
and  defined by the gauge interactions as opposed to the SM.

The potential (\ref{Higpot}), in accordance with supersymmetry, is
positive definite and stable. It has no nontrivial minimum
different from zero.  Indeed, let us write the minimization
condition for  the potential (\ref{Higpot})
\begin{eqnarray} \frac 12\frac{\delta V}{\delta
H_1}&=&m_1^2v_1 -m^2_3v_2+ \frac{g^2+g'^2}4(v_1^2-v_2^2)v_1=0,
\label{min1}
\\ \frac 12\frac{\delta V}{\delta H_2}&=&m_2^2v_2-m^2_3v_1+ \frac{
g^2+g'^2}4(v_1^2-v_2^2)v_2=0, \label{min2}
\end{eqnarray}
where we have  introduced the notation $$<H_1>\equiv v_1= v
\cos\beta , \ \  <H_2>\equiv v_2= v \sin\beta, \ \ \ \ v^2=
v_1^2+v_2^2,\ \ \  \tan\beta \equiv \frac{v_2}{v_1}.$$ Solution of
eqs.(\ref{min1}),(\ref{min2}) can be expressed in terms of $v^2$
and $\sin 2\beta$
\begin{equation} v^2=\frac{\displaystyle  4(m^2_1-m^2_2\tan^2 \beta
)}{\displaystyle (g^2+ g'^2)(\tan^2\beta -1)},\ \ \ \sin2\beta
=\frac{\displaystyle 2m^2_3}{\displaystyle m^2_1+m^2_2}.
\label{min}
\end{equation}
One can easily see from eq.(\ref{min}) that if
$m_1^2=m_2^2=m_0^2+\mu_0^2$, $v^2$ happens to be negative, i.e.
the minimum does not exist.  In fact, real positive solutions to
eqs.(\ref{min1}),(\ref{min2}) exist only if the following
conditions are satisfied \cite{haber}:
\begin{equation}
m_1^2+m_2^2 > 2 m_3^2, \ \ \ \  m_1^2m_2^2 < m_3^4 , \label{cond}
\end{equation}
which is not the case at the GUT scale. This means that
spontaneous breaking of the $SU(2)$  gauge invariance, which is
needed in the SM to give masses for all the particles, does not
take place in the MSSM.

This strong statement is valid, however, only at the GUT scale.
Indeed, going down with energy, the parameters of the potential
(\ref{Higpot}) are renormalized.  They become the ``running''
parameters with the energy scale dependence given by the RG
equations. The running of the parameters leads to a remarkable
phenomenon known as  {\em radiative spontaneous symmetry breaking}
to be discussed below.

Provided conditions (\ref{cond}) are satisfied, the mass matrices
at the tree level are \\ CP-odd components
 $P_1$ and $P_2$ :
\begin{equation}
{\cal M}^{odd} = \left.\frac{\partial^2 V}{\partial P_i \partial
P_j} \right |_{H_i=v_i} = \left( \begin{array}{cc}  \tan\beta &1
\\1& \cot\beta \end{array}\right) m_3^2,
\end{equation}
 CP-even neutral components $S_1$ and $S_2$:
\begin{equation}
{\cal M}^{even} = \left.\frac{\partial^2 V}{\partial S_i \partial
S_j} \right|_{H_i=v_i} = \left( \begin{array}{cc}  \tan\beta & -1
\\-1& \cot\beta \end{array}\right) m_3^2 +\left( \begin{array}{cc}
\cot\beta & -1 \\-1& \tan\beta \end{array}\right) M_Z\cos\beta
\sin\beta,
\end{equation}
Charged components $H^-$ and $H^+$:
\begin{equation}
{\cal M}^{charged} =\left.\frac{\partial^2 V}{\partial H^+_i
\partial H^-_j} \right|_{H_i=v_i} = \left( \begin{array}{cc}
\tan\beta &1 \\1& \cot\beta \end{array}\right)
 (m_3^2+M_W\cos\beta\sin\beta).
\end{equation} Diagonalizing the mass matrices, one gets the mass
eigenstates \cite{haber}:
 $$\begin{array}{l} \left\{
\begin{array}{lllr} G^0 \ &=& -\cos\beta P_1+\sin \beta P_2 , & \ \ \
Goldstone \ boson \ \to Z_0, \\ A \ &=& \sin\beta P_1+\cos \beta P_2 , & \
\ \ \ \ \ \ \ Neutral \ CP=-1 \ Higgs, \end{array}\right.\\  \\ \left\{
\begin{array}{lllr} G^+ &=& -\cos\beta (H^-_1)^*+\sin \beta H^+_2 , &\ \
Goldstone \ boson \ \to W^+, \\ H^+ &=& \sin\beta (H^-_1)^*+\cos \beta
H^+_2 , &\ \ \ Charged \ Higgs, \end{array}\right.\\ \\ \left\{
\begin{array}{lllr} h \ &=& -\sin\alpha S_1+\cos\alpha S_2 , & \ \ \ \ \ \
\ SM \ Higgs \ boson \ CP=1, \\ H \ &=& \cos\alpha S_1+\sin\alpha S_2 , &
\ \ \ \ \ \ \ Extra \ heavy \ Higgs \ boson , \end{array}\right.
\end{array}$$ where the mixing angle $\alpha $ is
given by $$ \tan 2\alpha = -\tan 2\beta
\left(\frac{m^2_A+M^2_Z}{m^2_A-M^2_Z}\right).$$ The physical Higgs bosons
acquire the following masses \cite{MSSM}:
 \begin{eqnarray} \mbox{CP-odd
neutral Higgs} \ \ A: && \ \ \ \ \ \ \ \ \ \ \ \ m^2_A = m^2_1+m^2_2,
\nonumber \\ \mbox{Charge Higgses} \ \ H^{\pm}: && \ \ \ \ \ \
 \ \ \ \ \ m^2_{H^{\pm}}=m^2_A+M^2_W ,
 \end{eqnarray}
CP-even neutral Higgses \ \ H, h:
\begin{equation} m^2_{H,h}=
\frac{1}{2}\left[m^2_A+M^2_Z \pm
\sqrt{(m^2_A+M_Z^2)^2-4m^2_AM_Z^2\cos^22\beta}\right],
\end{equation}
where, as usual,
 $$ M^2_W=\frac{g^2}{2}v^2, \ \
M^2_Z=\frac{g^2+g'^2}{2}v^2 .$$ This leads to the once celebrated
SUSY mass relations
\begin{equation}\begin{array}{c} m_{H^{\pm}} \geq M_W, \\[0.2cm]
m_h \leq m_A \leq M_H, \\[0.2cm] m_h \leq M_Z |\cos 2\beta| \leq
M_Z , \\[0.2cm]  m_h^2+m_H^2=m_A^2+M_Z^2.\end{array}\label{bound}
\end{equation}

Thus, the lightest neutral Higgs boson happens to be lighter than
the $Z$ boson, which clearly distinguishes it from the SM one.
Though we do not know the mass of the Higgs boson in the SM, there
are several indirect constraints leading to the lower boundary of
$m_h^{SM} \geq 135 $ GeV~\cite{bound1}. After including the
radiative corrections,  the mass of the lightest Higgs boson in
the MSSM, $m_h$, however  increases. We consider it in more detail
below.

\subsection{Renormalization group analysis}

To calculate the low energy values of the soft terms, we use the
corresponding RG equations. The one-loop RG equations for the
rigid MSSM couplings are \cite{Ibanez}
\begin{eqnarray}
\frac{d\tilde{\alpha}_i}{dt}&=&b_i \tilde{\alpha}_i^2, \ \ \ \ t\equiv \log
Q^2/M_{GUT}^2 \nonumber\\ \frac{dY_U}{dt} & = &
-Y_L\left(\frac{16}{3}\tilde{\alpha}_3 + 3\tilde{\alpha}_2 +
\frac{13}{15}\tilde{\alpha}_1-6Y_U-Y_D\right) , \nonumber \\ \frac{dY_D}{dt}
& = & -Y_D\left(\frac{16}{3}\tilde{\alpha}_3 + 3\tilde{\alpha}_2 +
\frac{7}{15}\tilde{\alpha}_1-Y_U-6Y_D-Y_L\right), \nonumber
\\ \frac{dY_L}{dt} & = &- Y_L\left( 3\tilde{\alpha}_2 +
  \frac{9}{5}\tilde{\alpha}_1-3Y_D-4Y_L\right), \label{eq}
\end{eqnarray}
where we use the notation $\tilde \alpha = \alpha/4\pi= g^2/16\pi^2,\ Y
=y^2/16\pi^2$.

 For the soft terms one finds
\begin{eqnarray}
\frac{dM_i}{dt} & = & b_i \tilde{\alpha}_iM_i . \nonumber\\ \frac{dA_U}{dt}
& = & \frac{16}{3}\tilde{\alpha}_3 M_3 + 3\tilde{\alpha}_2 M_2 +
\frac{13}{15}\tilde{\alpha}_1 M_1+6Y_UA_U+Y_DA_D, \nonumber
\\
 \frac{dA_D}{dt} & = & \frac{16}{3}\tilde{\alpha}_3 M_3 +
3\tilde{\alpha}_2 M_2 + \frac{7}{15}\tilde{\alpha}_1
M_1+6Y_DA_D+Y_UA_U+Y_LA_L, \nonumber \\ \frac{dA_L}{dt} & = &
3\tilde{\alpha}_2 M_2 + \frac{9}{5}\tilde{\alpha}_1 M_1+3Y_DA_D+4Y_LA_L,
\nonumber
\\ \frac{dB}{dt} & = & 3\tilde{\alpha}_2 M_2 +
\frac{3}{5}\tilde{\alpha}_1 M_1+3Y_UA_U+3Y_DA_D+Y_LA_L. \nonumber
\\
 \frac{d\tilde{m}^2_Q}{dt} & =&- \left[
(\frac{16}{3}\tilde{\alpha}_3M^2_3 + 3\tilde{\alpha}_2M^2_2 +
\frac{1}{15}\tilde{\alpha}_1M^2_1)
-Y_U(\tilde{m}^2_Q+\tilde{m}^2_U+m^2_{H_2}+A^2_U)\right. \nonumber
\\
 & &\left. -Y_D(\tilde{m}^2_Q+\tilde{m}^2_D+m^2_{H_1}+A^2_D)\right], \nonumber\\
\frac{d\tilde{m}^2_U}{dt} & = &- \left[(\frac{16}{3}\tilde{\alpha}_3M^2_3
+\frac{16}{15}\tilde{\alpha}_1M^2_1)
-2Y_U(\tilde{m}^2_Q+\tilde{m}^2_U+m^2_{H_2}+A^2_U)\right] , \nonumber\\
\frac{d\tilde{m}^2_D}{dt} & =
 &- \left[(\frac{16}{3}\tilde{\alpha}_3M^2_3+ \frac{4}{15}\tilde{\alpha}_1M^2_1)
-2Y_D(\tilde{m}^2_Q+\tilde{m}^2_D+m^2_{H_1}+A^2_D)\right],
\nonumber\\
  \frac{d\tilde{m}^2_L}{dt} & = &
-\left[3(
 \tilde{\alpha}_2M^2_2 + \frac{1}{5}\tilde{\alpha}_1M^2_1)
-Y_L(\tilde{m}^2_L+\tilde{m}^2_E+m^2_{H_1}+A^2_L)\right],
\nonumber
\\ \frac{d\tilde{m}^2_E}{dt} & = &-\left[ (
 \frac{12}{5}\tilde{\alpha}_1M^2_1)-2Y_L(
\tilde{m}^2_L+\tilde{m}^2_E+m^2_{H_1}+A^2_L)\right], \nonumber
\\
\frac{d\mu^2}{dt}&=&-\mu^2\left[3(\tilde{\alpha}_2+
\frac{1}{5}\tilde{\alpha}_1)-(3Y_U+3Y_D+Y_L)\right],\label{eq2}\\
\frac{dm^2_{H_1}}{dt} & = & -\left[3({\mbox a}_2M^2_2
+\frac{1}{5}{\mbox a}_1M^2_1)
-3Y_D(\tilde{m}^2_Q+\tilde{m}^2_D+m^2_{H_1}+A^2_D)\right.
\nonumber
\\ &&\left. - Y_L(\tilde{m}^2_L+\tilde{m}^2_E+m^2_{H_1}+A^2_L)\right] ,
\nonumber\\ \frac{dm^2_{H_2}}{dt} & = &-\left[ 3({\mbox a}_2M^2_2
+\frac{1}{5}{\mbox a}_1M^2_1)
-3Y_U(\tilde{m}^2_Q+\tilde{m}^2_U+m^2_{H_2}+A^2_U)\right].
\nonumber
\end{eqnarray}

\begin{figure}[hbt]\vspace{-0.5cm}
\begin{flushleft} \leavevmode
\epsfxsize=8cm \epsffile{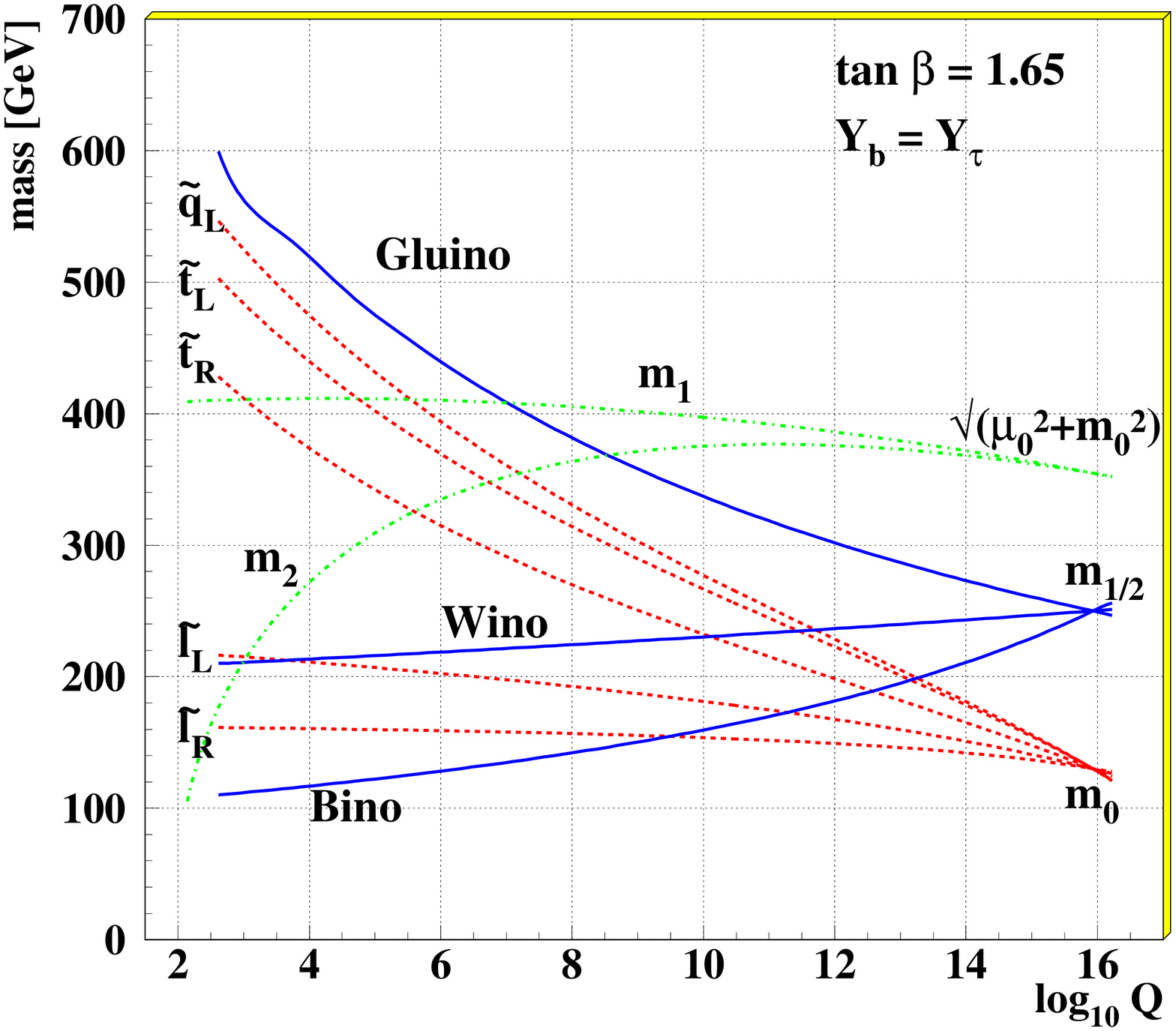}
\end{flushleft}
\vspace{-8.5cm}\hspace{8cm}
 \leavevmode
       \epsfxsize=8cm
\epsffile{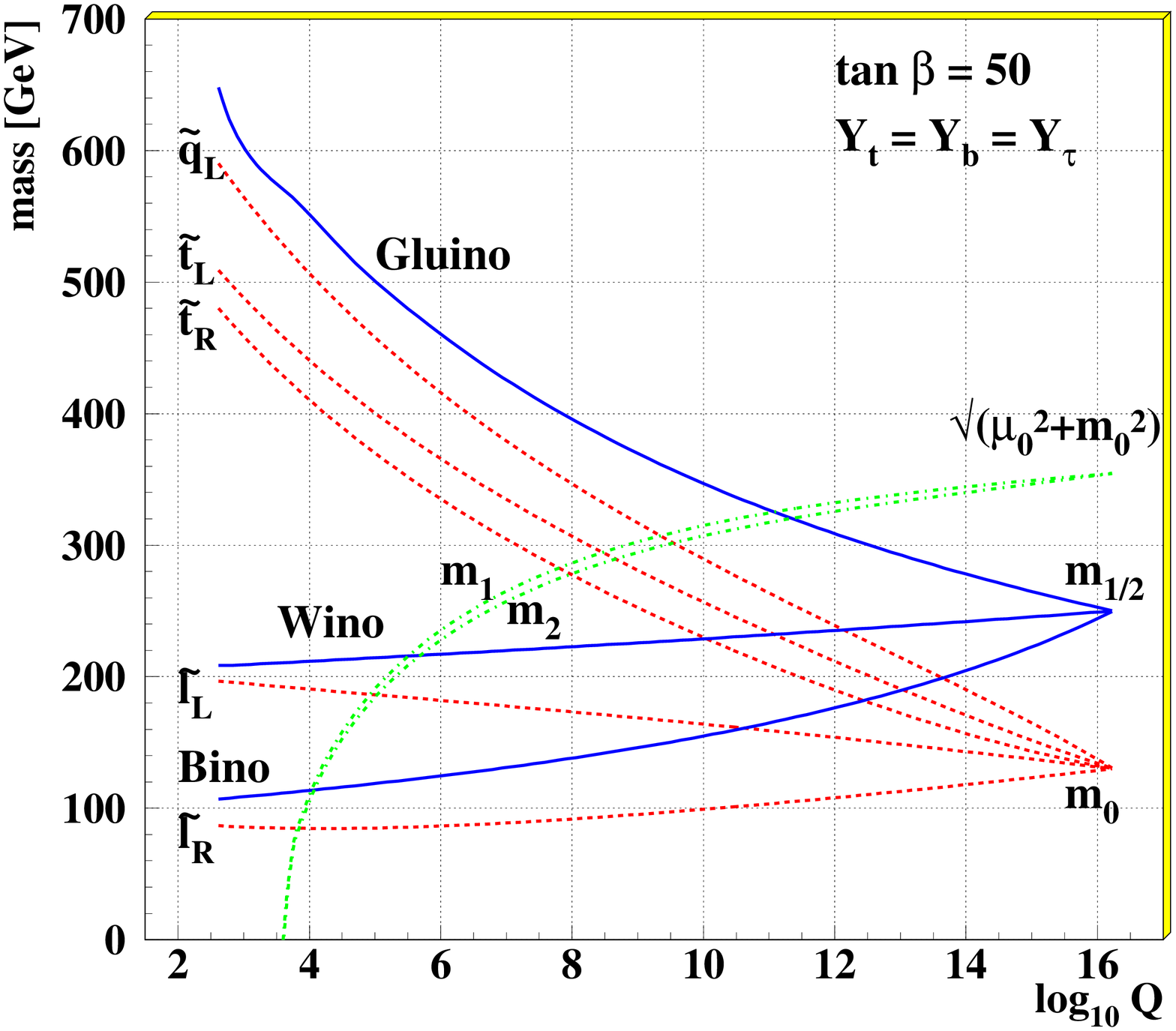}\vspace{-0.5cm}
\caption{An example of
evolution of sparticle masses and soft supersymmetry breaking parameters
$m_1^2=m^2_{H_1}+\mu^2$ and $m_2^2=m^2_{H_2}+\mu^2$ for low (left) and
high (right) values of $\tan\beta$ } \label{16}
\end{figure}

Having all the RG equations, one can now find the RG flow for the
soft terms.  To see what happens at lower scales, one has to run
the RG equations for the mass parameters in the opposite direction
from the GUT to the EW scale. Let us take some initial values of
the soft masses at the GUT scale in the interval between $10^2\div
10^3$ GeV consistent with the SUSY scale suggested by unification
of the gauge couplings (\ref{MSUSY}). This leads to the following
RG flow of the soft terms shown in
Fig.\ref{16}.~\cite{spectrum,BEK}
%
%

One should mention the following general features common to any
choice of initial conditions:

 i) The gaugino masses follow the running of the gauge couplings
 and split at low energies. The gluino mass is running faster
 than the others  and is usually the heaviest due to the strong interaction.

 ii) The squark and slepton masses also split at low energies, the
 stops (and sbottoms) being the lightest due to relatively big Yukawa couplings
 of the third generation.

  iii) The Higgs masses (or at least one of them) are running down
 very quickly and may even become negative.

 To calculate the masses one has also to take into account the
mixing between various states (see eqs.(\ref{neut},\ref{char},
\ref{stopmat}-\ref{staumat}).

Numerical solutions allow one to understand the significance of
different initial conditions for the evolution down to low
energies. As an example we present below the results of a
numerical solution to the RG equations for the soft terms in the
case of low values of $\tan\beta$. In this case, one can ignore
the bottom and tau Yukawa couplings and keep only the top one.
Taking $M_{GUT}=2.0\cdot 10^{16}$ GeV, $\alpha(M_{GUT})\approx
1/24.3, \ Y_t(M_{GUT}) \approx \tilde{\alpha}(M_{GUT}), \ tan\beta
=1.65$, one gets the following numerical results~\cite{BEK}:
%
%
 \begin{eqnarray*}
M_3(M_Z) &=& 2.7 \ m_{1/2}, \\ M_2(M_Z) &=& 0.8 \ m_{1/2}, \\ M_1(M_Z) &=&
0.4 \ m_{1/2}, \\ \mu(M_Z) &=& 0.63\ \mu_0, \\ A_t(M_Z)&=& 0.009\ A_t(0)
-1.7\ m_{1/2},
\\
  \tilde{m}^2_{E_{L}}(M_Z)    & = &
        m^2_0 + 0.52\ m^2_{1/2} - 0.27\ \cos(2\beta) M_Z^2, \\
  \tilde{m}^2_{\nu_{L}}(M_Z)  & = &
        m^2_0 + 0.52\ m^2_{1/2} + 0.5\ \cos(2\beta) M_Z^2, \\
  \tilde{m}^2_{E_{R}}(M_Z)    & = &
        m^2_0 + 0.15\ m^2_{1/2} - 0.23\ \cos(2\beta) M_Z^2, \\
  \tilde{m}^2_{U_{L}}(M_Z)    & = &
        m^2_0 + 6.6\ m^2_{1/2} + 0.35\ \cos(2\beta) M_Z^2, \\
  \tilde{m}^2_{D_{L}}(M_Z)    & = &
        m^2_0 + 6.6\ m^2_{1/2} - 0.42\ \cos(2\beta) M_Z^2, \\
  \tilde{m}^2_{U_{R}}(M_Z)    & = &
        m^2_0 + 6.2\ m^2_{1/2} + 0.15\ \cos(2\beta) M_Z^2, \\
  \tilde{m}^2_{D_{R}}(M_Z)    & = &
        m^2_0 + 6.1\ m^2_{1/2} - 0.07\ \cos(2\beta) M_Z^2, \\
  \tilde{m}^2_{b_{R}}(M_Z) & = & \tilde{m}^2_{D_{R}}, \\
  \tilde{m}^2_{b_{L}}(M_Z) & = &
       \tilde{m}^2_{D_{L}} - 0.48\ m^2_0 - 1.21\ m^2_{1/2}, \\
  \tilde{m}^2_{t_{R}}(M_Z) & = &
       \tilde{m}^2_{U_{R}} - 0.96\ m^2_0 - 2.42\ m^2_{1/2}, \\
  \tilde{m}^2_{t_{L}}(M_Z) & = &
       \tilde{m}^2_{U_{L}}  - 0.48\ m^2_0 - 1.21\ m^2_{1/2},\\
  m^2_1(M_Z)&=&m_0^2+0.40\ \mu_0^2+0.52\ m^2_{1/2}, \\
  m^2_2(M_Z)&=&-0.44\ m_0^2+0.40\ \mu_0^2 - 3.11\ m^2_{1/2}- 0.09\
  A_0m_{1/2}- 0/02\ A_0^2.
 \end{eqnarray*}

\begin{figure}[ht]\vspace{0.2cm}
 \leavevmode
  \epsfxsize=7.5cm
 \epsffile{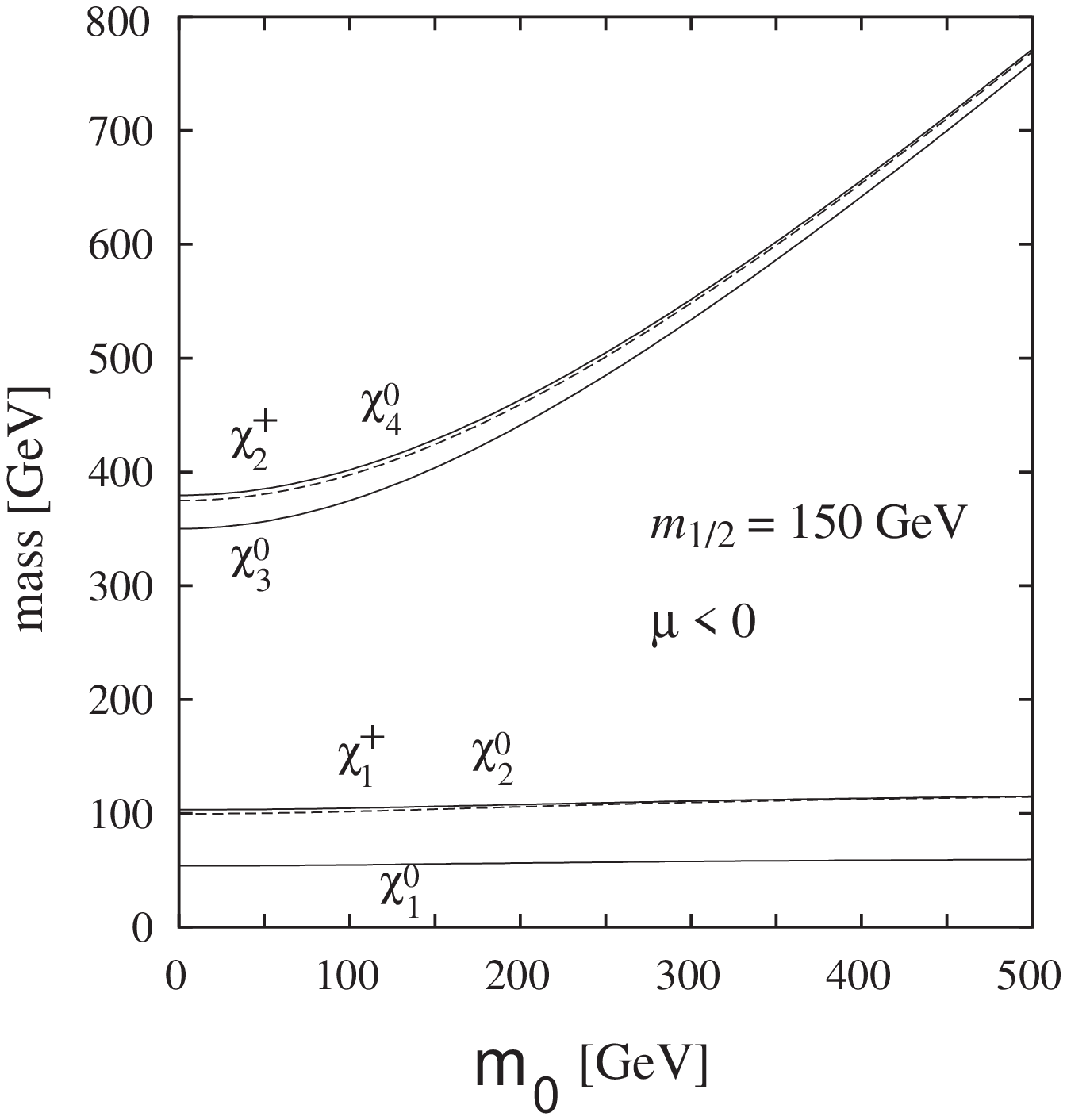}
 \hspace*{0.3cm}\vspace{-1.5cm}
 \epsfxsize=8cm\epsffile{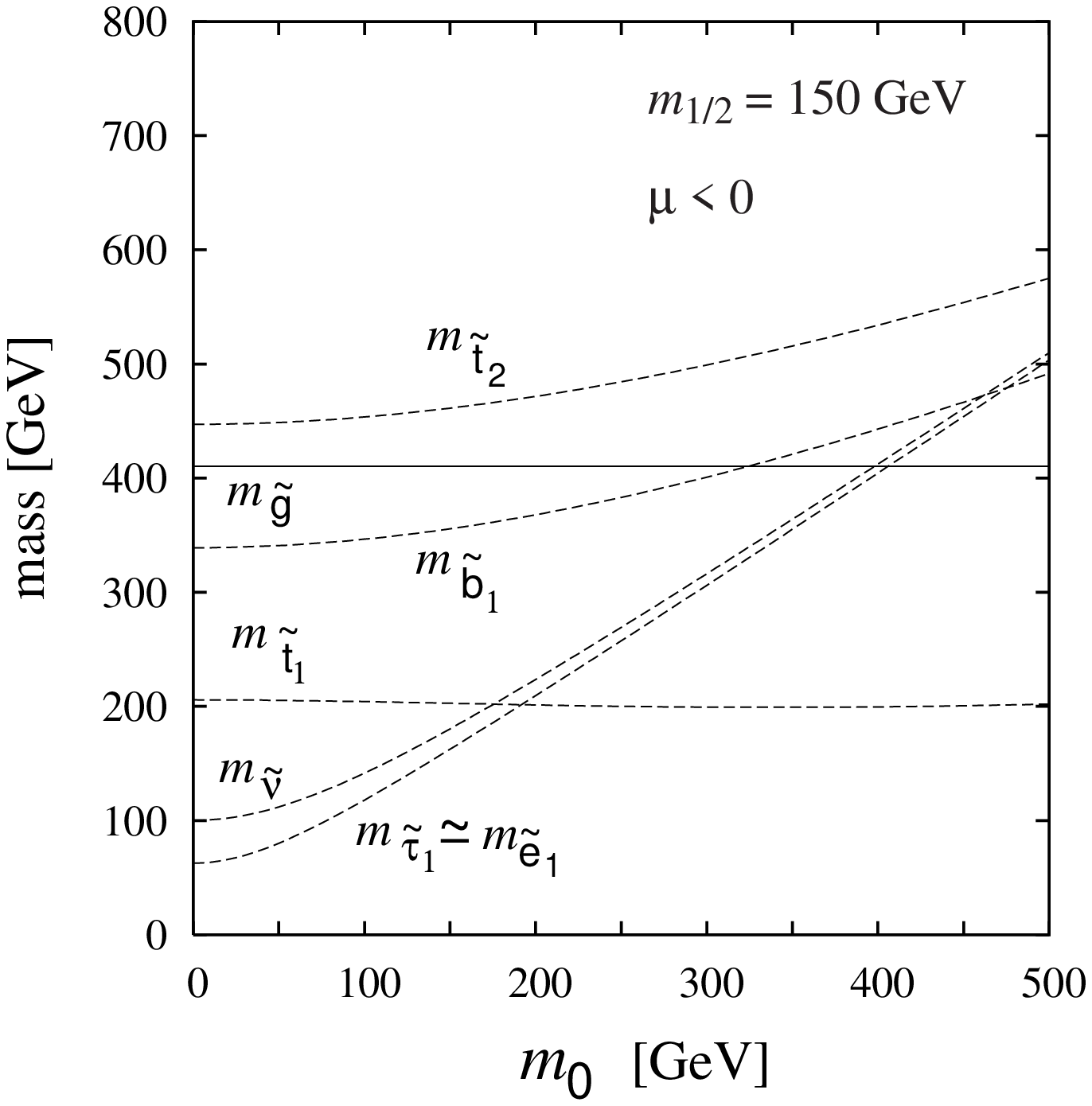}
\vspace{1cm}\caption{The masses of sparticles as functions of the initial
value $m_0$} \label{fig:barger}
\end{figure}

Typical dependence of the mass spectra on the initial conditions
($m_0$) is also shown in Fig.\ref{fig:barger} ~\cite{Barger}. For
a given value of $m_{1/2}$ the masses of the lightest particles
are practically independent of $m_0$, while the heavier ones
increase with it monotonically  as it follows also from the
numerical solutions given above. One can see that the lightest
neutralinos and charginos as well as the stop squark may be rather
light.

\subsection{Radiative electroweak symmetry breaking}

The running of the Higgs masses leads to the phenomenon known as
{\em radiative electroweak symmetry breaking}. By this we mean the
following: at the GUT energy scale both the Higgs mass parameters
$m_1^2$ and $m_2^2$ are positive, and the Higgs potential has no
nontrivial minima. However, when running down to the EW scale due
to the radiative corrections they may change the sign so that the
potential develops  a nontrivial minimum. At this minimum the
electroweak symmetry happens to be spontaneously broken. Thus,
contrary to the SM, where one has to choose the negative sign of
the Higgs mass squared  "by hand", in the MSSM the effect of
spontaneous symmetry breaking is triggered by the radiative
corrections.

Indeed, one can see in Fig.\ref{16}  that $m_2^2$ (or both $m_1^2$
and $m_2^2$) decreases when going down from the GUT scale to the
$M_Z$ scale and  can even become negative. This is the effect of
the large top (and bottom)  Yukawa couplings in the RG equations.
As a result, at some value of $Q^2$  the conditions (\ref{cond})
are satisfied, so that the nontrivial minimum appears. This
triggers spontaneous breaking of the $SU(2)$ gauge invariance. The
vacuum expectations of the Higgs fields acquire nonzero values and
provide masses to  quarks, leptons and $SU(2)$ gauge bosons, and
additional masses to their superpartners.

In this way one also obtains the explanation of why the two scales
are so much different. Due to the logarithmic running of the
parameters, one needs  a long "running time" to get $m_2^2$ (or
both $m_1^2$ and $m_2^2$) to be negative when starting from a
positive value of the order of $M_{SUSY}\sim 10^2 \div 10^3$ GeV
at the GUT scale.

\section{Constrained MSSM}
 \setcounter{equation} 0

\subsection{Parameter space of the MSSM}

The Minimal Supersymmetric Standard Model has the following free
parameters:
\begin{itemize}
\item Three gauge couplings $\alpha_i$.
\item The matrices of the Yukawa couplings $y^i_{ab}$, where $i = L,
  U, D$.
\item The Higgs field mixing parameter  $\mu $.
\item The soft supersymmetry breaking parameters.
\end{itemize}
Compared to the SM there is an additional Higgs mixing parameter,
but the Higgs self-coupling, which is arbitrary in the SM,  is
fixed by supersymmetry. The main uncertainty comes from the
unknown soft terms.

With universality hypothesis one is left with the following set of
5 free parameters defining the mass scales
 $$ \mu, \ m_0, \ m_{1/2}, \ A \ \mbox{and}\  B. $$
Parameter $B$ is usually traded for $\tan\beta$, the ratio of the
v.e.v.s of the two Higgs fields.

In particular models, like in SUGRA or gauge and anomaly
mediation, some of soft parameters may be related to each other.
However, since the mechanism of SUSY breaking is unknown, in what
follows we consider them as free phenomenological parameters to be
fitted by experiment. The experimental constraints are sufficient
to determine these parameters, albeit with large uncertainties.
The statistical analysis yields the probability for every point in
the SUSY parameter space, which allows one to calculate the cross
sections for the expected new physics of the MSSM at the existing
or future accelerators (LEP II, Tevatron, LHC).

While choosing parameters and making predictions, one has two
possible ways to proceed:

 i) take the low-energy parameters as input, impose the constraints,
define the allowed parameter space and calculate the spectrum and
cross-sections as functions of these parameters. They might be the
superparticle masses $\tilde{m}_{t1},\tilde{m}_{t2}, m_A$,
$\tan\beta$, mixings $X_{stop},\mu$, etc.

 ii) take the high-energy parameters as input, run the RG
equations, find the low-energy values,  then impose the constrains
and define the allowed parameter space for initial values. Now the
calculations can be carried out in terms of the initial
parameters. They might be, for example, the above mentioned 5 soft
parameters.

Both the ways are used in a phenomenological analysis. We show
below how it works in practice.

\subsection{The choice of constraints}

Among the constraints that we are going to impose on the MSSM
model are those which follow from the comparison of the SM with
experimental data, from the experimental limits on the masses of
as yet unobserved particles, etc, and also those that follow from
the ideas of unification and from SUSY GUT models. Some of them
look very obvious while the others depend on a choice. Perhaps,
the most remarkable fact is that all of them can be fulfilled
simultaneously. The only model where one can do it is proved to be
the MSSM.

In our analysis we impose the following constraints on the
parameter space of the MSSM:

$\bullet$ Gauge coupling constant unification; \\ This is one of
the most  restrictive constraints, which we have discussed in Sect
2. It fixes the scale of SUSY breaking of an order of 1 TeV.

$\bullet$ $M_Z$ from electroweak symmetry breaking;\\ Radiative
corrections trigger spontaneous symmetry breaking in the
electroweak sector.  In this case, the Higgs potential does not
have its minimum for all fields equal to zero, but the minimum is
obtained for nonzero vacuum expectation values of the fields.
Solving $M_Z$ from eq.(\ref{min}) yields
\begin{equation}
\label{defmz} M_Z^2=2\frac{m_1^2-m_2^2\tan^2\beta}{\tan^2\beta-1}
.
\end{equation}
To get the right value of $M_Z$ requires proper adjustment of
parameters. This condition determines the value of $\mu$ for given
values of $m_0$ and $m_{1/2}$.

$\bullet$ Yukawa coupling constant unification;\\ The masses of top,
bottom and $\tau$ can be obtained from the low energy values of the
running Yukawa couplings \begin{equation} m_t=y_t\ v\sin\beta, \ \
m_b=y_b\ v\cos\beta, \ \ m_\tau=y_\tau \ v\cos\beta . \label{yuk}
\end{equation} Eq.(\ref{yuk}) is written for the so-called running masses.
They can be translated to the pole masses with account taken of
the radiative corrections. For the pole masses of the third
generation the following values are taken~\cite{mtop}, \cite{SM}
\begin{eqnarray}
 M_t    & = & 174.3     \pm 5.1~     \mbox{GeV}/c^2,
\nonumber \\  M_b    & = &   4.94  \pm  0.15~ \mbox{GeV}/c^2,\\
 M_\tau & = &   1.7771\pm  0.0005~\mbox{GeV}/c^2.
\nonumber
\end{eqnarray}

\begin{figure}[ht]
\begin{center}
 \leavevmode
  \epsfxsize=10cm
 \epsffile{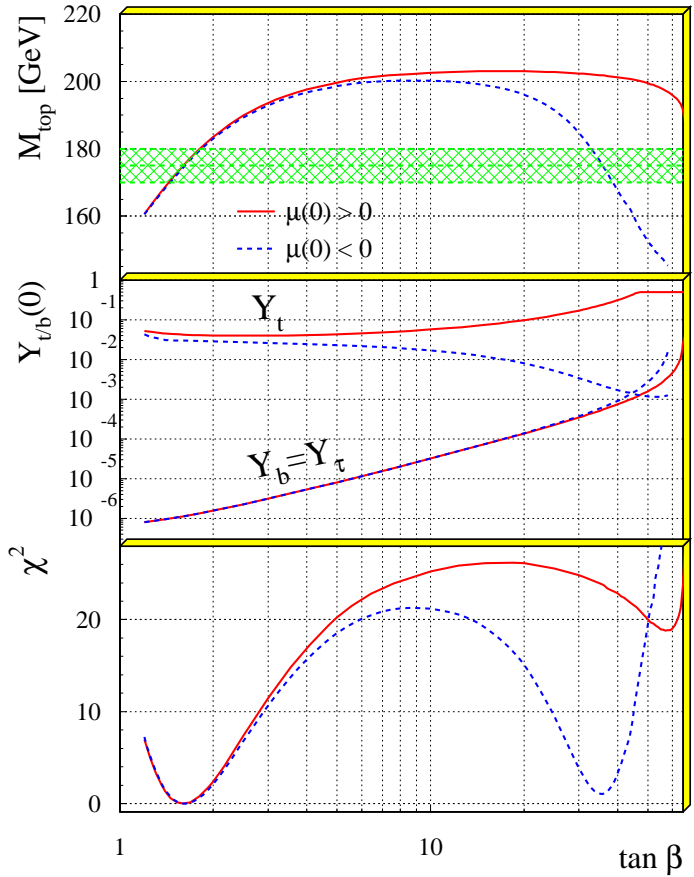}
\end{center}\vspace{-1cm}
\caption{The upper part shows the top quark mass as a function of
$\tan\beta$ for $m_0= 600$ GeV, $m_{1/2}=400$ GeV. The middle part
shows the corresponding values of the Yukawa couplings at the GUT
scale and the lower part of the $\chi^2$ values.} \label{fig:tb}
\end{figure}

The requirement of bottom-tau Yukawa coupling unification strongly
restricts the possible solutions in  $m_t$ versus $\tan\beta$
plane~\cite{bbog}-\cite{ara91} as it can be seen from
Fig.\ref{fig:tb}.

$\bullet$ Branching ratio $BR(b\to s \gamma)$; \\ The branching
ratio $BR(b \to s\gamma)$ has been measured by the
CLEO~\cite{CLEO} collaboration and later by ALEPH~\cite{ALBSG} and
yields the world average of $BR(b\to s \gamma)=(3.14\pm0.48)\cdot
10^{-4}$. The Standard Model contribution to this process comes
from the $W-t$ loop and gives a prediction which is very close to
the experimental value leaving little space for SUSY. In the MSSM,
this flavour changing neutral current (FCNC) receives  additional
contributions from the $H^\pm-t$, $\tilde{\chi}^\pm - \tilde{t}$
and $\tilde{g}-\tilde{q}$ loops. The $\tilde{\chi}^0 -\tilde{t}$
loops are much smaller~\cite{borz,bsgamm3}. In the leading order,
SUSY contribution may be rather big, exceeding the experimental
value by several standard deviations. However, the NLO corrections
are essential.

This requirement imposes severe restrictions on the parameter
space, especially for the case of large $\tan\beta$.

$\bullet$ Experimental lower limits on SUSY masses; \\ SUSY
particles have not been found so far and from the searches at LEP
one knows  the lower limit on the charged lepton and chargino
masses of about  half of the centre of mass energy~\cite{LEPSUSY}.
The lower limit on the neutralino masses  is smaller. The lower
limit on the Higgs mass is roughly given by the c.m.e. minus the
Z-boson mass. These limits restrict the  minimal values for the
SUSY mass parameters. There exist also limits on squark and gluino
masses from the hadron colliders~\cite{TEVSUSY}, but these limits
depend on the assumed decay modes.  Furthermore, if one takes the
limits given above into account, the constraints from the limits
on all other particles are usually fulfilled, so they do not
provide additional reductions of the parameter space in the case
of the minimal SUSY model.

$\bullet$ Dark Matter constraint; \\ Abundant evidence of the
existence of nonrelativistic, neutral, nonbaryonic dark matter
exists in our Universe~\cite{borner,kolb}. The lightest
supersymmetric particle (LSP) is supposedly stable and would be an
ideal candidate for dark  matter.

The present lifetime of the universe is at least $10^{10}$ years,
which implies an upper limit on the expansion rate
 and correspondingly on the total relic abundance.
Assuming $h_0>0.4$ one finds that  the contribution of
 each relic particle species $\chi$  has to obey~\cite{kolb}
$$\Omega_\chi h^2_0<1,$$ where $\Omega_\chi h^2$ is the ratio of
the relic particle density of particle $\chi$ and the critical
density, which overcloses the Universe. This bound can only be
met, if most of the LSP's   annihilated into fermion-antifermion
pairs, which in turn would annihilate into  photons again.

Since the neutralinos are mixtures of gauginos and higgsinos, the
annihilation can occur both, via s-channel exchange of the $Z^0$
and Higgs bosons and
 t-channel exchange of a scalar particle, like a selectron~\cite{relic}.
This constrains the parameter space, as discussed by many
groups~\cite{relictst}-\cite{bop}.

$\bullet$ Proton life-time constraint; \\ There are two sources of
proton decay in SUSY GUTs. The first one is the same as in
non-SUSY theories and is related to the s-channel exchange of
heavy gauge bosons. To avoid contradiction with  experiment, the
unification scale has to be above $10^{15}$ GeV which is usually
satisfied in any SUSY GUT.

The second source is more specific to SUSY models. The proton
decay in this case takes place due to the loop diagrams with the
exchange of heavy higgsino triplets. The preferable decay mode in
this case is $p \to \bar \nu  K$ or $p \to \mu^+ K$ instead of $p
\to  e^+\pi$ in non-SUSY GUTs. The decay rate in this case depends
on a particular GUT model and it is not  so easy to satisfy the
experimental requirements.

Having in mind the above mentioned constraints one can try to fix
the arbitrariness in the parameters. In a kind of a statistical
analysis, in which all the constraints are implemented in a
$\chi^2$ definition, one can find the most probable region of the
parameter space by minimizing the $\chi^2$ function. For the
purpose of this analysis the following $\chi^2$ definition is used
\cite{BEK}:
\begin{eqnarray} \chi^2 & = &
{\sum_{i=1}^3\frac{(\alpha_i^{-1}(M_Z)-\alpha^{-1}_{MSSM_i}(M_Z))^2}
{\sigma_i^2}}\nonumber\\ & &
+\frac{(M_Z-91.18)^2}{\sigma_Z^2}+{\frac{(M_t -
174)^2}{\sigma_t^2}} \nonumber  \\ & &
+\frac{(M_b-4.94)^2}{\sigma_b^2}+\frac{(M_\tau-1.7771)^2}{\sigma_\tau^2}
\nonumber    \\ & & +{\frac{(Br(b\to s\gamma)-3.14 \times
10^{-4})^2} {\sigma(b\to s\gamma)^2}}  \\
 & &+{\frac{(\Omega h^2-1)^2}{\sigma^2_\Omega}}\qquad(for ~\Omega h^2 > 1)
\nonumber   \\
 & &+{\frac{(\tilde{M}-\tilde{M}_{exp})^2}{\sigma_{\tilde{M}}^2}}
 \qquad{(for~\tilde{M} < \tilde{M}_{exp})}  \nonumber \\
 & &+{\frac{(\tilde{m}_{LSP}-\tilde{m}_{\chi})^2}{\sigma_{LSP}^2}}
 \qquad{(for~ \tilde{m}_{LSP}~charged )}.\nonumber   \label{chi2}
\end{eqnarray}
The first six terms are used to enforce gauge coupling
unification, electroweak symmetry breaking and $b-\tau$ Yukawa
coupling unification, respectively. The following two terms impose
the constraints from $b \to s\gamma$ and the relic density, while
the  last terms require the SUSY masses to be above the
experimental lower limits and the lightest supersymmetric particle
(LSP) to be  a neutralino since a charged stable LSP would have
been observed. The input and fitted output variables have been
summarized in Table \ref{t1}.
\begin{table*}[htb]
\begin{center}
\normalsize
\begin{tabular}{|c||c||c|c|}
\hline
                     &&
                       \multicolumn{2}{|c|}{Fit parameters}      \\
\cline{3-4} exp.~input data &$\Rightarrow$ & low $\tan\beta$
                                               & high $\tan\beta$     \\
\hline $\alpha_1,\alpha_2,\alpha_3$ && $M_{GUT},~\alpha_{GUT}$ &
$M_{GUT}, ~\alpha_{GUT}$ \\
$m_t$                   &   & $Y_t^0,~Y_b^0=Y_\tau^0$ &
                                   $Y_t^0=Y_b^0=Y_\tau^0$ \\
$m_b$           & minimize & $m_0, m_{1/2}$ & $m_0, m_{1/2}$  \\
 $m_\tau$        & $\chi^2$ & $\tan\beta$ &    $\tan\beta$      \\
$M_Z$                          &   & $\mu$ &  $\mu$      \\
$b \to s\gamma$                        &   & $(A_0)$ & $A_0$ \\
 $\tau_{universe}$           &   &             &               \\
\hline
\end{tabular} \end{center}
\caption[]{\label{t1}Summary of fit input and output variables.}
\end{table*}

The five-dimensional parameter space of the MSSM is big enough to
be represented in a two- or three-dimensional picture. To make our
analysis more clear, we consider various  projections of the
parameter space.

We first choose the value of the Higgs mixing parameter $\mu$ from
the requirement of radiative EW symmetry breaking, then we take
the values of $\tan\beta$ from the requirement of Yukawa coupling
unification (see Fig.\ref{fig:tb}). One finds two possible
solutions: low $\tan\beta$ solution corresponding to $\tan\beta
\approx 1.7$ and high $\tan\beta$ solution corresponding to
$\tan\beta \approx 30\div 60$. In what follows, we refer to these
two solutions as low and high $\tan\beta$ scenarios, respectively.

What is left are the values of the soft parameters $A,\ m_0$ and
$m_{1/2}$. However, the role of the trilinear coupling $A$ is not
essential since at low energies it runs to the infra-red fixed
point and is almost independent of initial conditions. Therefore,
imposing the above-mentioned constraints, the parameter space of
the MSSM  is reduced to a two dimensional one. In what follows, we
consider the plane $m_0,m_{1/2}$ and find the allowed region in
this plane. Each point at this plane  corresponds to a fixed set
of parameters and allows one to calculate the spectrum, the
cross-sections and  other quantities of interest.

\begin{figure}[ht]
\begin{center}\vspace{-0.5cm}
 \leavevmode
  \epsfxsize=13cm
 \epsffile{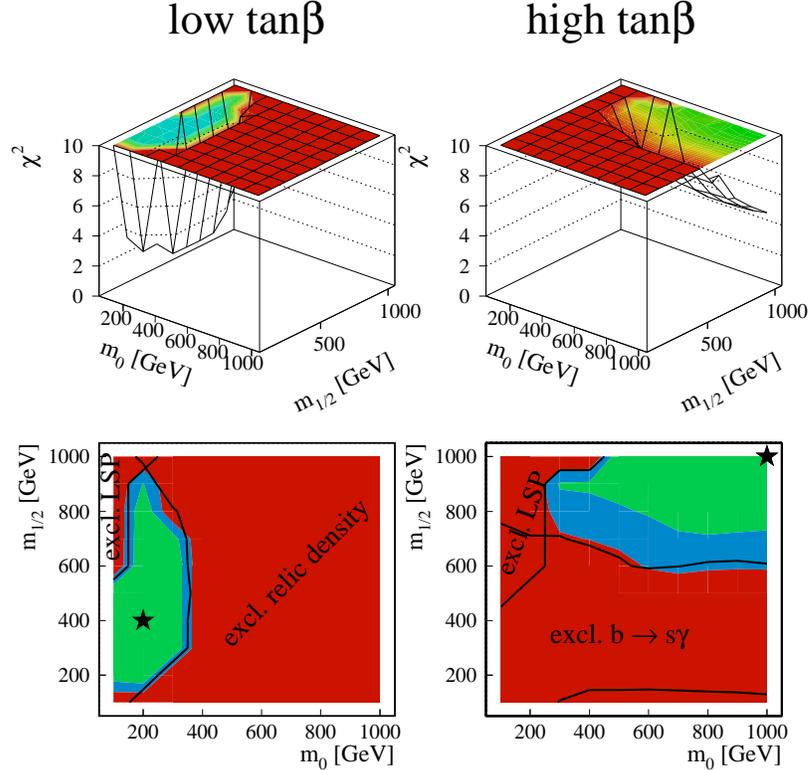}
\end{center}\vspace{-0.8cm} \caption{The $\chi^2$-distribution  for low
and high $\tan\beta$ solutions. The different shades in the
projections indicate steps of $\Delta \chi^2=4$, so basically only
the light shaded region is allowed. The stars indicate the optimum
solution. Contours enclose domains by the particular constraints
used in the analysis.} \label{chi}
 \end{figure}

\begin{figure}[ht] \begin{center}\vspace{-1cm}
 \leavevmode
  \epsfxsize=13cm
 \epsffile{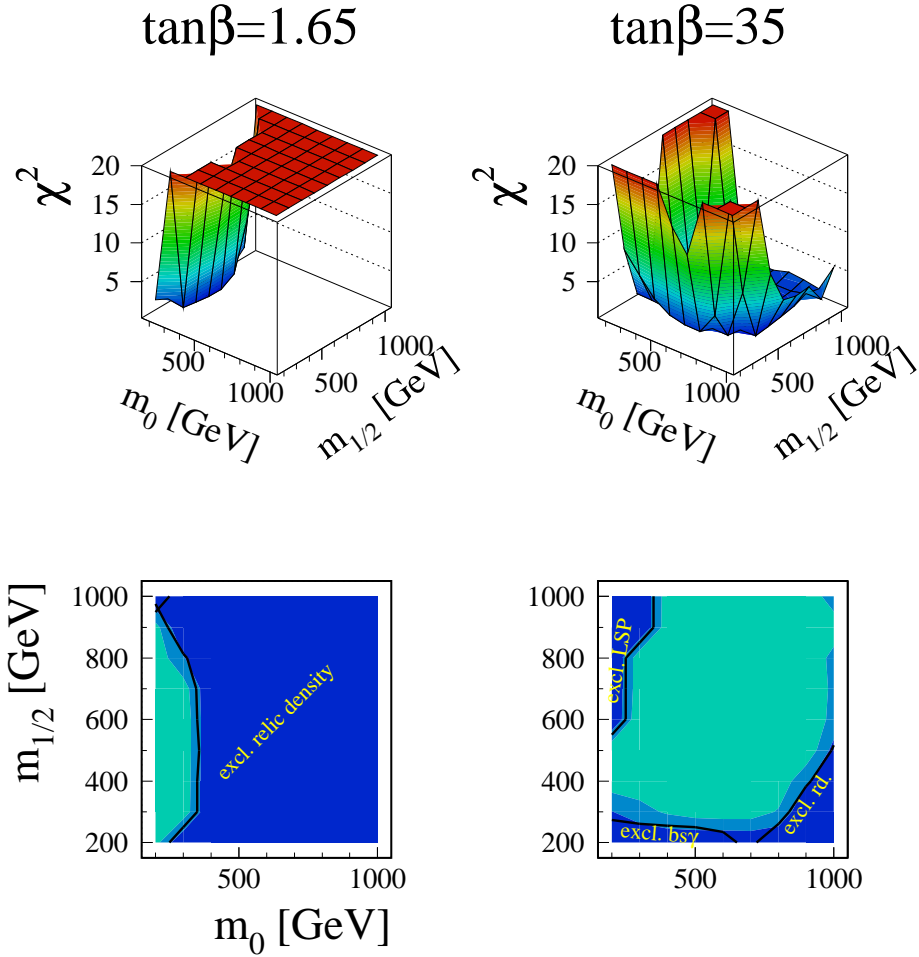}
\end{center}\vspace{-1cm} \caption{The same as Fig.\ref{chi} but with the
$b\to s\gamma$ constraint released with account taken of  the
higher order corrections \cite{BHGK}.} \label{ho}
 \end{figure}

We present the allowed regions of the parameter space for low and
high $\tan\beta$ scenarios in Fig.\ref{chi}. This plot
demonstrates the role of various constraints in the $\chi^2$
function. The contours enclose domains by the particular
constraints used in the analysis \cite{BGGK}.
 In case when the requirement of the
$b\to s\gamma$ decay rate is not taken into account (due to
uncertainties of the high order contributions), the allowed region
of parameter space becomes much wider, as it is illustrated in
Fig.\ref{ho}. Now much lower values of $m_0$ and $m_{1/2}$ are
allowed which lead to lower values of sparticle masses.

\subsection{The mass spectrum of superpartners}

When the parameter set is fixed, one can calculate the mass
spectrum of superpartners. Below we show  the set of parameters
and the predicted mass spectrum corresponding to the best fit
values indicated by stars in Fig.\ref{chi}~\cite{BEK}.
\begin{table}[htb]
\renewcommand{\arraystretch}{1.20}
\begin{center}
\normalsize
\begin{tabular}{|c|r|r|}
\hline
 \multicolumn{3}{|c|}{ Fitted SUSY parameters }                       \\
\hline \hline Symbol & \makebox[3.0cm]{\bf{low $\tan\beta$}} &
\makebox[3.0cm]{\bf{high $\tan\beta$}}\\ \hline
 $\tan\beta$    & 1.71 &  35.0 \\ \hline
 $m_0$          &  200          &  600 \\
\hline
 $m_{1/2}$      &  500          &  400  \\
\hline
 $\mu(0)$       & 1084         &  -558\\
 \hline
  $A(0)$     & 0             &  0\\
\hline
 $1/\alpha_{GUT}$ & 24.8          &  24.8\\
\hline
 $M_{GUT}$          & $1.6\;10^{16}$&  $1.6\;10^{16}$\\
\hline \hline
\end{tabular} \end{center}
\caption[]{\label{t2a}Values of the fitted SUSY parameters
             for low and high $\tan\beta$ (in GeV, when applicable).
            }
\end{table}

\begin{table}[htb]
\renewcommand{\arraystretch}{1.10}
\begin{center}
\normalsize \vspace{-0.5cm}
\begin{tabular}{|c|r|r|}
\hline
 \multicolumn{3}{|c|}{SUSY masses in [GeV]}             \\
\hline \hline Symbol & \makebox[3.0cm]{{low $\tan\beta$}} &
\makebox[3.0cm]{{high $\tan\beta$}}\\ \hline \hline
 $\tilde{\chi}^0_1(\tilde{B})$, $\tilde{\chi}^0_2(\tilde{W}^3)$ &214, 413 & 170, 322\\
\hline $ \tilde{\chi}^0_3(\tilde{H}_1)$,$
\tilde{\chi}^0_4(\tilde{H}_2)$ &1028, 1016  & 481, 498\\ \hline
 $\tilde{\chi}^{\pm}_1(\tilde{W}^\pm)$, $\tilde{\chi}^{\pm}_2(\tilde{H}^{\pm})$
  &  413, 1026  & 322, 499\\
\hline \hline
  $\tilde{g}$                   &  1155  & 950\\
\hline  \hline
  $\tilde{e}_L$, $\tilde{e}_R$   &  303, 270  & 663, 621\\
\hline
  $\tilde{\nu}_L$               &  290  & 658\\
\hline  \hline
  $\tilde{q}_L$, $\tilde{q}_R$  &  1028, 936  & 1040, 1010\\
\hline
  $\tilde{\tau}_1$, $\tilde{\tau}_2$ &  279, 403  & 537, 634\\
\hline
  $\tilde{b}_1$,  $\tilde{b}_2$ &  953, 1010  & 835, 915\\
\hline
  $\tilde{t}_1$,  $\tilde{t}_2$ &  727, 1017  & 735, 906\\
\hline        \hline
  $       h $, $H$   &  95, 1344   & 119, 565\\
\hline
  $       A $,  $       H ^{\pm}$ &  1340, 1344  & 565, 571\\
\hline  \hline
\end{tabular} \end{center}
\caption[]{\label{t2}Values of the  SUSY
            mass spectra for the low and high $\tan\beta$ solutions given in
            Table \ref{t2a}.}
\end{table}

To demonstrate the dependence of masses of the lightest particles
on the choice of parameters, we show below in
Figs.\ref{fig:susyall},\ref{f5} their values in the whole
$m_0,m_{1/2}$ plane for the case of low and high $tan\beta$
solutions, respectively~\cite{BGGK}. One can see that the masses
of gauginos (charginos and neutralinos) and Higgses basically
depend on $m_{1/2}$, while those of squarks and sleptons on $m_0$.
\begin{figure}[ht]
\begin{center}\vspace{-0.5cm}
 \leavevmode
  \epsfxsize=15cm \epsfysize=13cm
 \epsffile{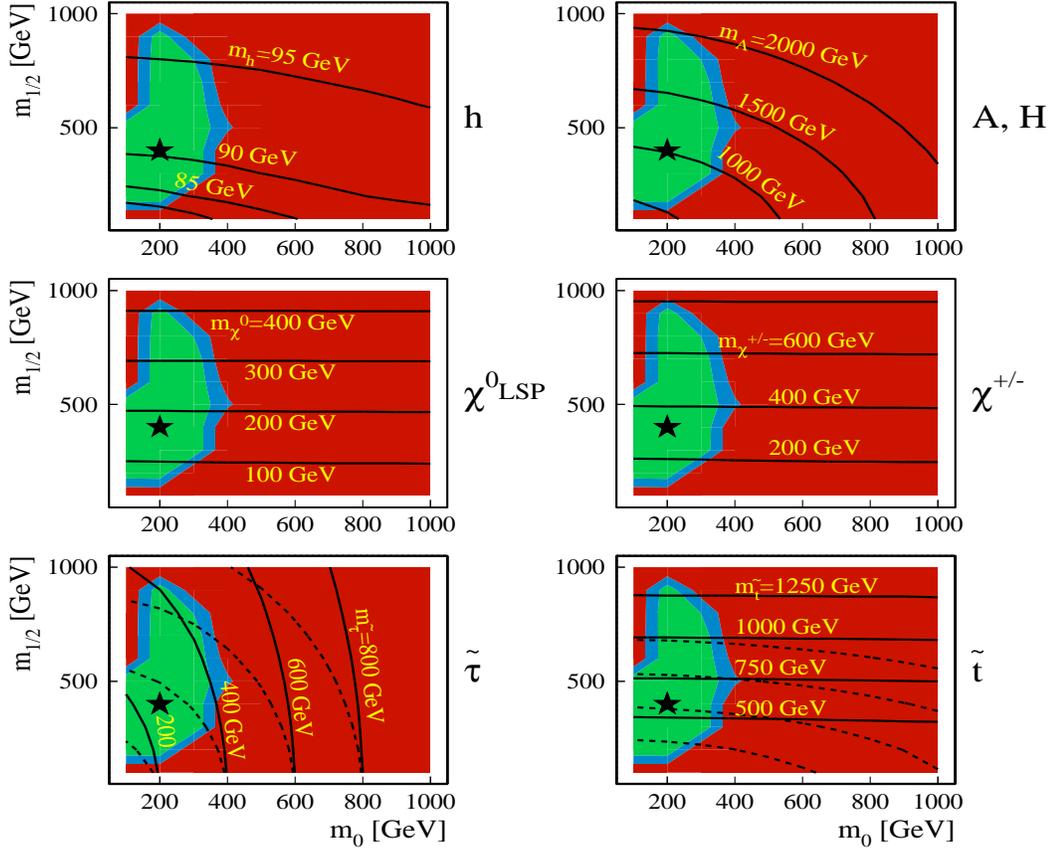}
\end{center}\vspace{-1cm}
\caption{The masses of the lightest particles in the CMSSM for the
low $\tan\beta$ scenario. The contours show the fixed mass values
of the corresponding particles.} \label{fig:susyall}
\end{figure}

\begin{figure}[t] \vspace{-1.5cm} \begin{center}
\epsfig{file=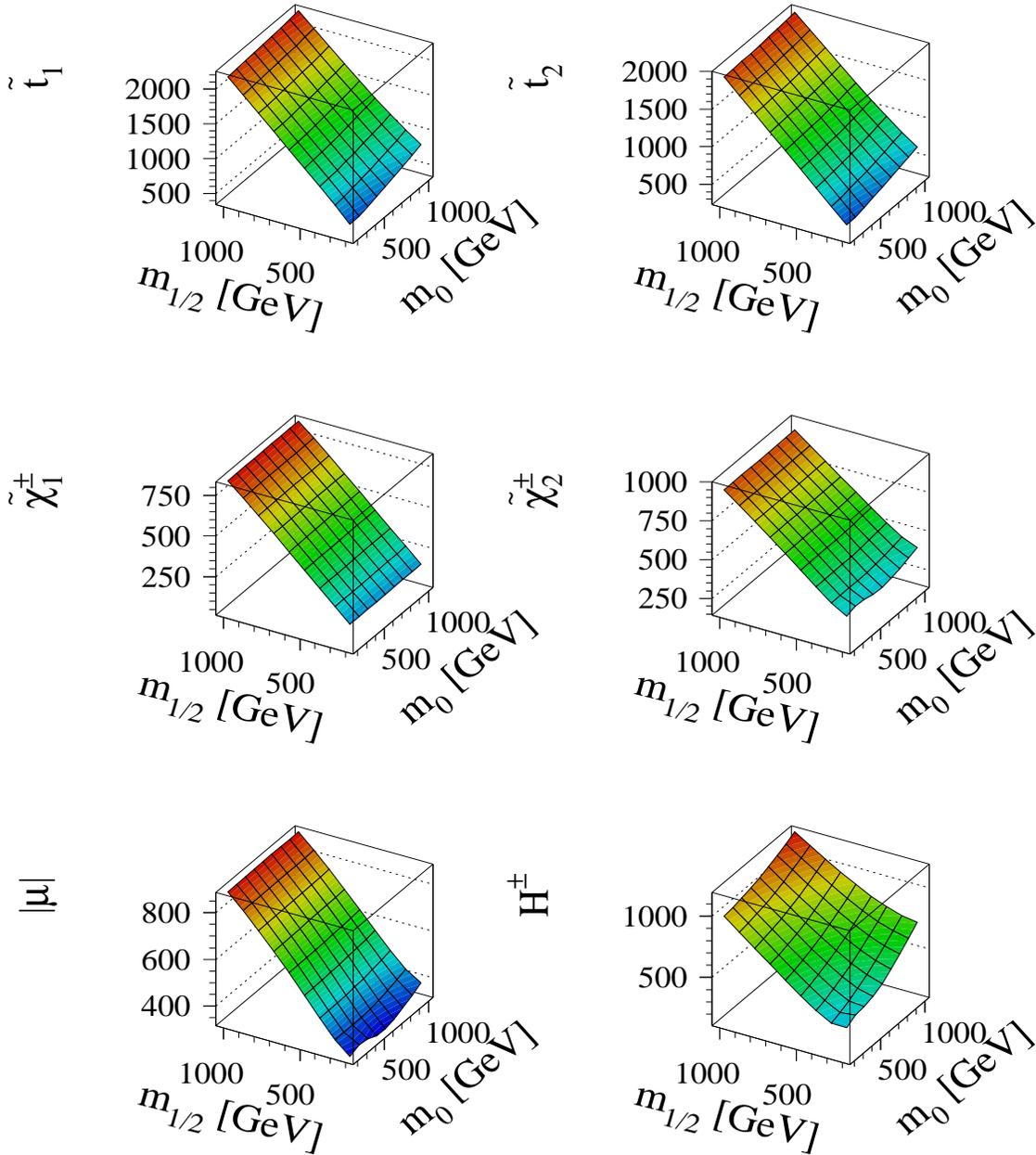,width=\textwidth,height=\textheight}
\vspace{-2.cm}
\caption[]{\label{f5} The masses of the particles and the
Higgs mixing parameter $\mu$ for $\tan\beta = 35, \mu < 0$.}
\end{center}
\end{figure} \clearpage

\subsection{Experimental signatures at $e^+e^-$ colliders}

Experiments are finally beginning to push into a significant
region of supersymmetry parameter space. We know the sparticles
and their couplings, but we do not know their masses and mixings.
Given the mass spectrum one can calculate the cross-sections and
consider the possibilities of observing  new particles at modern
accelerators. Otherwise, one can get  restrictions on unknown
parameters.

\begin{figure}[ht]
\begin{center}\vspace{-0.3cm}
 \leavevmode
  \epsfxsize=9cm
 \epsffile{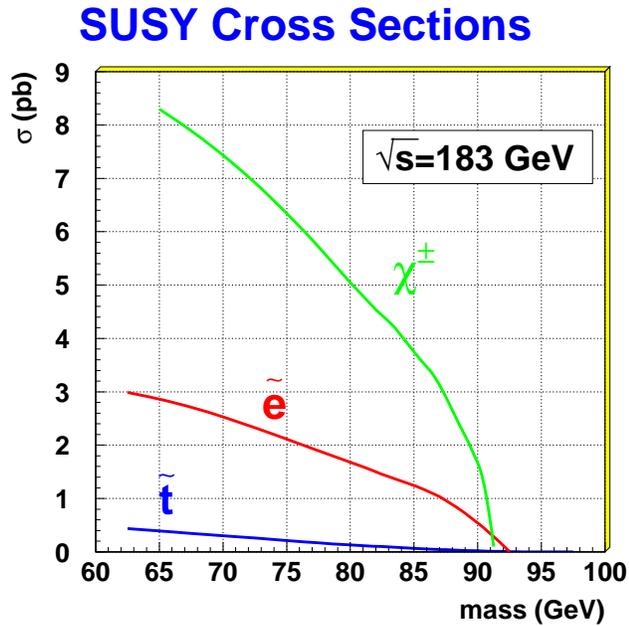}
\end{center}\vspace{-4cm} \caption{The cross-section of sparticle production
at LEP II as functions of sparticle masses} \label{fig:xs}
\end{figure}

We start with $e^+e^-$ colliders and, first of all, with LEP II.
In the leading order creation of superpartners is given by the
diagrams shown in Fig.\ref{creation} above. For a given center of
mass energy the cross-sections depend on the mass of created
particles and vanish at the kinematic  boundary. For a  sample
example of  c.m. energy of LEP II equal to 183 GeV, they are shown
at Fig.\ref{fig:xs}.

Experimental signatures are defined by the decay modes which vary
with the mass spectrum. The main ones are summarized below.

$$\begin{array}{lll}
\mbox{\underline{Production}}&\mbox{\underline{Key Decay
Modes}}&~~ \mbox{\underline{Signatures}} \\ && \\ \bullet
~~~~~\tilde{l}_{L,R}\tilde{l}_{L,R}~~~~~~~~~~~ &\tilde{l}^\pm_R
\to l^\pm \tilde{\chi}^0_i \searrow  \mbox{cascade}~~~~~ &
\mbox{acomplanar pair of}
\\ &  \tilde{l}^\pm_L \to l^\pm \tilde{\chi}^0_i \nearrow \mbox{decays} &
\mbox{charged leptons} + \Big/ \hspace{-0.3cm E_T} \\ \bullet
~~~~~\tilde{\nu}\tilde{\nu}& \tilde{\nu}\to l^\pm \tilde{\chi}^0_1
 & \Big/ \hspace{-0.3cm E_T}\\
\bullet  ~~~~~\tilde{\chi}^\pm_1\tilde{\chi}^\pm_1 &\tilde{\chi}^\pm_1 \to
 \tilde{\chi}^0_1 l^\pm \nu, \ \tilde{\chi}^0_1q \bar q' &
 \mbox{isolated lepton + 2 jets} + \Big/ \hspace{-0.3cm E_T} \\
&  \tilde{\chi}^\pm_1 \to \tilde{\chi}^0_2 f \bar f' &\mbox{pair
of acomplanar}\\
 &  \tilde{\chi}^\pm_1 \to l \tilde{\nu}_l \to
l\nu_l\tilde{\chi}^0_1 &\mbox{leptons} + \Big/ \hspace{-0.3cm E_T}
\\ & \tilde{\chi}^\pm_1 \to \nu_l \tilde{l} \to \nu_l
l\tilde{\chi}^0_1&\mbox{4 jets} + \Big/ \hspace{-0.3cm E_T}
\end{array}$$
$$\begin{array}{lll}
 \bullet  ~~~~~\tilde{\chi}^0_i\tilde{\chi}^0_j &
\tilde{\chi}^0_i \to \tilde{\chi}^0_1 X, \tilde{\chi}^0_j  \to
\tilde{\chi}^0_1 X' & X=\nu_l \bar \nu_l \ \mbox{invisible} \\ &&
~~= \gamma,2l,\mbox{2 jets} \\ && 2l + \Big/ \hspace{-0.3cm E_T},
l+2j + \Big/ \hspace{-0.3cm E_T} \\ \bullet
~~~~~\tilde{t}_i\tilde{t}_j & \tilde{t}_1 \to c \tilde{\chi}^0_1 &
\mbox{2 jets}+ \Big/ \hspace{-0.3cm E_T} \\ & \tilde{t}_1 \to b
\tilde{\chi}^\pm_1 \to b f¿\bar f'\tilde{\chi}^0_1 & \mbox{2 b
jets}+ \mbox{2 leptons} + \Big/ \hspace{-0.3cm E_T} \\ && \mbox{2
b jets}+ \mbox{2 jets}+\mbox{lepton} + \Big/ \hspace{-0.3cm E_T}
\\ \bullet  ~~~~~\tilde{b}_i\tilde{b}_j & \tilde{b}_i \to b
\tilde{\chi}^0_1 & \mbox{2 b jets}+ \Big/ \hspace{-0.3cm E_T} \\ &
\tilde{b}_i \to b \tilde{\chi}^0_2 \to b f¿\bar f'\tilde{\chi}^0_1
& \mbox{2 b jets}+ \mbox{2 leptons} + \Big/ \hspace{-0.3cm E_T} \\
&& \mbox{2 b jets}+ \mbox{2 jets}+ \Big/ \hspace{-0.3cm E_T}
 \end{array} $$
A characteristic feature of all possible signatures is the missing
energy and transverse momenta, which is a trade mark of a new
physics.

\begin{figure}[ht]
\vspace{-0.9cm}
 \leavevmode
  \epsfxsize=8.5cm
 \epsffile{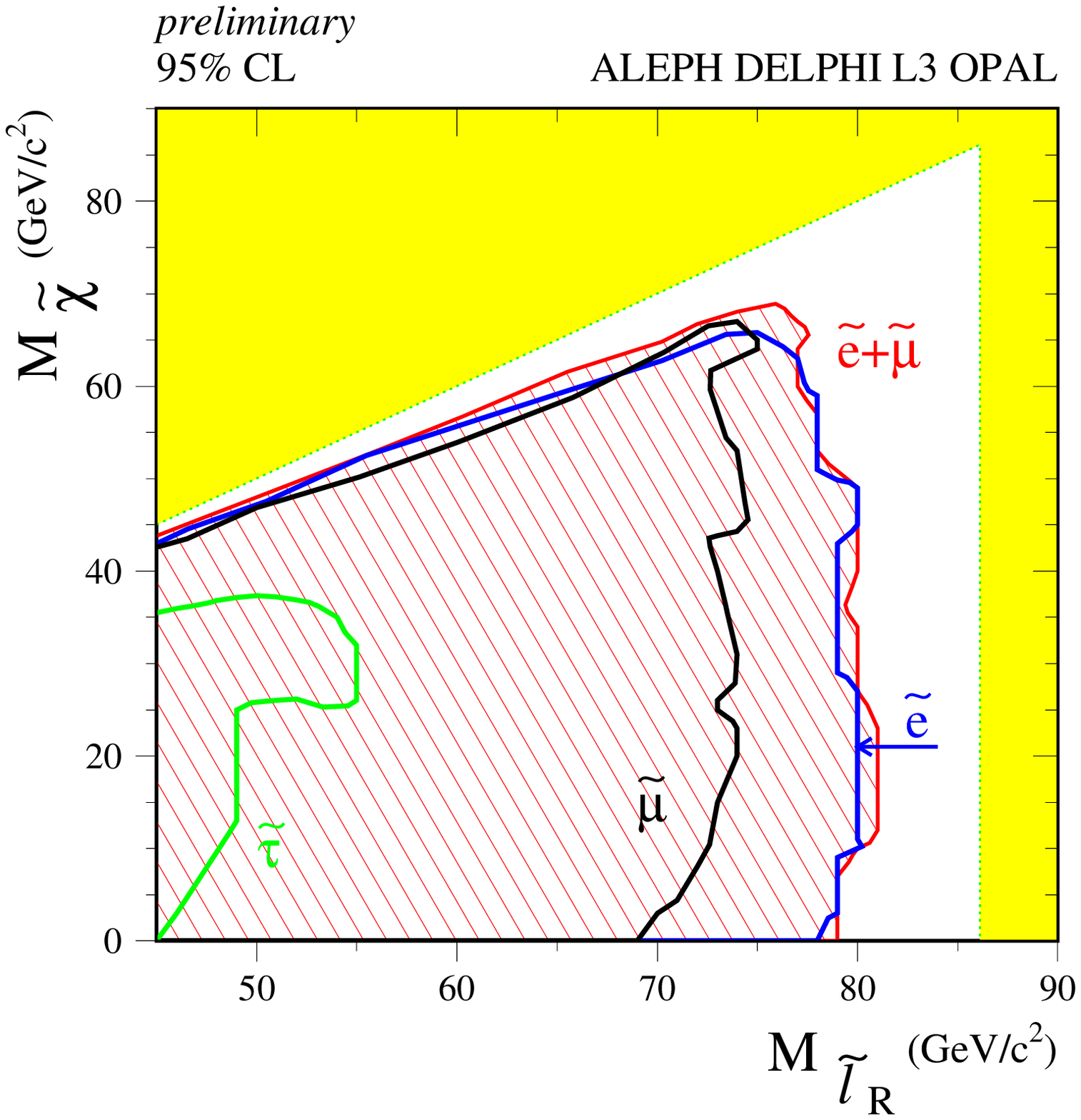}
 \vspace{-10.5cm}

 \epsfxsize=7.0cm
 \hspace*{9cm}\epsffile{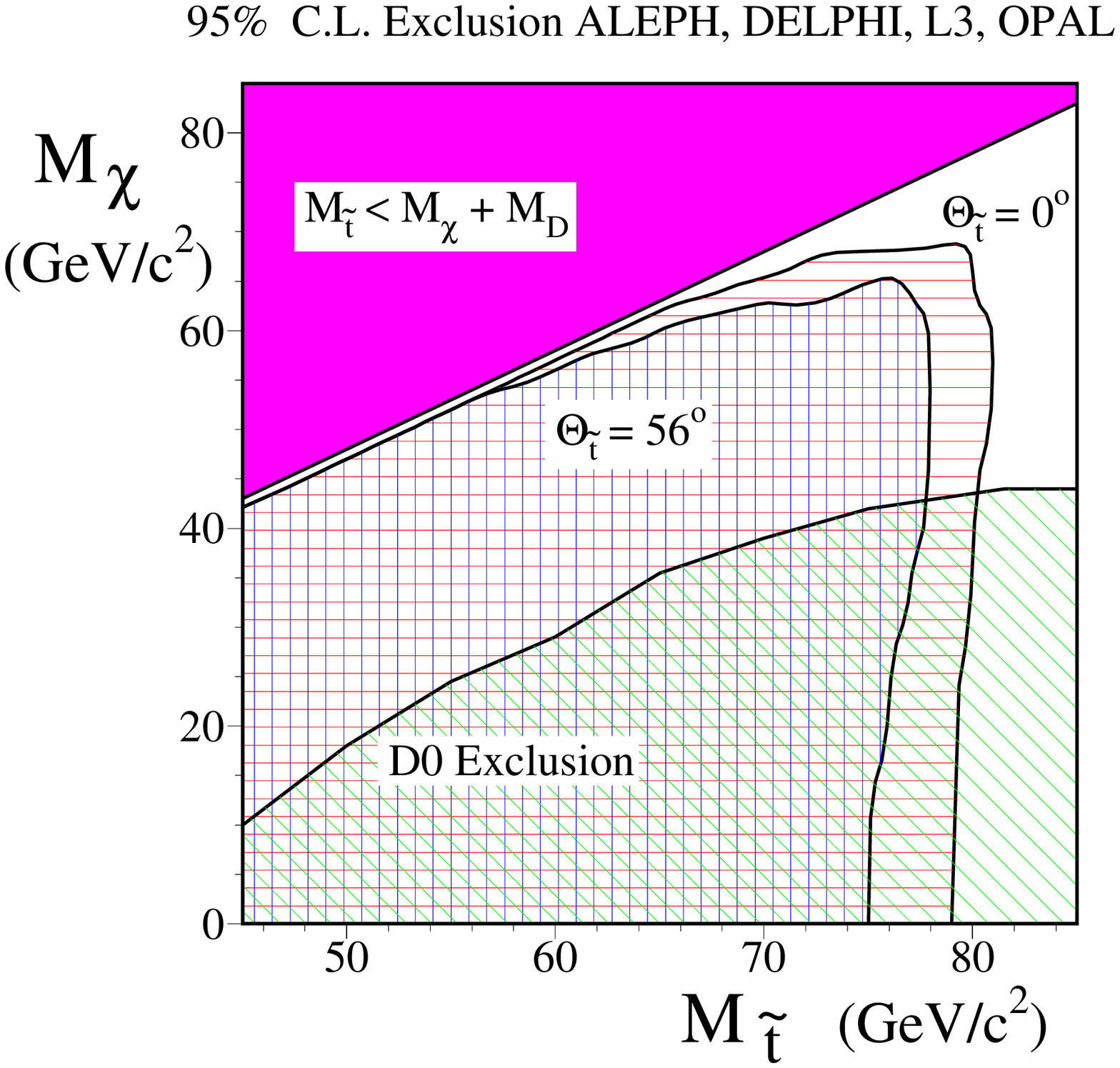}
 \vspace{-0cm}\caption{The excluded region in chargino-slepton and
 chargino-stop mass plane } \label{fig:slepton}
 \end{figure}

\begin{figure}[ht]
\begin{center}
 \leavevmode
  \epsfxsize=11cm
 \epsffile{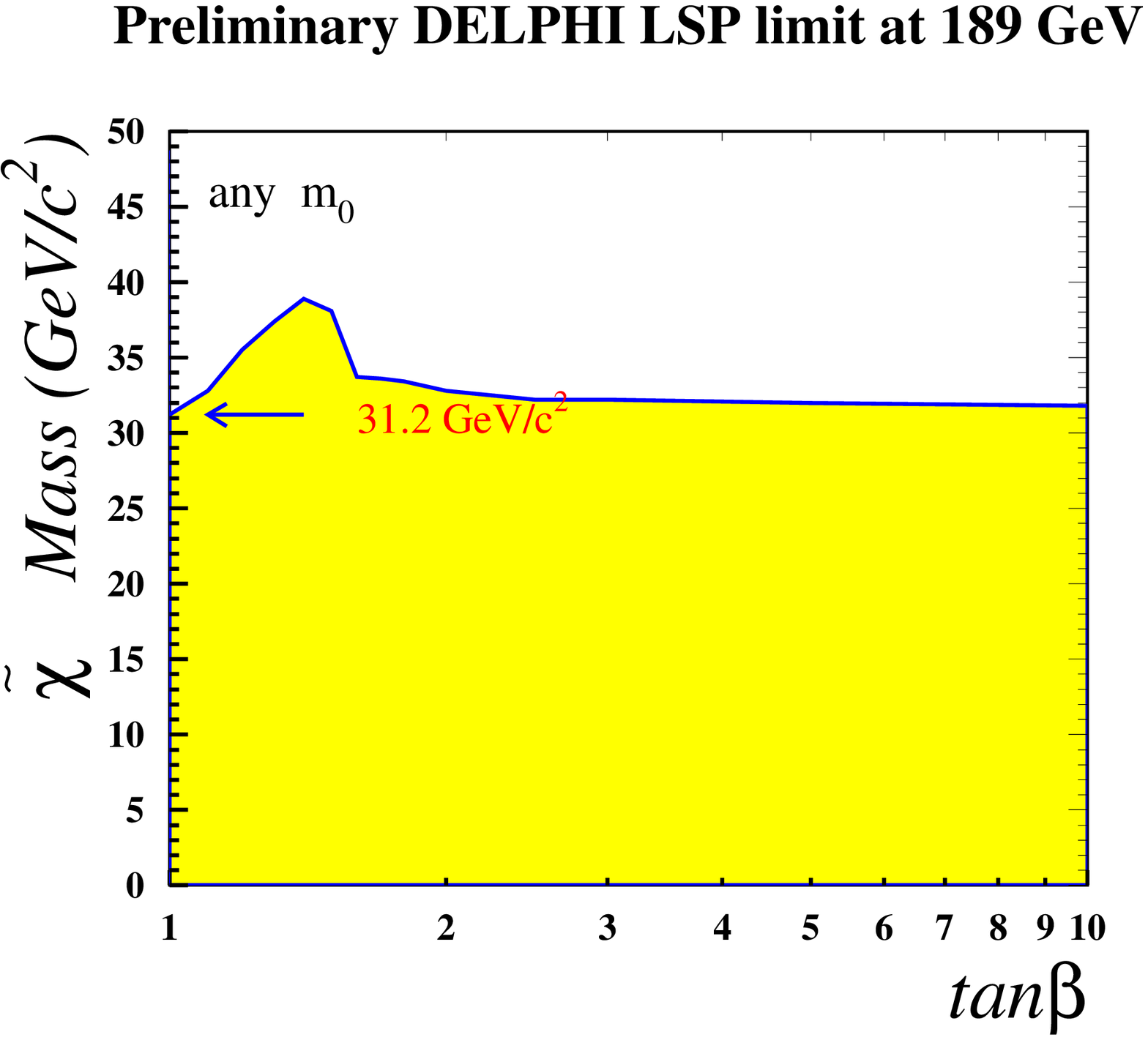}
\end{center} \vspace{-0.5cm}
\caption{Cross section of chargino production at LEP and
experimental limits on chargino mass} \label{fig:ch}
 \end{figure}

 Numerous attempts to find  superpartners at LEP II
gave no positive result thus imposing the lower bounds on their
masses~\cite{LEPSUSY}.
 They are shown on the parameter plane in Figs.\ref{fig:slepton},\ref{fig:ch}.

In  the case of stop masses, the result depends on the stop mixing
angle $\Theta_{\tilde t}$ calculated from the stop mixing matrix.
It defines the mass eigenstates basis $\tilde t_1$ and $\tilde
t_2$ $$\left(\begin{array}{c}\tilde t_1 \\ \tilde t_2
\end{array}\right)= \left(\begin{array}{cc}\cos \Theta_{\tilde t} & \sin
\Theta_{\tilde t} \\ -\sin \Theta_{\tilde t} &\cos \Theta_{\tilde
t}\end{array}\right)\ \left(\begin{array}{c}\tilde t_L \\ \tilde t_R
\end{array}\right).$$

Nonobservation of charginos at the maximal LEP II energy defines
the lower limit on chargino masses as shown in
Fig.\ref{fig:ch}~\cite{LEPSUSY}.

Typical LEP II limits on the masses of superpartners are
\begin{equation}\begin{array}{lll}
 m_{\chi^0_1} > 40 \ GeV & m_{\tilde e_{L,R}}>105\ GeV &
 m_{\tilde t}> 90\ GeV \\
 m_{\chi^\pm_1} > 100 \ GeV & m_{\tilde \mu_{L,R}}>100\ GeV &
 m_{\tilde b}> 80\ GeV \\
 & m_{\tilde \tau_{L,R}}>80\ GeV &
\end{array} \nonumber\end{equation}

\subsection{Experimental signatures at hadron colliders}

\begin{figure}[hbt]
\begin{center}
\leavevmode
  \epsfxsize=14cm
 \epsffile{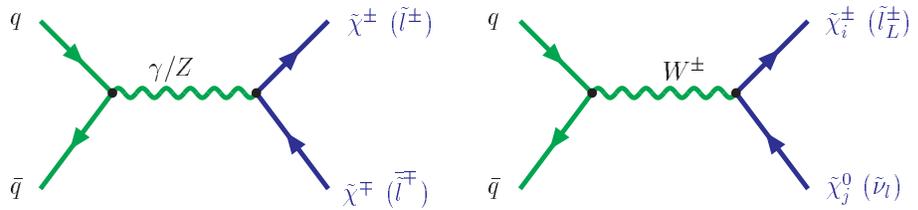}
\end{center}\vspace{-1cm}
\caption{Annihilation channel}\label{annihil}
\end{figure}

\begin{figure}[ht]\vspace{-1cm}
\begin{center}
\leavevmode
  \epsfxsize=16cm
 \epsffile{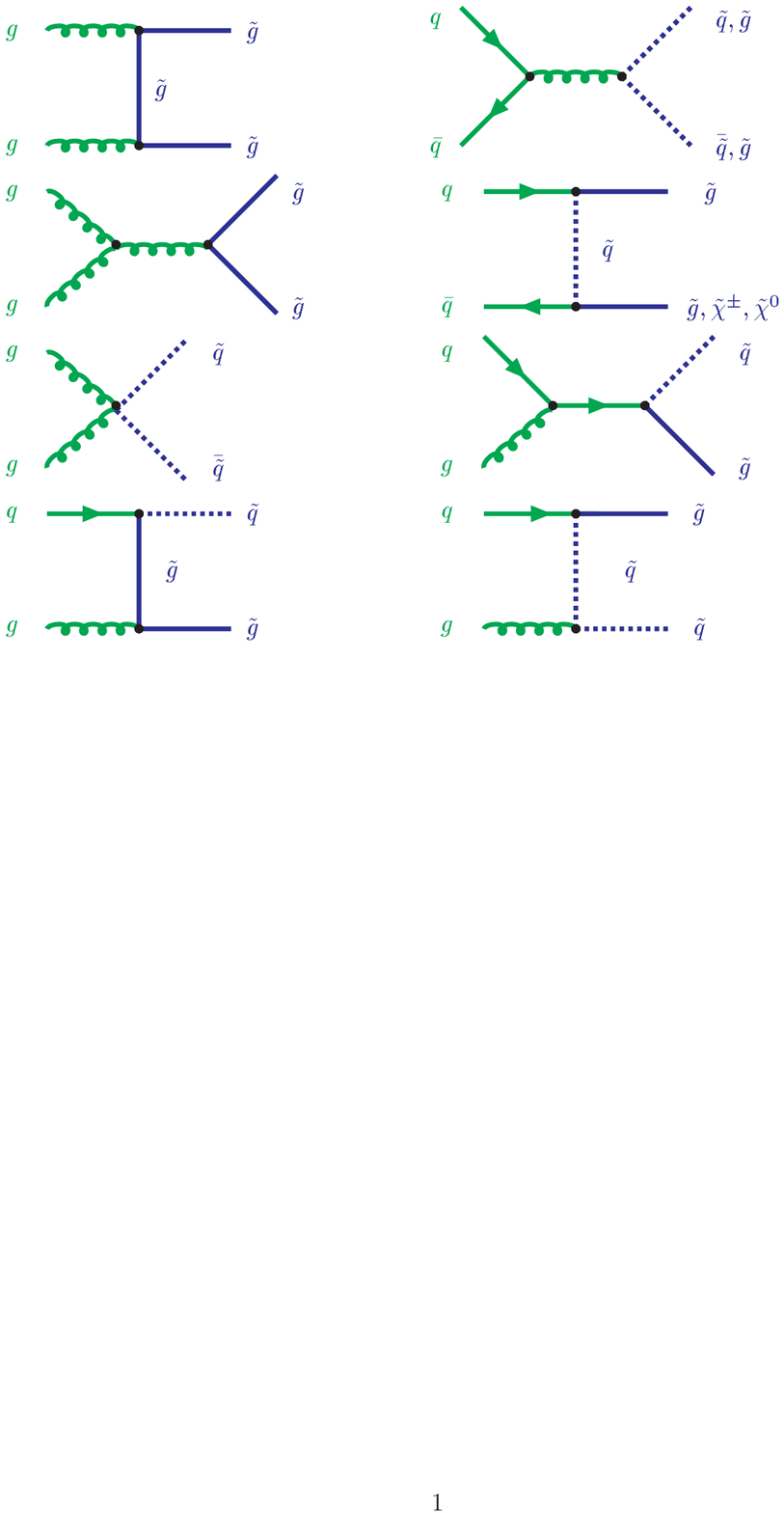}
\end{center}\vspace{-1.5cm} \caption{Gluon fusion, $q\bar q$ scattering,
quark-gluon scattering}\label{fusion} \end{figure}

Experimental signatures at hadron colliders are similar to those
at $e^+e^-$ machines; however, here one has much wider
possibilities. Besides the usual annihilation channel identical to
$e^+e^-$ one with the obvious replacement of electrons by quarks
(see Fig.\ref{annihil}), one has numerous processes of gluon
fusion, quark-antiquark and quark-gluon scattering (see
Fig.\ref{fusion}).

The final states depend on gluino decay modes. If squarks are
heavier, i.e. $m_{\tilde q} > m_{\tilde g}$, then the main gluino
decay modes are
 $$\tilde g \to t+\bar t + \tilde{\chi}^0_i, \ \ \ \tilde g
 \to t+\bar b + \tilde{\chi}^-_i, \ \ \ \tilde g \to t+ b +
 \tilde{\chi}^+_i ,$$
 otherwise gluino can decay into quarks and squarks with further
 decay of the latter.

Experimental SUSY signatures at the Tevatron (and LHC) are
 $$\begin{array}{lll}
\mbox{\underline{Production}}&\mbox{\underline{Key Decay
Modes}}&~~ \mbox{\underline{Signatures}} \\ && \\ \bullet
~~~~~\tilde{g}\tilde{g}, \tilde{q}\tilde{q},
\tilde{g}\tilde{q}~~~~~~~~~ ~~ &\left.\begin{array}{l} \tilde{g}
\to q\bar q \tilde{\chi}^0_1   \\
 ~~~~~ q\bar q' \tilde{\chi}^\pm_1  \\
 ~~~~~ g\tilde{\chi}^0_1 \end{array} \right\}
 m_{\tilde{q}}>m_{\tilde{g}} &
\begin{array}{c} \Big/ \hspace{-0.3cm E_T} + \mbox{multijets}\\
 (+\mbox{leptons}) \end{array} \\
& \left.\begin{array}{l}\tilde{q} \to q \tilde{\chi}^0_i \\
    \tilde{q} \to q' \tilde{\chi}^\pm_i \end{array} \right\}
  m_{\tilde{g}}>m_{\tilde{q}}&
\end{array}$$
$$\begin{array}{lll}
 \bullet
~~~~~\tilde{\chi}^\pm_1\tilde{\chi}^0_2 &\tilde{\chi}^\pm_1 \to
 \tilde{\chi}^0_1 l^\pm \nu, \ \tilde{\chi}^0_2 \to
 \tilde{\chi}^0_1 ll &
 \mbox{Trilepton} + \Big/ \hspace{-0.3cm E_T} \\
&  \tilde{\chi}^\pm_1 \to \tilde{\chi}^0_1 q \bar q', \tilde{\chi}^0_2 \to
\tilde{\chi}^0_1 ll,&\mbox{Dilepton + jet} + \Big/ \hspace{-0.3cm E_T}\\
 \bullet  ~~~~~\tilde{\chi}^+_1\tilde{\chi}^-_1 &
\tilde{\chi}^+_1 \to l \tilde{\chi}^0_1 l^\pm \nu & \mbox{Dilepton} + \Big/
\hspace{-0.3cm E_T} \\ \bullet  ~~~~~\tilde{\chi}^0_i\tilde{\chi}^0_i &
\tilde{\chi}^0_i \to \tilde{\chi}^0_1 X, \tilde{\chi}^0_i  \to
\tilde{\chi}^0_1 X' & \Big/ \hspace{-0.3cm E_T} + \mbox{Dilepton + (jets) +
(leptons)}\\
 \bullet  ~~~~~\tilde{t}_1\tilde{t}_1 & \tilde{t}_1 \to c
\tilde{\chi}^0_1 & \mbox{2 acollinear jets}+ \Big/ \hspace{-0.3cm E_T} \\ &
\tilde{t}_1 \to b \tilde{\chi}^\pm_1, \tilde{\chi}^\pm_1 \to
\tilde{\chi}^0_1 l^\pm \nu , \tilde{\chi}^\pm_1 \to \tilde{\chi}^0_1 q\bar
q' &
 \mbox{single lepton} + \Big/ \hspace{-0.3cm E_T}  + b's\\
 &\tilde{t}_1 \to b \tilde{\chi}^\pm_1,\tilde{\chi}^\pm_1 \to \tilde{\chi}^0_1
l^\pm \nu , \tilde{\chi}^\pm_1 \to \tilde{\chi}^0_1  l^\pm \nu
& \mbox{Dilepton} + \Big/ \hspace{-0.3cm E_T} + b's\\
\bullet  ~~~~~\tilde{l}\tilde{l},\tilde{l}\tilde{\nu},\tilde{¿nu}\tilde{\nu}  &
 \tilde{l}^\pm \to l\pm \tilde{\chi}^0_i,\tilde{l}^\pm \to \nu_l
 \tilde{\chi}^\pm_i& \mbox{Dilepton}+ \Big/ \hspace{-0.3cm E_T} \\
& \tilde{\nu} \to \nu \tilde{\chi}^0_1& \mbox{Single lepton} +
\Big/ \hspace{-0.3cm E_T} +(jets) \\
&& \Big/ \hspace{-0.3cm E_T}
 \end{array}$$
Note again the characteristic missing energy and transverse momenta
events.

Contrary to  $e^+e^-$ colliders, at hadron machines the background
is extremely rich and essential.

\subsection{The lightest superparticle}

One of the crucial questions is the properties of the lightest
superparticle. Different SUSY breaking scenarios lead to different
experimental signatures and different LSP.

$\bullet$ Gravity mediation

In this case, the LSP is the lightest neutralino
$\tilde{\chi}^0_1$, which is almost 90\% photino for a low
$\tan\beta$ solution and contains more higgsino admixture for high
$\tan\beta$. The usual signature for LSP is missing energy;
$\tilde{\chi}^0_1$ is stable and is the best candidate for the
cold dark matter in the Universe. Typical processes, where the LSP
is created, end up with jets + $\Big/ \hspace{-0.3cm}E_T$, or
leptons + $\Big/ \hspace{-0.3cm}E_T$, or both jest + leptons +
$\Big/ \hspace{-0.3cm}E_T$.

$\bullet$ Gauge mediation

In this case the LSP is the  gravitino $\tilde G$ which also leads
to missing energy. The actual question here is what the NLSP, the
next lightest particle, is. There are two possibilities:

i) $\tilde{\chi}^0_1$ is the NLSP. Then the decay modes are
 $$\tilde{\chi}^0_1 \to \gamma \tilde G, \ h \tilde G, \ Z \tilde
 G.$$
 As a result, one has two hard photons + $\Big/
 \hspace{-0.3cm}E_T$, or jets + $\Big/ \hspace{-0.3cm}E_T$.

ii) $\tilde l_R$ is the NLSP. Then the decay mode is  $\tilde l_R
\to \tau \tilde G$ and the signature is a charged lepton and the
missing energy.

$\bullet$ Anomaly mediation

In this case, one also has two possibilities:

i) $\tilde{\chi}^0_1$ is the LSP and wino-like. It is almost
degenerate with the NLSP.

ii) $\tilde \nu_L$ is the LSP. Then it appears in the decay of
chargino $\tilde \chi^+ \to \tilde \nu l$ and the signature is the
charged lepton and the missing energy.

$\bullet$ R-parity violation

In this case, the LSP  is no longer stable and decays into the SM
particles. It may be charged (or even colored) and may lead to
rare decays like neutrinoless double $\beta$-decay, etc.

\begin{figure}[ht]
 \leavevmode
  \epsfxsize=8.2cm \epsfysize=8cm
 \epsffile{delphi_lsp.eps}
\vspace{-7.7cm}

 \epsfxsize=8.0cm
  \hspace*{8cm}\epsffile{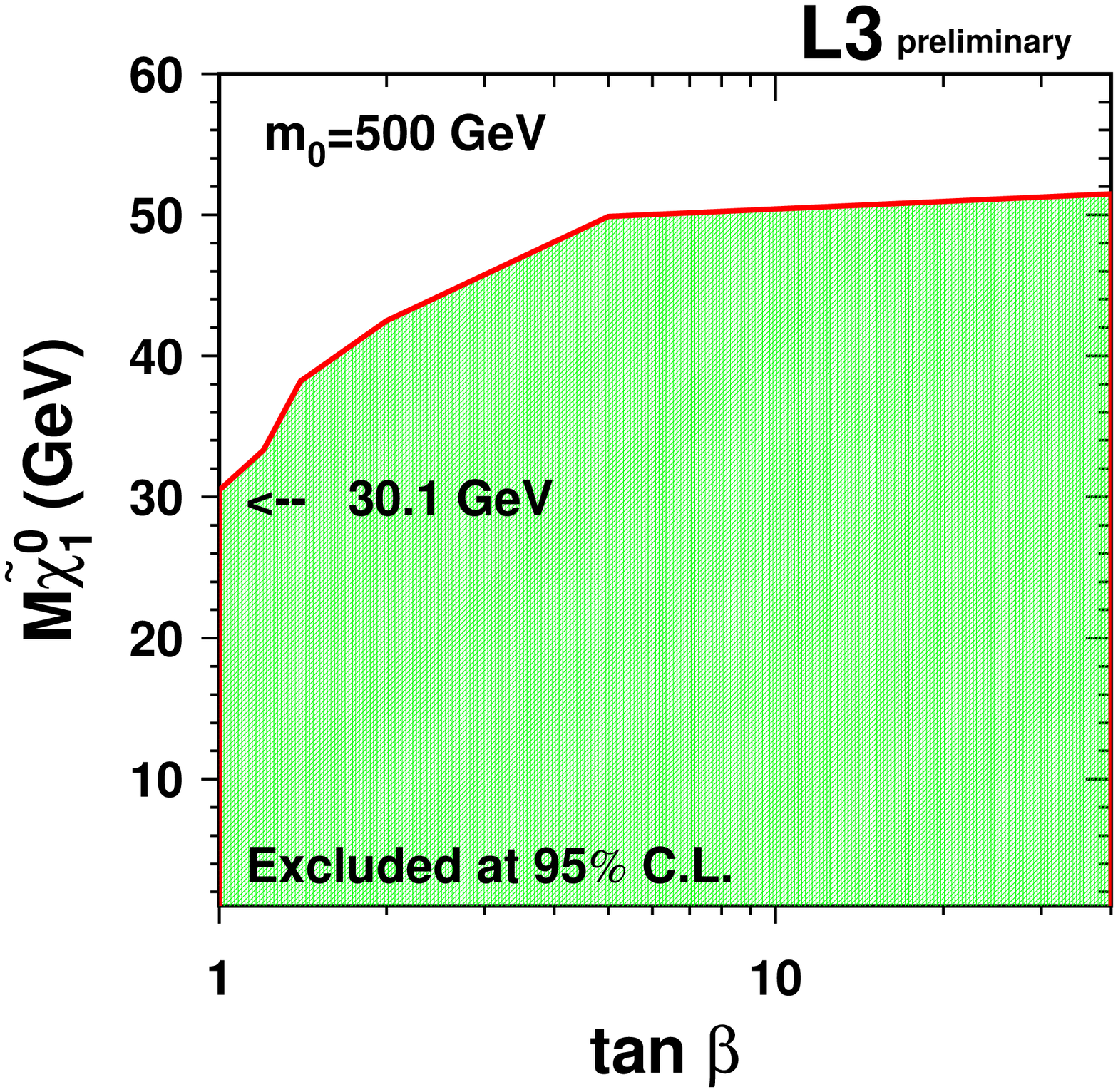}
\vspace{-1cm}
 \caption{The LSP mass limits within the MSSM ~\cite{LEPSUSY}} \label{fig:lsp}
 \end{figure}

Experimental limits on the LSP mass follow from non-observation of the
corresponding events. Modern low limit is around 40 GeV (see
Fig.\ref{fig:lsp}).

\section{The Higgs boson in the SM and the MSSM}
 \setcounter{equation} 0

One of the hottest topics in the SM now is the search for the
Higgs boson. It  is also a window to a new physics. Below we
consider the situation with the Higgs boson search and the
properties of the Higgs boson in the MSSM.

\subsection{Allowed mass range in the SM}
The last unobserved particle from the Standard Model is the Higgs
boson~\cite{higgs}. Its discovery would allow one to complete the
SM paradigm and confirm the mechanism of spontaneous symmetry
breaking. On the contrary, the absence of the Higgs boson would
awake doubts about the whole picture and would require new
concepts.

\begin{figure}[ht]
\begin{center}\vspace{-1.3cm}
 \leavevmode
  \epsfxsize=11cm
 \epsffile{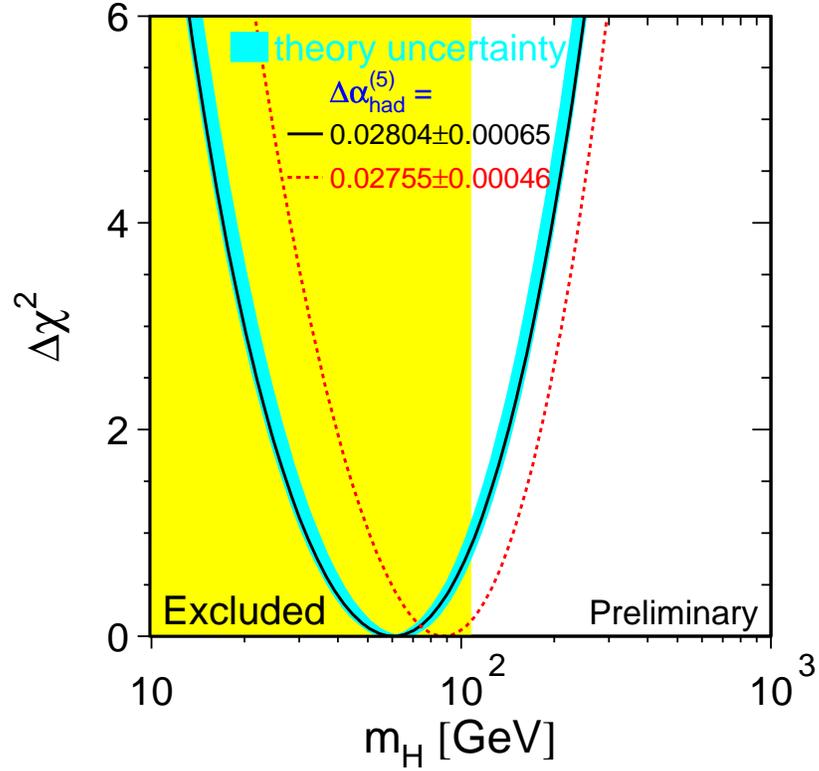}
\end{center}\vspace{-1cm}
\caption{The $\chi^2$ distribution as a function of the Higgs mass
from the SM fit to the electroweak precision observables and the
top mass. The shaded area is excluded by the direct searches.}
\label{fig:1}
\end{figure}

Experimental limits on the Higgs boson mass come from a direct
search at LEP II and Tevatron and from indirect fits of
 electroweak precision data, first of all from the radiative
corrections to the W and top quark masses. A combined fit of
modern experimental data gives~\cite{EWWG}
\begin{equation}
 m_h=90^{+55}_{-47}\; {\rm GeV},
\end{equation}
 which at the 95\% confidence level leads to
the upper bound of 200 GeV (see Fig.\ref{fig:1}). At the same
time, recent direct searches at LEP II for the c.m. energy of 209
GeV give the lower limit of  113.4 GeV\cite{EWWG}. From a
theoretical point of view a low Higgs mass could be a hint for
physics beyond the SM, in particular, for the supersymmetric
extension of the SM. 

Within the Standard Model the value of the Higgs mass $m_h$ is not
predicted. However, one can get the bounds on the Higgs mass
\cite{bound1, bound2}.  They follow from the behaviour of the
quartic coupling which is related to the Higgs mass by
eqs.(\ref{vac},\ref{mass}) $m_h^2=2\lambda v$ and obeys the
following renormalization group equation describing the change of
$\lambda$ with a scale:
\begin{equation}\label{betalambda}
   \frac{d \lambda}{d t} = \frac{1}{16 \pi^2} \left(
6\lambda^2  + 6\lambda y_t^2 - 6y_t^4  + \mbox{gauge  terms}
\right)
\end{equation}
 with $t= \ln (Q^2/\mu^2)$. Here $y_t$ is the top-quark Yukawa
coupling.

 Since the quartic coupling grows with rising energy
infinitely and reaches the Landau pole, the upper bound on $m_h$
follows from the requirement that the theory be valid up to the
scale $M_{Planck}$ or up to a given cut-off scale $\Lambda$ below
$M_{Planck}$ \cite{bound1}. The scale $\Lambda$ could be
identified with the scale at which the Landau pole develops.  The
upper bound on $m_h$ depends mildly on the top-quark mass through
the impact of the top-quark Yukawa coupling on the running of the
quartic coupling $\lambda$ in eq.(\ref{betalambda}).
 \begin{figure}[ht]\vspace{-1cm}
 \begin{center} \leavevmode
  \epsfxsize=16cm
 \epsffile{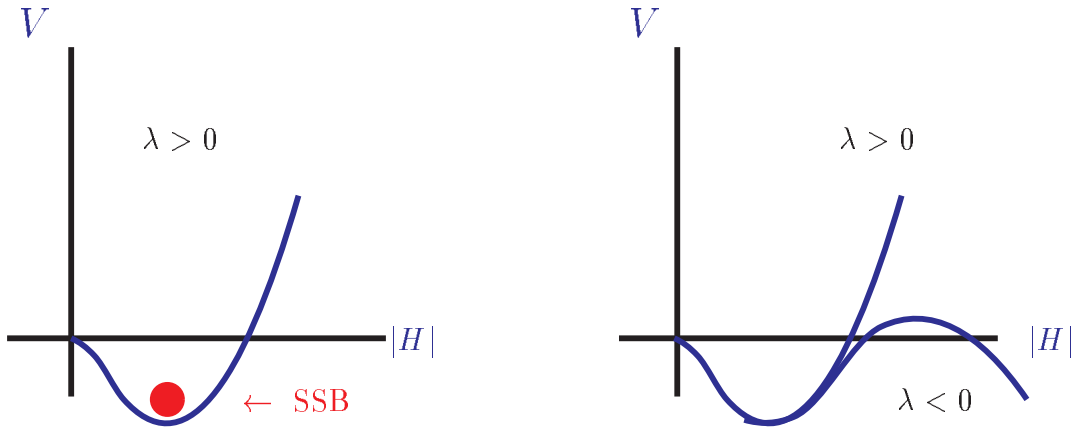}
 \end{center}\vspace{-1.5cm}
 \caption{The shape of the Higgs potential}\label{higpot}
 \end{figure}

On the other hand, the requirement of vacuum stability in the SM
(positivity of $\lambda$) imposes a lower bound on the Higgs boson
mass, which crucially depends on both the top-quark mass and the
cut-off $\Lambda$ \cite{bound1,bound2}. Again, the dependence of
this lower bound on $m_t$ is due to the effect of the top-quark
Yukawa coupling on the quartic coupling in eq.(\ref{betalambda}),
which drives $\lambda$ to negative values at large scales, thus
destabilizing the standard electroweak vacuum (see
Figs.\ref{higpot}).

\begin{figure}[htb]\vspace{0.9cm}
 \begin{center}
 \leavevmode
  \epsfxsize=10cm
 \epsffile{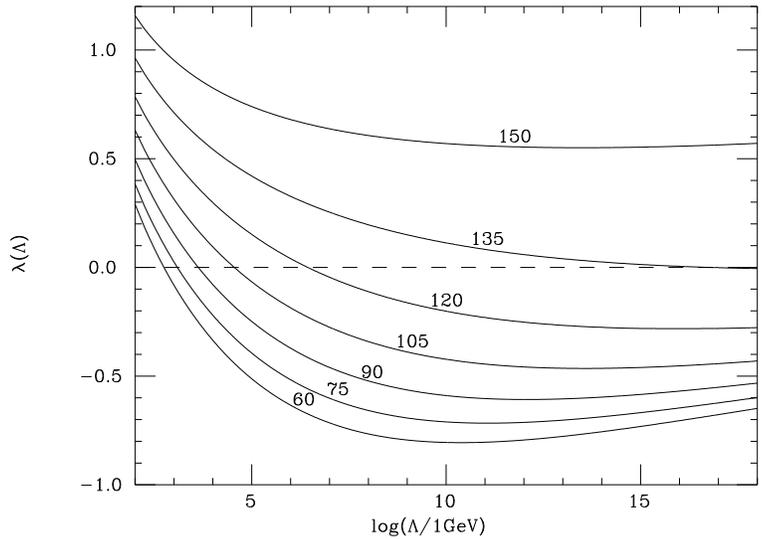}
 \end{center} \caption{The running of the Higgs quartic coupling.
 Numbers shown above the lines indicate the value of the Higgs mass in GeV.}
\label{fig:stab}
 \end{figure}

From the point of view of LEP and Tevatron physics, the upper
bound on the SM Higgs boson mass does not pose any relevant
restriction. The lower bound on $m_h$, instead, is particularly
important in view of the search for a Higgs boson at LEP II and
Tevatron. For $m_t\sim  174$ GeV and $\alpha_s(M_Z)=0.118$ the
running of the Higgs quartic coupling is shown in
Fig.\ref{fig:stab}.
The results at $\Lambda=10^{19}$ GeV or at $\Lambda=1$ TeV can be given by
the approximate formulae \cite{bound2}
 \begin{eqnarray}
m_h&>&135
+2.1[m_t-174]-4.5\left[\frac{\alpha_s(M_Z)-0.118}{0.006}\right], \
\ \ \ \Lambda=10^{19} \  {\rm GeV}, \label{19G}\\ m_h&>&\ 72
+0.9[m_t-174]-1.0\left[\frac{\alpha_s(M_Z)-0.118}{0.006}\right], \
\ \ \ \ \ \Lambda=1 \ {\rm TeV}, \label{1T}
 \end{eqnarray}
where the masses are in units of GeV.

\begin{figure}[htb]
\begin{center}
\epsfig{file=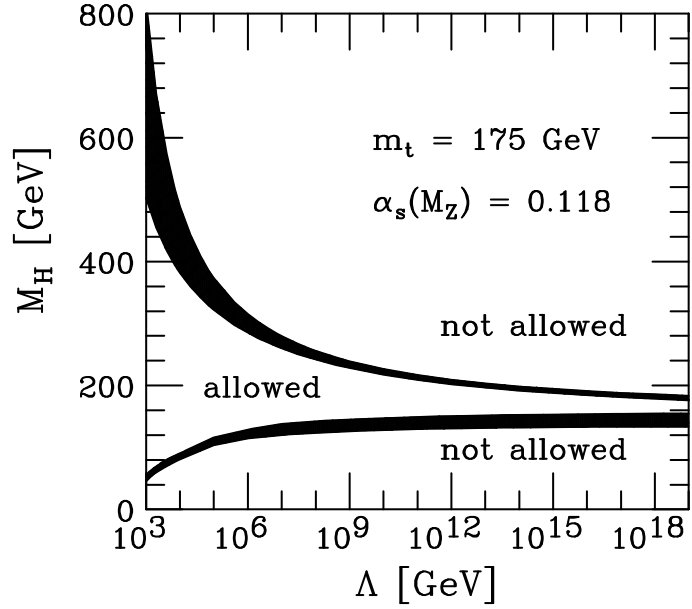,height=9cm,width=8cm,angle=90 }
\end{center}
 \caption{ Strong interaction and stability bounds on the SM Higgs
  boson mass. $\Lambda$ denotes the energy scale up to which the SM is valid.}
\label{mariano}
\end{figure}

Fig.\ref{mariano}~\cite{HR} shows the perturbativity and stability
bounds on the Higgs boson mass of the SM for different values of
the cut-off $\Lambda$ at which new physics is expected.
%
%
We see from Fig.\ref{mariano} and eqs.(\ref{19G},\ref{1T}) that
indeed for $m_t\sim 174$ GeV the discovery of a Higgs particle at
LEP II would imply that the Standard Model breaks down at a scale
$\Lambda$ well below $M_{GUT}$ or $M_{Planck}$, smaller for
lighter Higgs.  Actually, if the SM is valid up to $\Lambda \sim
M_{GUT}$ or $M_{Planck}$, for $m_t\sim 174$ GeV only a small range
of values is allowed: $134<m_h<\sim 200$ GeV.  For $m_t$ = 174 GeV
and $m_h < 100$~GeV [i.e. in the LEP II range] new physics should
appear below the scale $\Lambda \sim$ a few to 100~TeV. The
dependence on the top-quark mass however is noticeable. A lower
value, $m_t \simeq$ 170 GeV, would relax the previous requirement
to $\Lambda \sim 10^3 $ TeV, while a heavier value $ m_t \simeq$
180 GeV would demand new physics at an energy scale as low as
10~TeV.

\subsection{SM Higgs production at LEP}

\begin{figure}[h]
\begin{center}\vspace{-0.5cm}
 \leavevmode
  \epsfxsize=10cm
 \epsffile{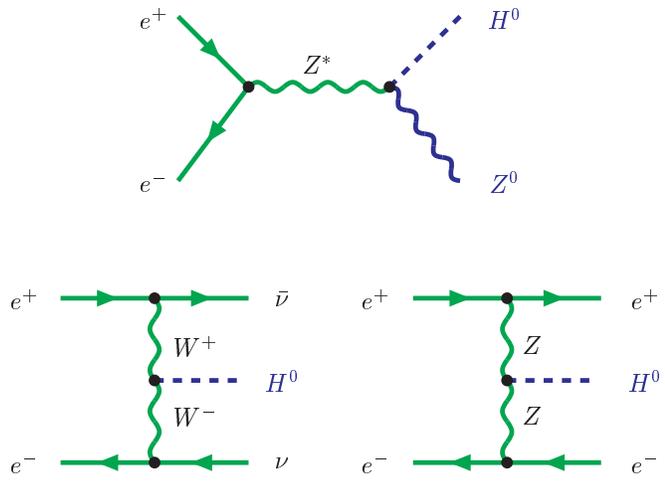}
\end{center}\vspace{-1cm} \caption{SM Higgs production at LEP:
Higgsstrahlung (above) and WW- and ZZ- fusion (below) } \label{higprod}
 \end{figure}

\begin{figure}[ht] \begin{center}\vspace{-0.5cm}
 \leavevmode
\epsfxsize=13cm \epsfysize=7cm\epsffile{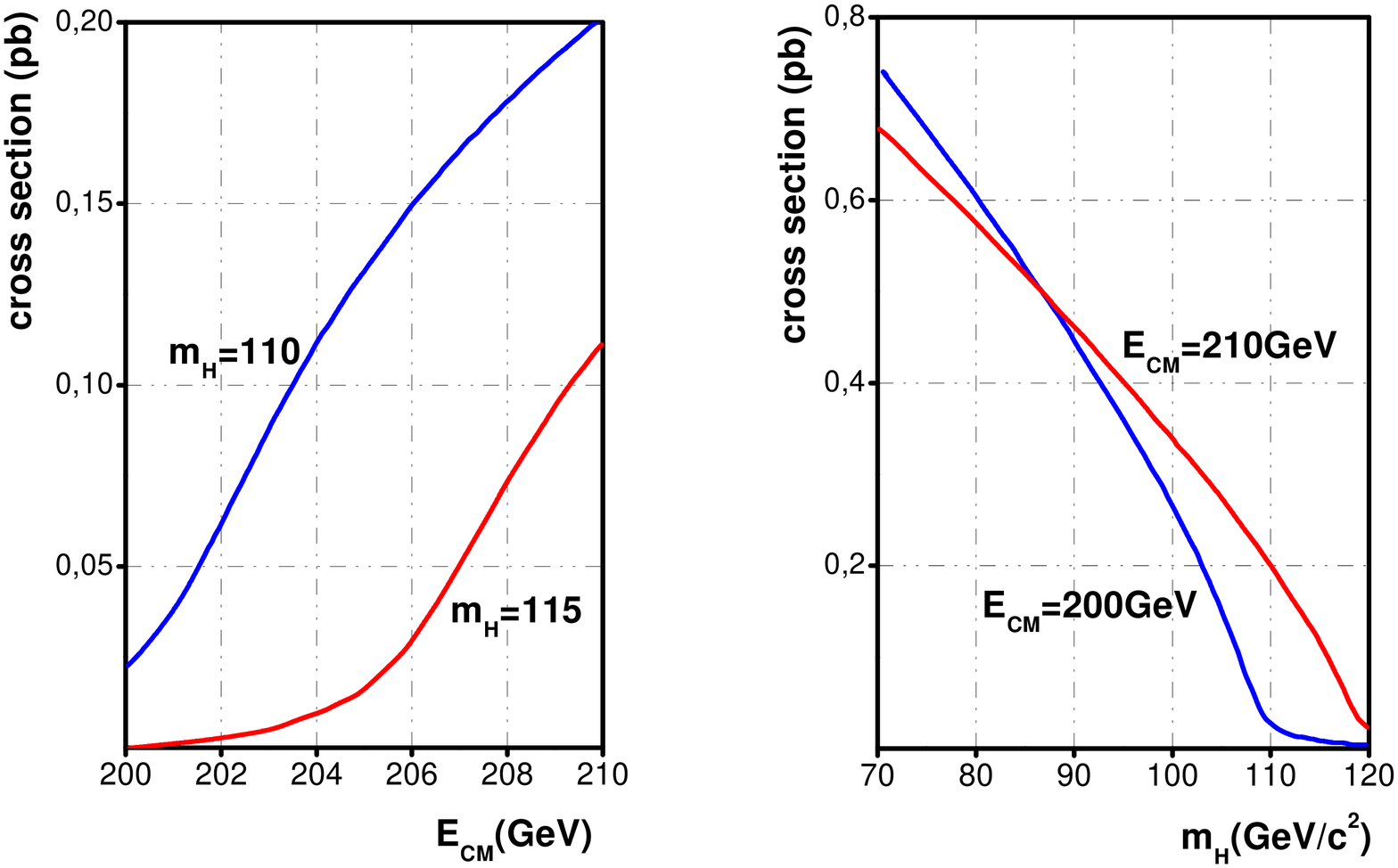}
 \end{center}\vspace{-0.3cm}
\caption{The cross section of the Higgs production at LEP II}
\label{higxsec}
 \end{figure}

The dominant mechanism for the Higgs boson production at LEP is
the Higgsstrahlung. The Higgs boson is produced together with the
$Z^0$ boson. A small contribution to the cross section comes also
from the WW- and ZZ- fusion processes (see Fig.\ref{higprod}).
The cross section depends on the Higgs boson mass and decreases
with increase of the latter. On the other hand, it grows with the
centre of mass energy, as shown in
Fig.\ref{higxsec}~\cite{nielsen}. Kinematic limit on the Higgs
production is given by the c.m. energy  minus the $Z$-boson mass.

\begin{figure}[h]
 \begin{center}\vspace{-0.8cm}
 \leavevmode
  \epsfxsize=11cm
 \epsffile{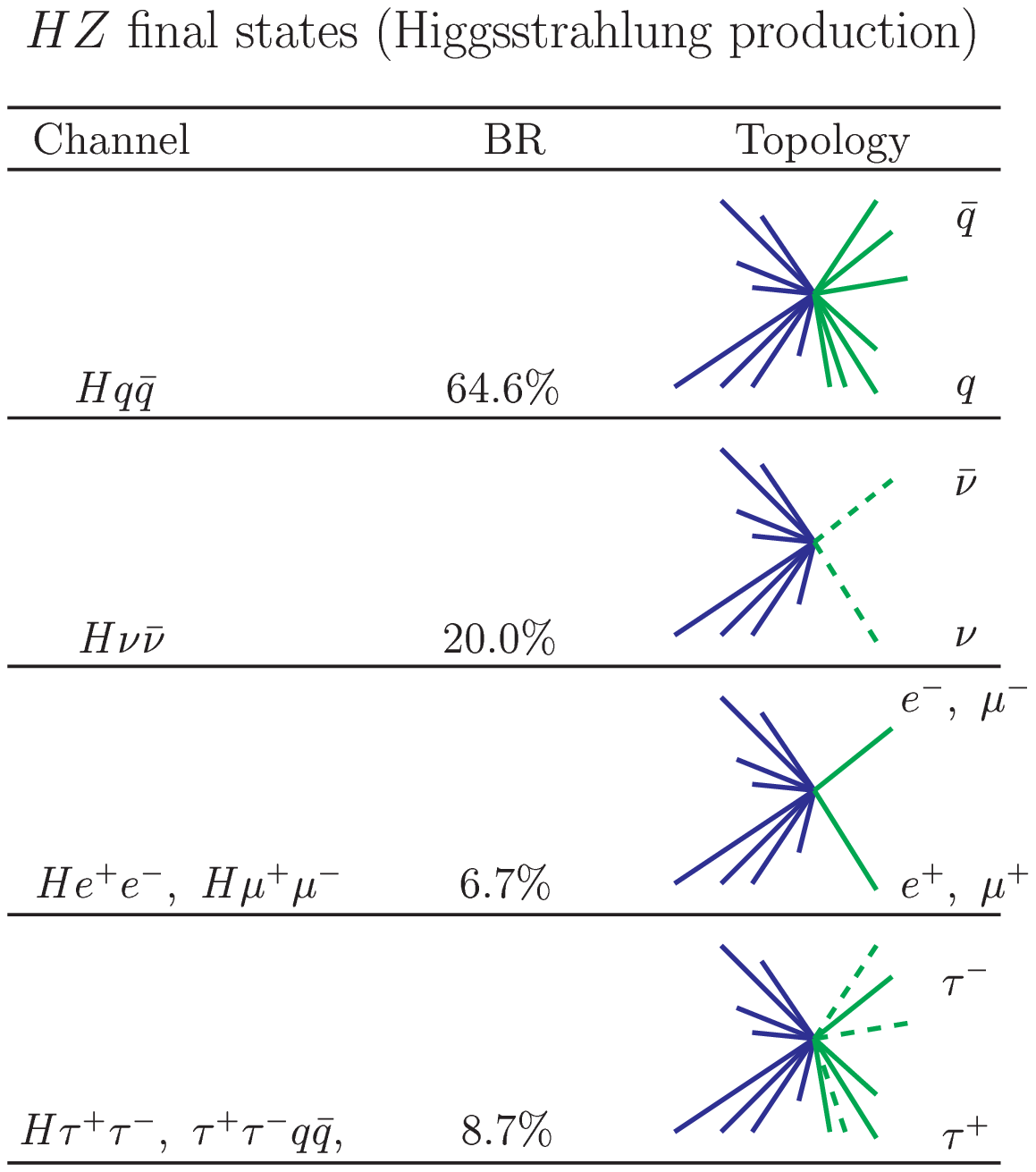}
\end{center}\vspace{-1cm}\caption{The final states of the Higgs boson decay with
the branching ratios} \label{final}
 \end{figure}

However, one of the main problems is to distinguish the final
products of the Higgs boson decay from the background, mainly the
$ZZ$ pair production. The branching ratios for the Higgs boson
decay are shown in Fig.\ref{final}. The $Z$ boson has the same
decay modes with different branchings.
In the final states, one has either four hadronic jets, or two
jets and two leptons, or four leptons. The most probable is the
four jet configuration, which is the most difficult from the point
of view of unwanted background. A two-jet and two-lepton final
state is more clean though less probable.

\begin{figure}[t]\vspace{0.3cm}
\begin{center}
 \leavevmode
 \epsfxsize=12cm
 \epsffile{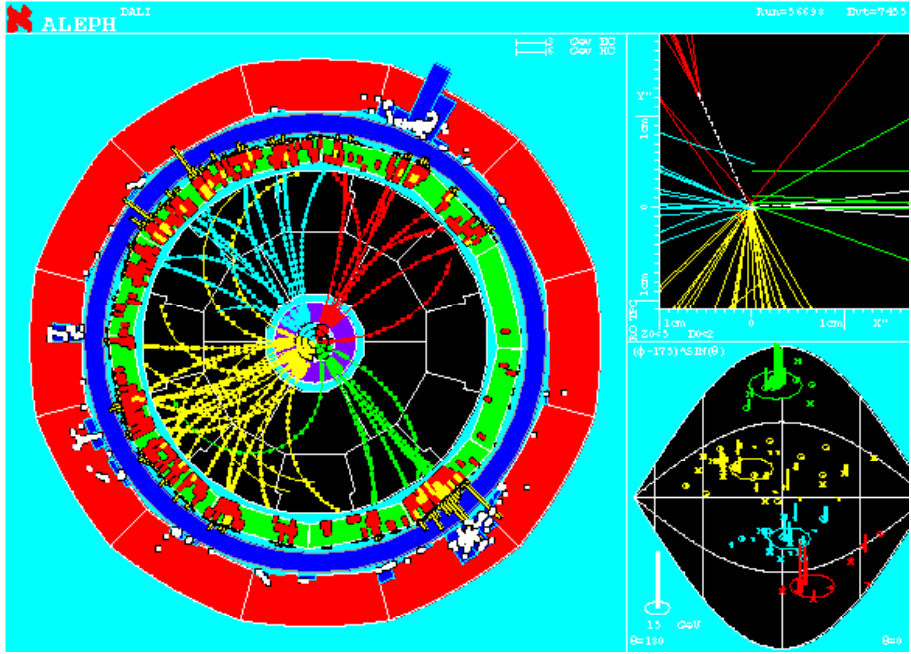}
\end{center} \caption{Typical four jet event} \label{event}
 \end{figure}

\begin{figure}[ht]
\begin{center}
 \leavevmode
  \epsfxsize=8cm
 \epsffile{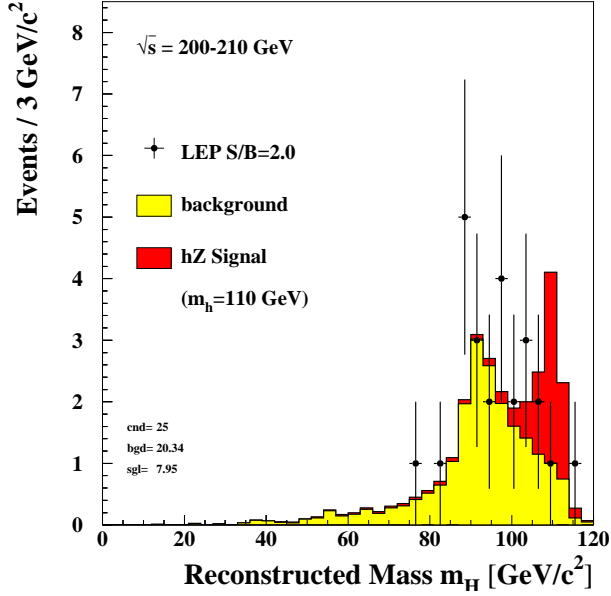}
\end{center}\caption{Reconstructed Higgs mass for four jet events. The peak
shown in red corresponds to a would be Higgs boson with mass of
110 GeV} \label{massdis}
 \end{figure}

Attempts to find the Higgs boson have not met success so far. All
the data are consistent with the background. An interesting
four-jet event is shown in Fig.\ref{event} and is most likely a
$ZZ$ candidate~\cite{ALEPH}. A reconstructed invariant mass of two
jets does not show noticeable deviation from background
expectation. For the 68.1 background events expected, there are 70
events observed. The reconstructed Higgs mass for four-jet events
is shown in Fig.\ref{massdis}. At this kind of plots the real
Higgs boson should give a peak above the background, as is shown
for a would be Higgs mass of 110 GeV in
Fig.\ref{massdis}~\cite{ALEPH}.

\begin{figure}[ht]
\vspace{0.5cm}
 \leavevmode
  \epsfxsize=11.3cm \epsfysize=6.9cm
\hspace*{-2.8cm} \epsffile{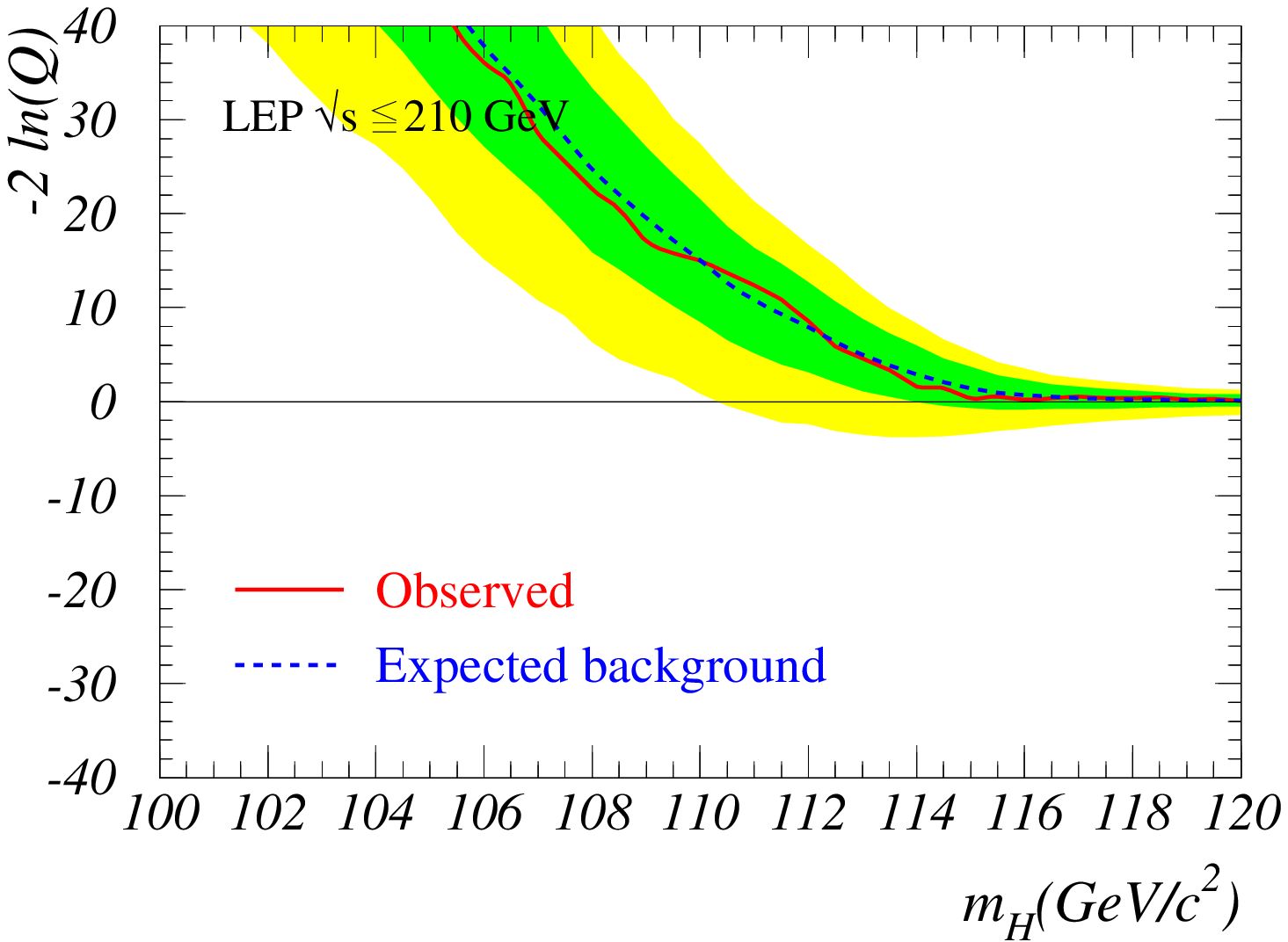}\vspace{-6.9cm}

 \epsfxsize=7.3cm  \epsfysize=7cm
\hspace*{8cm} \epsffile{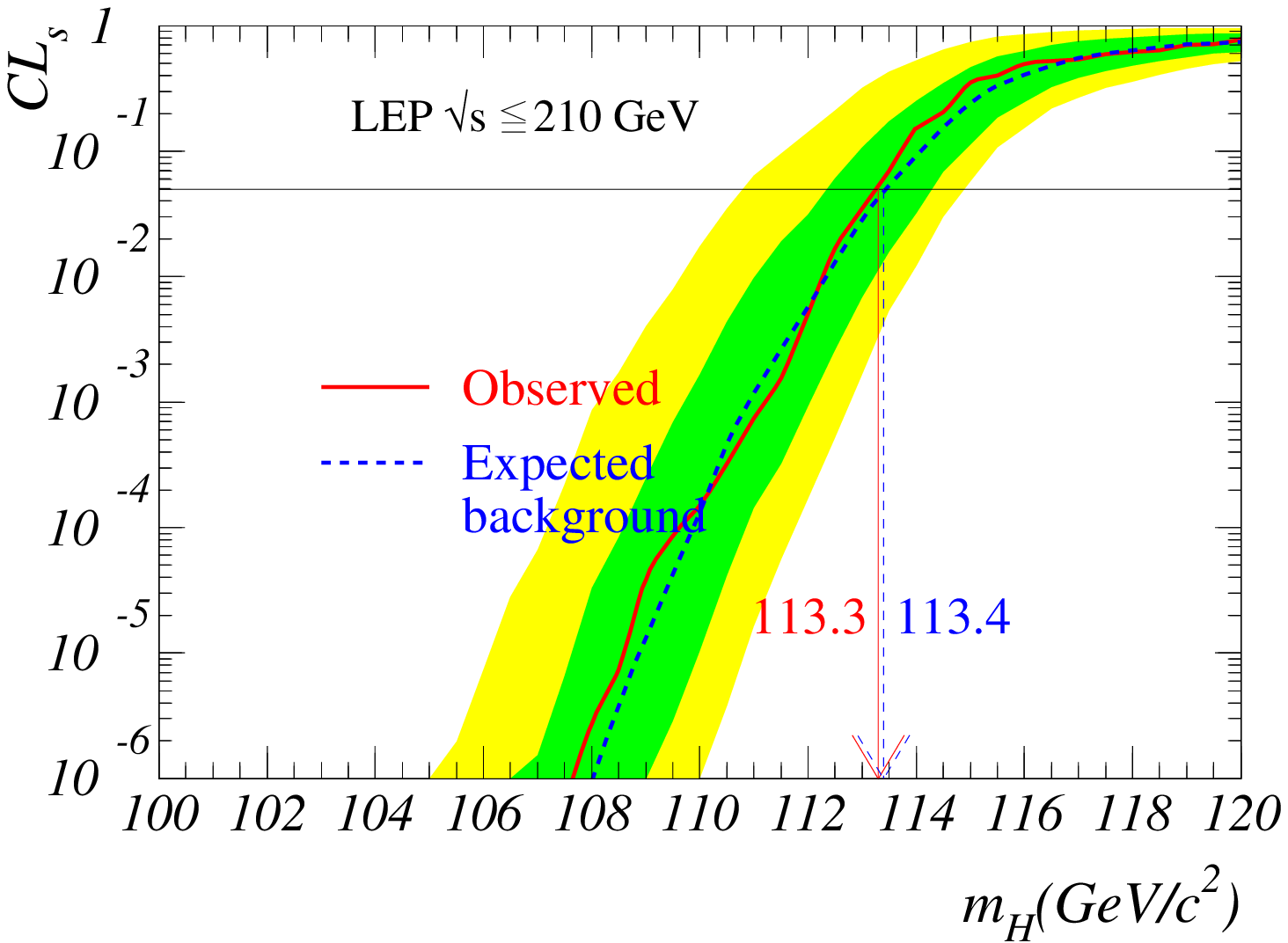}
 \caption{Combined Confidence Level plots for the Higgs searches at LEP} \label{CL}
\end{figure}

\begin{figure}[ht]\vspace{-0.5cm}\Large
 \begin{center}
 \hspace*{-5cm}{\bf Events/3 GeV/c$^2$}\vspace{0.3cm}

 \leavevmode
  \epsfxsize=10cm \epsfysize=13cm
 \hspace*{-2cm}\epsffile{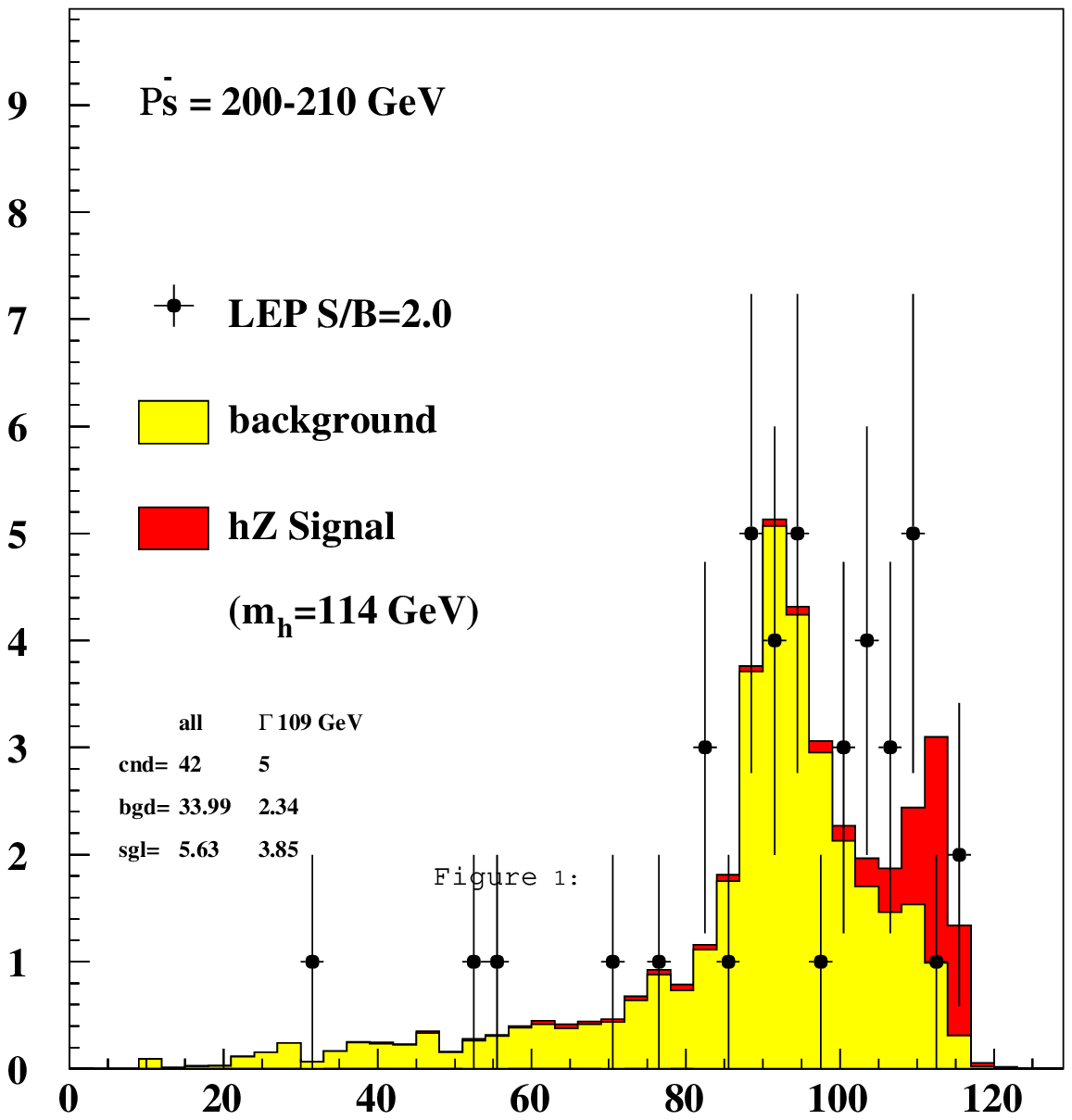}

 \vspace{-4.2cm}{\bf Reconstructed Mass m$_H$  [GeV/c$^2$]}\vspace{0.5cm}

\begin{tabular}{lrrr}& Data & Backg & Signal \\
\hline
All $m_{rec}$ & 42 & 34.0 & 5.6 \\
$m_{rec}>$ 109 GeV & 5 & 2.3 & 3.9 \\
\hline
\end{tabular}
\end{center}
\caption{Higgs Candidate for 114 GeV/$c^2$} \label{higcand}
\end{figure}

  Combined results from four LEP collaborations (ALEPH, DELPHI, L3 and OPAL)
 in the energy interval $\sqrt s =200-210$ GeV allow one to get a lower
limit on the Higgs mass. As it follows from Fig.\ref{CL}, at the
95\% confidence level it is \cite{EWWG}
\begin{equation}
  m_h > 113.3\ {\rm GeV/c}^2 \ \ \  \mbox{@}\  95\% \ C.L.
\end{equation}
Recent hot news from the LEP II accelerator show slight excess of
events in hadronic channels. For the hard cuts keeping only
"really good" events one can achieve the signal/background ratio
of 2 with a few signal events indicating the 114 GeV Higgs boson
(see Fig.\ref{higcand}). Deviation from the background achieves
2.9 standard deviations and is better seen in the confidence level
plots~\cite{ALEPH}. There are also some events in leptonic
channel~\cite{L3}. However, statistics is not enough to make
definite conclusions.
 \normalsize

\subsection{The Higgs boson mass in the MSSM}

It has already been mentioned that in the MSSM the mass of the
lightest Higgs boson is predicted to be less than the $Z$-boson
mass. This is, however, the tree level result and the masses
acquire the radiative corrections.

With account taken of the radiative corrections, the effective
Higgs bosons potential is
\begin{equation}
V_{Higgs}^{eff} = V_{tree} + \Delta V,
\end{equation}
where $V_{tree}$ is given by eq.(\ref{Higpot}) and in the one-loop
order
\begin{equation}
 \Delta V_{1 loop}= \sum_k \frac{1}{64\pi^2}(-1)^{J_k}(2J_k+1)c_km_k^4
 \left(\log \frac{m_k^2}{Q^2}- \frac{3}{2}\right).
\end{equation}
Here the sum is taken  over all the particles in the loop, $J_k$
is the spin and $m_k$ is the field dependent mass of a particle at
the scale $Q$.

\begin{figure}[ht]\vspace{-1cm}
\begin{center}
\leavevmode
  \epsfxsize=12cm
 \epsffile{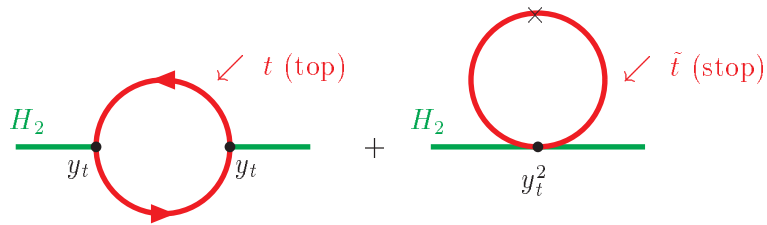}
\end{center}\vspace{-1.5cm}
\caption{Corrections to the Higgs boson self-energy from the
top(stop) loops} \label{topcor}
\end{figure}

 The main contribution comes from the diagrams shown in Fig.\ref{topcor}.
These radiative corrections vanish when supersymmetry is not
broken and are positive in the softly broken case. They are
proportional to the mass squared of top (stop) quarks and depend
on the values of the soft breaking parameters.  Contributions from
the other particles are much smaller~\cite{radcorr,carena,G}.
 The leading contribution comes from (s)top loops
\begin{equation}
 \Delta V_{1 loop}^{stop}= \frac{3}{32\pi^2}\left[\tilde{m}_{t_1}^4
 (\log \frac{\tilde{m}_{t_1}^2}{Q^2}- \frac{3}{2})+\tilde{m}_{t_2}^4
 (\log \frac{\tilde{m}_{t_2}^2}{Q^2}- \frac{3}{2})-2m_t^4
 (\log \frac{m_t^2}{Q^2}- \frac{3}{2})\right].
\end{equation}

These corrections lead to the following modification of  the
tree-level relation for the lightest Higgs mass
\begin{equation}
m_h^2 \approx M_Z^2\cos^2 2\beta + \frac{3g^2 m_t^4}{16\pi^2M_W^2}
\log\frac{ \tilde{m}_{t_1}^2\tilde{m}_{t_2}^2}{m_t^4}. \label{rad}
\end{equation}
One finds that the one-loop correction is positive and increases
the mass value. Two loop  corrections have the opposite effect but
are smaller and result in slightly lower value of  the Higgs mass
~\cite{CW,BGGK,feynhiggs}.

To find out  numerical values of these corrections, one has to
determine the masses of all superpartners.
\begin{figure}[ht] \begin{center}
 \leavevmode
  \epsfxsize=12cm
 \epsffile{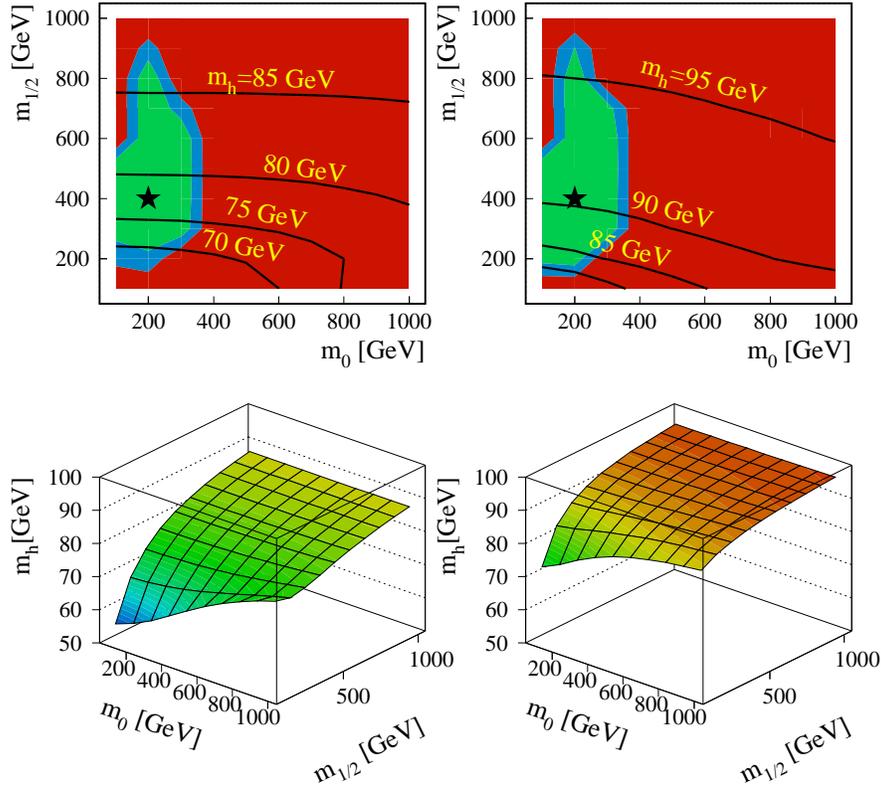}
\end{center}
\caption{The mass of the lightest Higgs boson for the low
$\tan\beta$ solution as a function of $m_0$ and $m_{1/2}$. The
contours at the upper plots correspond to fixed values of the
Higgs mass. The lower plots demonstrate the saturation of the mass
at high values of mass parameters.} \label{fig:mh}
\end{figure}
Within the Constrained MSSM, imposing various constraints, one can
define the allowed region in the parameter space and calculate the
spectrum of superpartners and, hence, the radiative corrections to
the Higgs boson mass (see Figs.\ref{fig:mh}, \ref{hi35}).

The Higgs mass depends mainly on the following parameters: the top
mass, the squark masses, the mixing in the stop sector, the
pseudoscalar Higgs mass and $\tan\beta$. As will be shown below,
the maximum Higgs mass is obtained for large $\tan\beta$, for a
maximum value of the top and squark masses and a minimum value of
the stop mixing.

Note that in the CMSSM the Higgs mixing parameter $\mu$ is
determined by the requirement of EWSB, which yields large values
for $\mu$~\cite{BEK}. Given that the pseudoscalar Higgs mass
increases rapidly with $\mu$, this mass is always much larger than
the lightest Higgs mass and thus decouples. This decoupling is
effective for all regions of the CMSSM parameter space, i.e. the
lightest Higgs has the couplings of the SM Higgs within a few per
cent.
\begin{figure}[b]
\epsfig{file=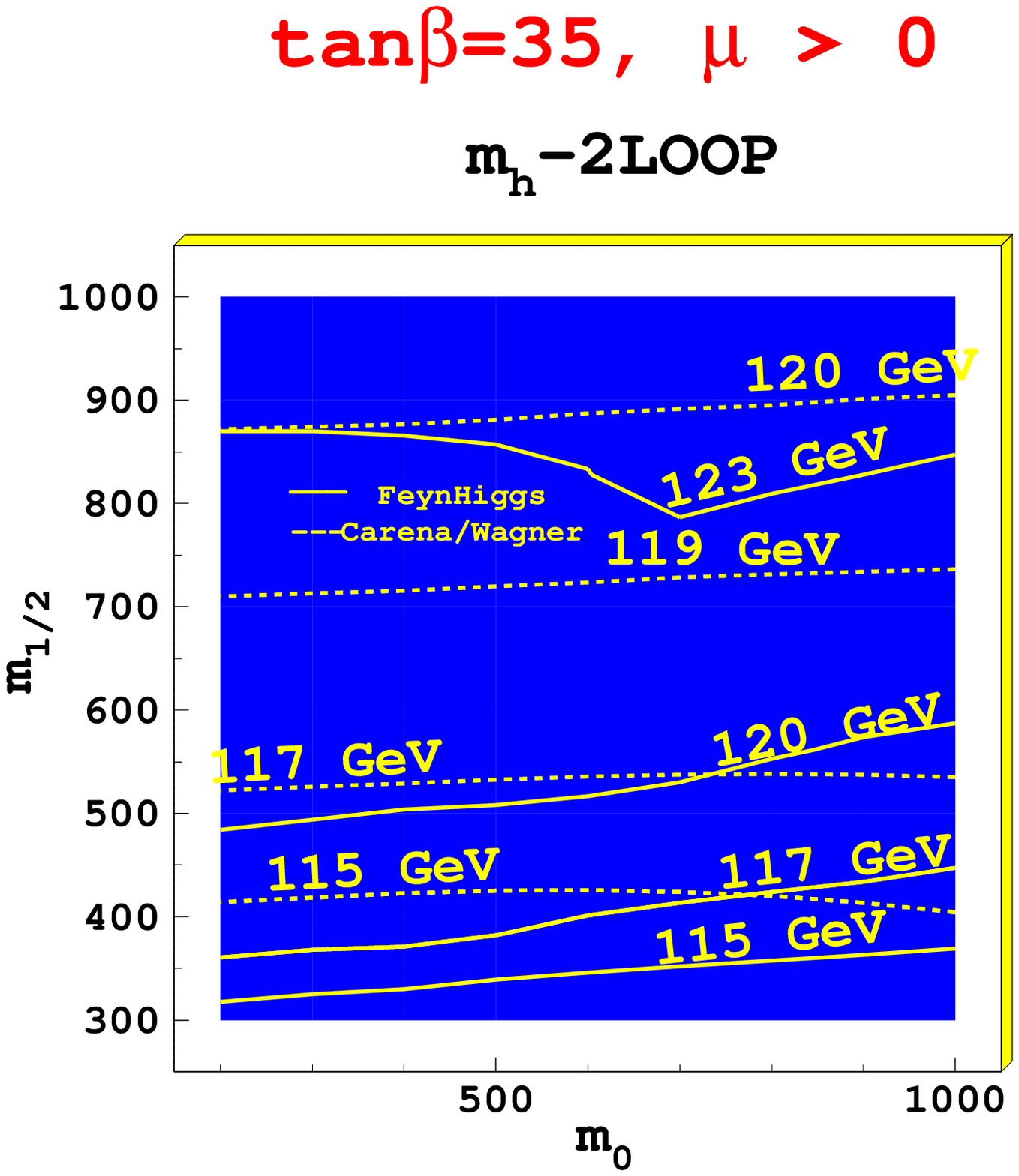,width=.49\textwidth}
\epsfig{file=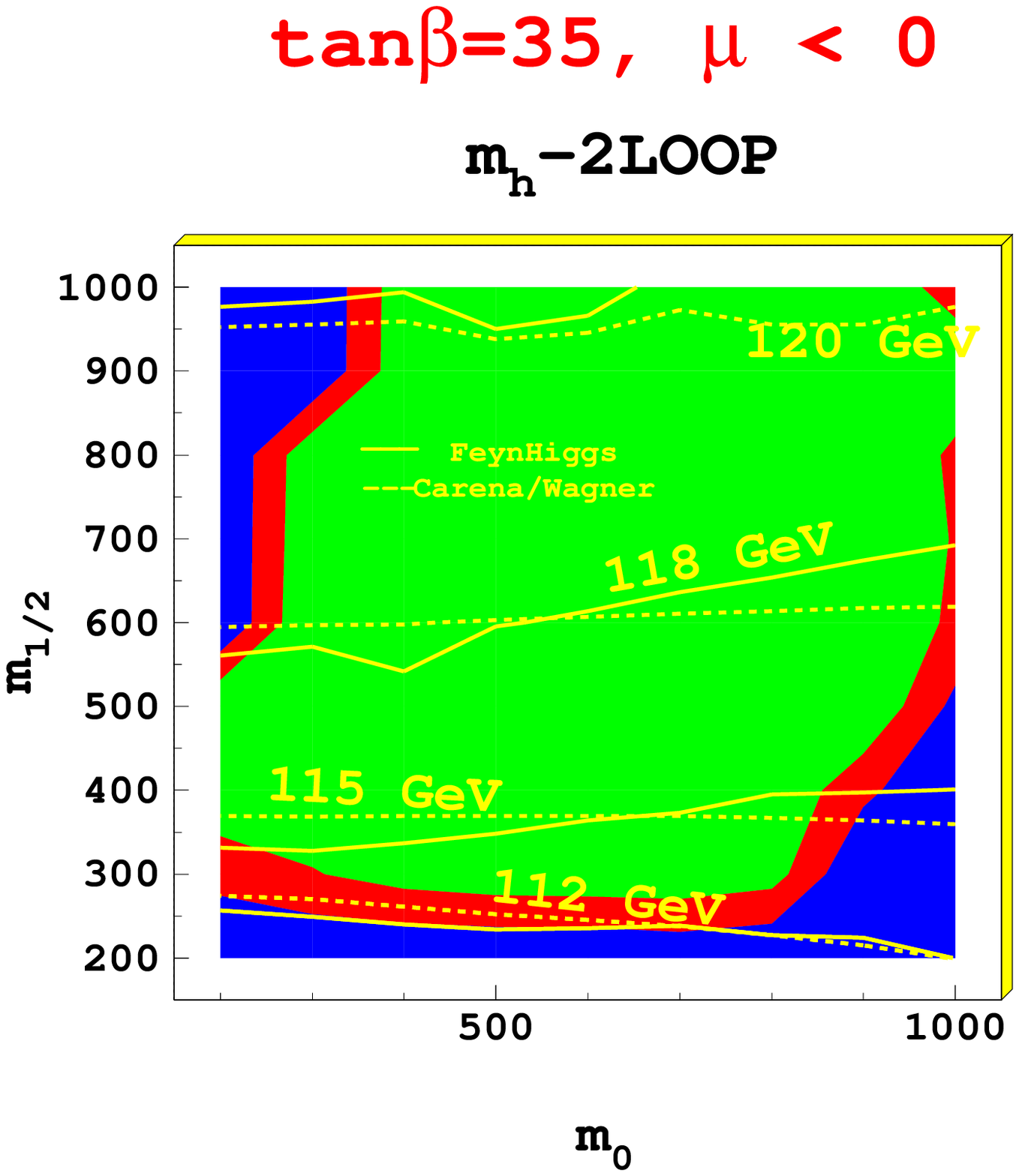,width=.49\textwidth}
\epsfig{file=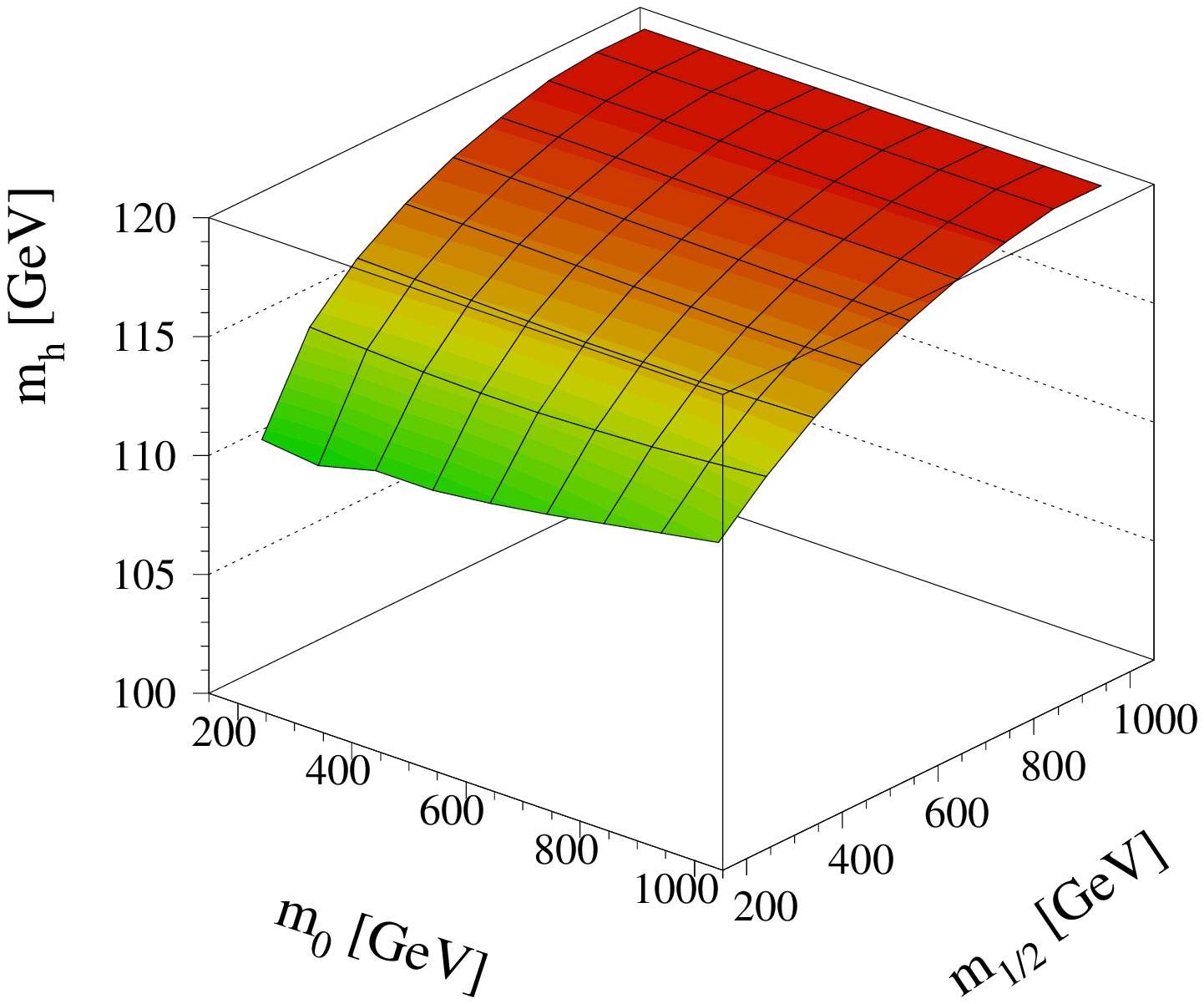,width=.49\textwidth}
\epsfig{file=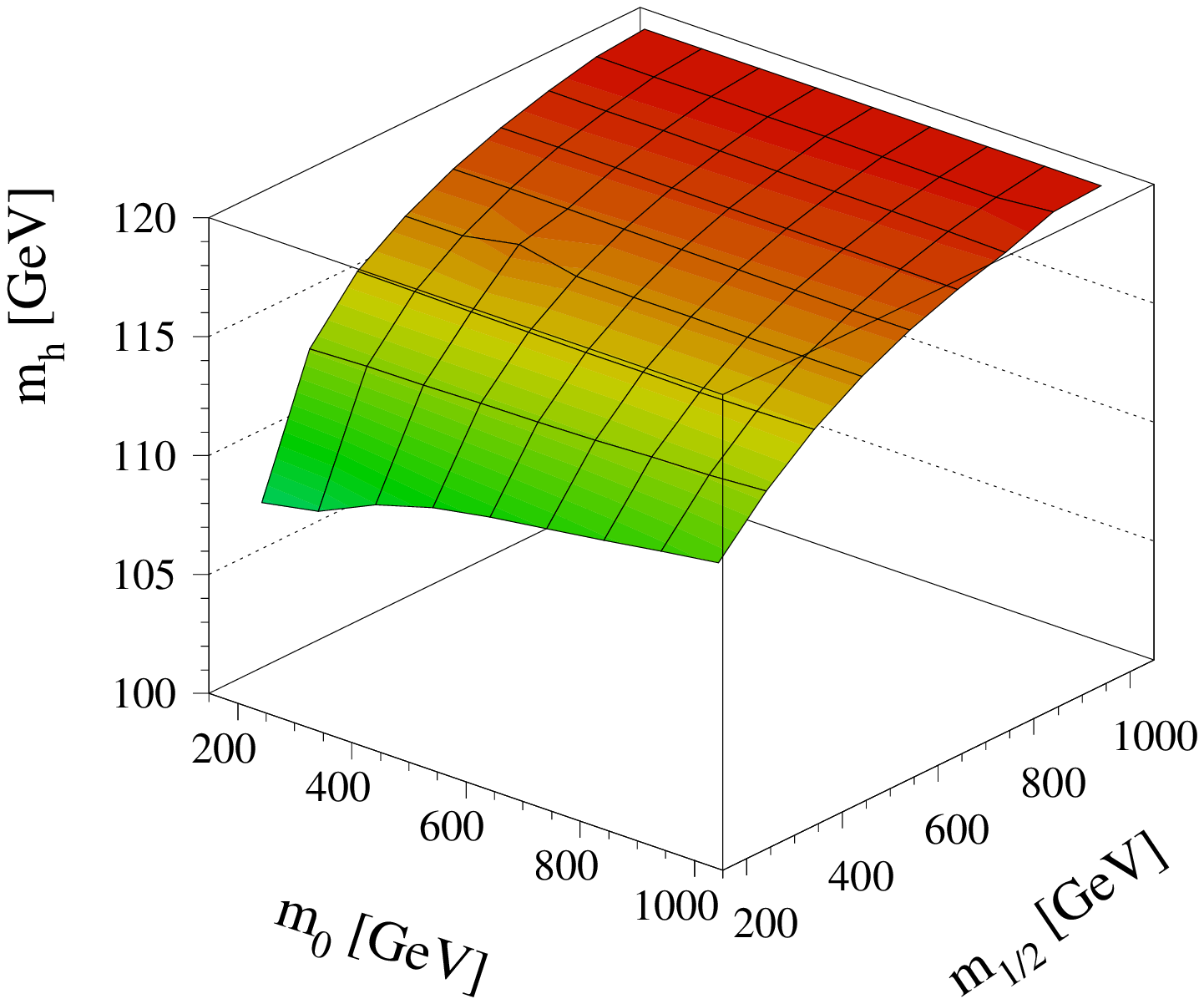,width=.49\textwidth}
\caption[]{\label{hi35} The same as in Fig.~\ref{fig:mh} but for
the high tan $\beta$ solution $\tan\beta=35$.}
\end{figure}
We present the value of the lightest Higgs mass in the whole
$m_0,m_{1/2}$ plane for low and high $\tan\beta$ solutions,
respectively~\cite{BHGK} in Figs.\ref{fig:mh}, \ref{hi35}. One can
see that it is practically constant in the whole plane and is
saturated for high values of $m_0$ and $m_{1/2}$.

\begin{figure}[ht]
\begin{center}
 \leavevmode
  \epsfxsize=10cm \epsfysize=9cm
 \epsffile{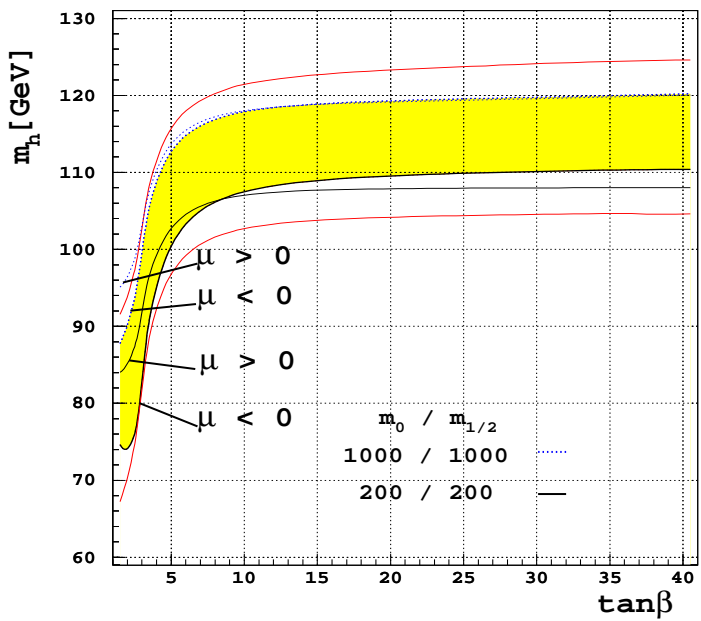}
\end{center}\vspace{-1cm}
\caption{The mass of the lightest Higgs boson as a function of
$\tan\beta$} \label{fig:mhtb}
\end{figure}

 The lightest
Higgs boson mass $m_h$ is shown as a function of $\tan\beta$ in
Fig.~\ref{fig:mhtb} \cite{BHGK}. The shaded band corresponds to
the uncertainty from the stop mass and stop mixing for $m_t=175$
GeV. The upper and lower lines correspond to $m_t$=170 and 180
GeV, respectively.

The parameters used for the calculation of the upper limit are:
$m_t=180$ GeV, $A_0=-3m_0$ and $m_0=m_{1/2}=1000$ GeV. The lowest
line of the same figure gives the minimal values  of $m_h$. For
high $\tan\beta$  the values of $m_h$ range from 105 GeV 125 GeV.
At present, there is no preference for any of the values in this
range but it can be seen that the 95\% C.L. lower limit on the
Higgs mass~\cite{EWWG} of 113.3 GeV excludes $\tan\beta<3.3$.

In order to better understand the Higgs mass uncertainties,
 the relevant parameters were varied one by one.
The largest uncertainty on the light Higgs mass originates from
the stop masses. The Higgs mass varies between 110 and 120 GeV, if
$m_0$ and $m_{1/2}$ are varied between 200 and 1000 GeV, which
implies stop masses varying between 400 and 2000 GeV. Since at
present there is no preference for any of the values between 110
and 120 GeV, the variance for a flat probability distribution is
10/$\sqrt{12}$=3 GeV, which we take as an error estimate.

The remaining uncertainty of the Higgs mass originates from the
mixing in the stop sector when one leaves $A_0$ as a free
parameter. The mixing is determined by the off-diagonal element in
the stop mass matrix $X_t=A_t-\mu/\tan\beta$. Its influence on the
Higgs mass is quite small in the CMSSM since the low energy value
$A_t$ tends to a fixed point so that the stop mixing parameter
$X_t=A_t-\mu/\tan\beta$ is not strongly dependent on $A_0$.
Furthermore, the $\mu$ term is not important at large $\tan\beta$.
If we vary $A_0$ between $\pm3m_0$,
 the error from the stop mixing in the Higgs boson mass
is estimated to be $\pm 1.5$ GeV. The values of $m_0=m_{1/2}=370$ GeV yield
the central value of $m_h=115$ GeV.

Given the uncertainty on the top mass of 5.2 GeV~\cite{mtop} leads
to the uncertainty for the Higgs mass at large $\tan\beta$ of
$\pm$ 5 GeV.

The uncertainties from the higher order calculations (HO) is
estimated to be 2 GeV from a comparison of the full diagrammatic
method~\cite{feynhiggs} and the effective potential
approach~\cite{CW}. So combining all the uncertainties discussed
before the results  for the  Higgs mass in the CMSSM can be
summarized as follows:
\begin{itemize}
 \item The low $\tan\beta$ scenario ($\tan\beta <3.3$) of the
CMSSM is excluded by the lower limit on the Higgs mass of  113.3
GeV~\cite{EWWG}.
\item For the high $\tan\beta$ scenario the  Higgs mass is found to be in
the range from 110 to 120 GeV for $m_t=175$ GeV.
 The  central value  is found to be~\cite{BHGK}:
 \begin{equation}
 m_h=115\pm3~ ({\rm stop mass})~\pm1.5~({\rm stop
mixing})~\pm2~({\rm theory})~\pm5~({\rm top mass})~\rm GeV,
 \end{equation}
 where the errors  are the estimated standard deviations
around the central value. This prediction is independent of
$\tan\beta$ for $\tan\beta >20$ and decreases for lower
$\tan\beta$.
 \end{itemize}

\begin{figure}[h]
\vspace{-0.7cm}
\begin{center}
\leavevmode \epsfxsize=13cm \epsfysize=8cm
 \epsffile{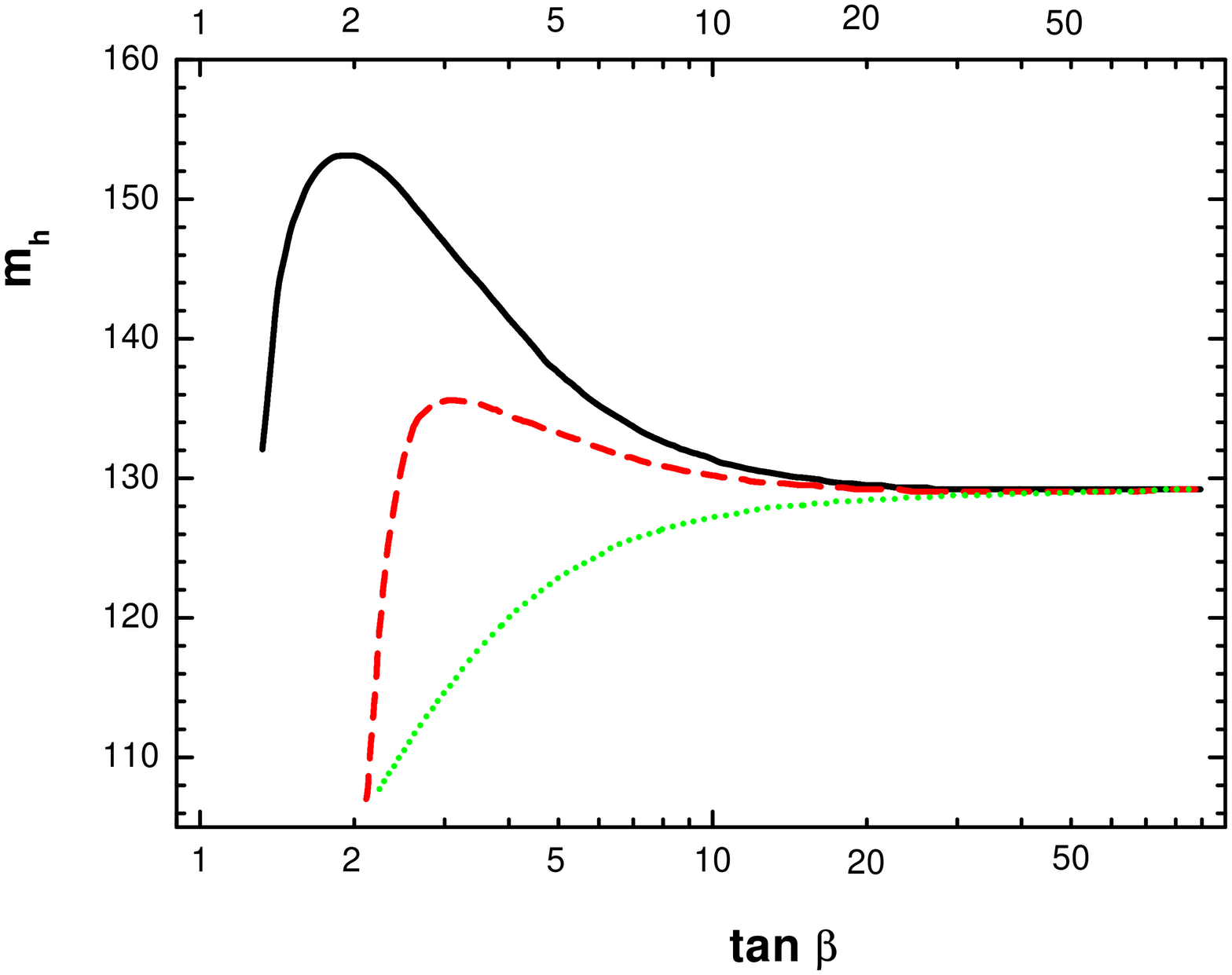}
\vspace{-0.5cm} \caption[]{ Dependence of the upper bound on the
lightest Higgs boson mass on $\tan\beta$ in MSSM (lower curve),
NMSSM (middle curve) and extended SSM (upper curve)\label{f3} }
\end{center}
\end{figure}

 However, these SUSY limits on the Higgs mass may not be so restricting if
non-minimal SUSY models are considered. In a SUSY model extended
by a singlet, the so-called Next-to-Minimal model,
eq.(\ref{bound}) is modified and at the tree level the upper bound
looks like~\cite{pomarol}
\begin{equation}
  m_h^2 \simeq M_Z^2\cos^2 2\beta+ \lambda^2v^2\sin^2 2\beta,
\end{equation}
where $\lambda$ is an additional singlet Yukawa coupling. This
coupling being unknown brings us back to the SM situation, though
its influence is reduced by $\sin 2\beta$. As a result, for low
$\tan\beta$ the upper bound on the Higgs mass is slightly modified
(see Fig.\ref{f3}).

Even more dramatic changes are possible in models containing
non-standard fields at intermediate scales. These fields appear in
scenarios with gauge mediated supersymmetry breaking. In this
case, the upper bound on the Higgs mass may increase up to 155
GeV~\cite{pomarol} (the upper curve in Fig.\ref{f3}), though it is
not necessarily saturated.  One should notice, however, that these
more sophisticated models do not change the generic feature of
SUSY theories, the presence of the light Higgs boson.

\subsection{Perspectives of observation}

\underline{LEP}
 \vspace{0.3cm}

\begin{figure}[ht]
 \begin{center}
 \leavevmode
  \epsfxsize=13cm
 \epsffile{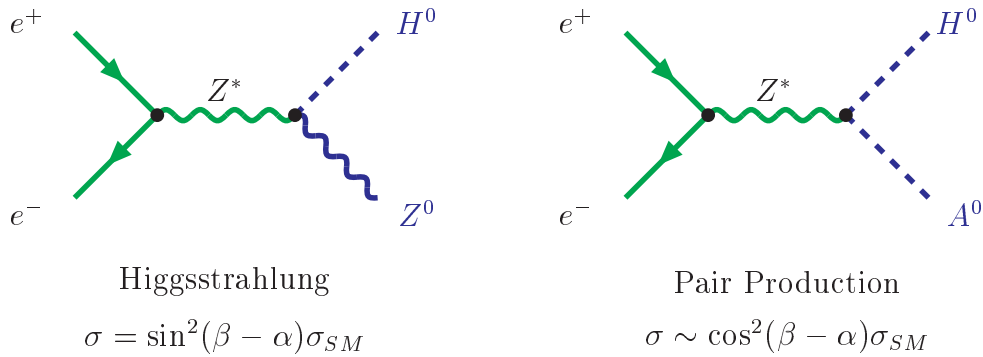}\vspace{-0.5cm}
\end{center} \caption{MSSM Higgs production at LEP: complimentary diagrams}
\label{susyprod}
 \end{figure}

\begin{figure}[h]\vspace{0.5cm}
 \begin{center}
 \leavevmode
  \epsfxsize=8cm
\epsffile{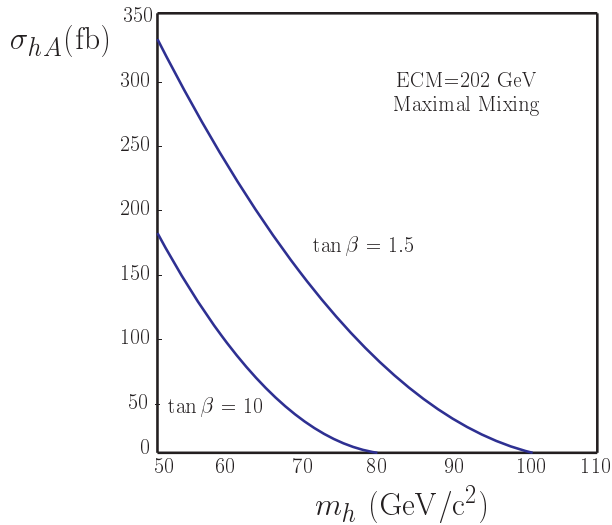}\vspace{-0.5cm}
\end{center} \caption{$hA$ pair production cross section in fb as a
 function of $m_h$ and $\tan\beta$}
\label{pair}
 \end{figure}

In the case of supersymmetry, contrary to the SM, there are two
competing processes for neutral Higgs production. Besides the
usual Higgsstrahlung diagram there is also the pair production one
when two Higgs bosons (the usual one and the pseudoscalar boson
$A$) are produced. The cross-sections of these two processes are
complimentary and related to the SM one by a simple formula (see
Fig.\ref{susyprod}).
Thus, the cross-section for Higgs production in the MSSM is
usually lower than that of the SM. Therefore,  searches for pair
production are limited by a low cross-section rather than by a
threshold (see Fig.\ref{pair}).

Non-observation of the Higgs boson at LEP in general gives lower
bound on the Higgs boson mass than that in the SM. Modern
experimental limits on the MSSM Higgs bosons are~\cite{nielsen}
 \begin{equation}
  m_h > 90.5\ {\rm GeV/c}^2, \ \ \ m_A > 90.5\ {\rm GeV/c}^2
   \ \ \ {\mbox @}\ \ 95\% \ \ C.L.
 \end{equation}

\begin{figure}[ht]
 \leavevmode \hspace*{-1cm}
  \epsfxsize=8.5cm
 \epsffile{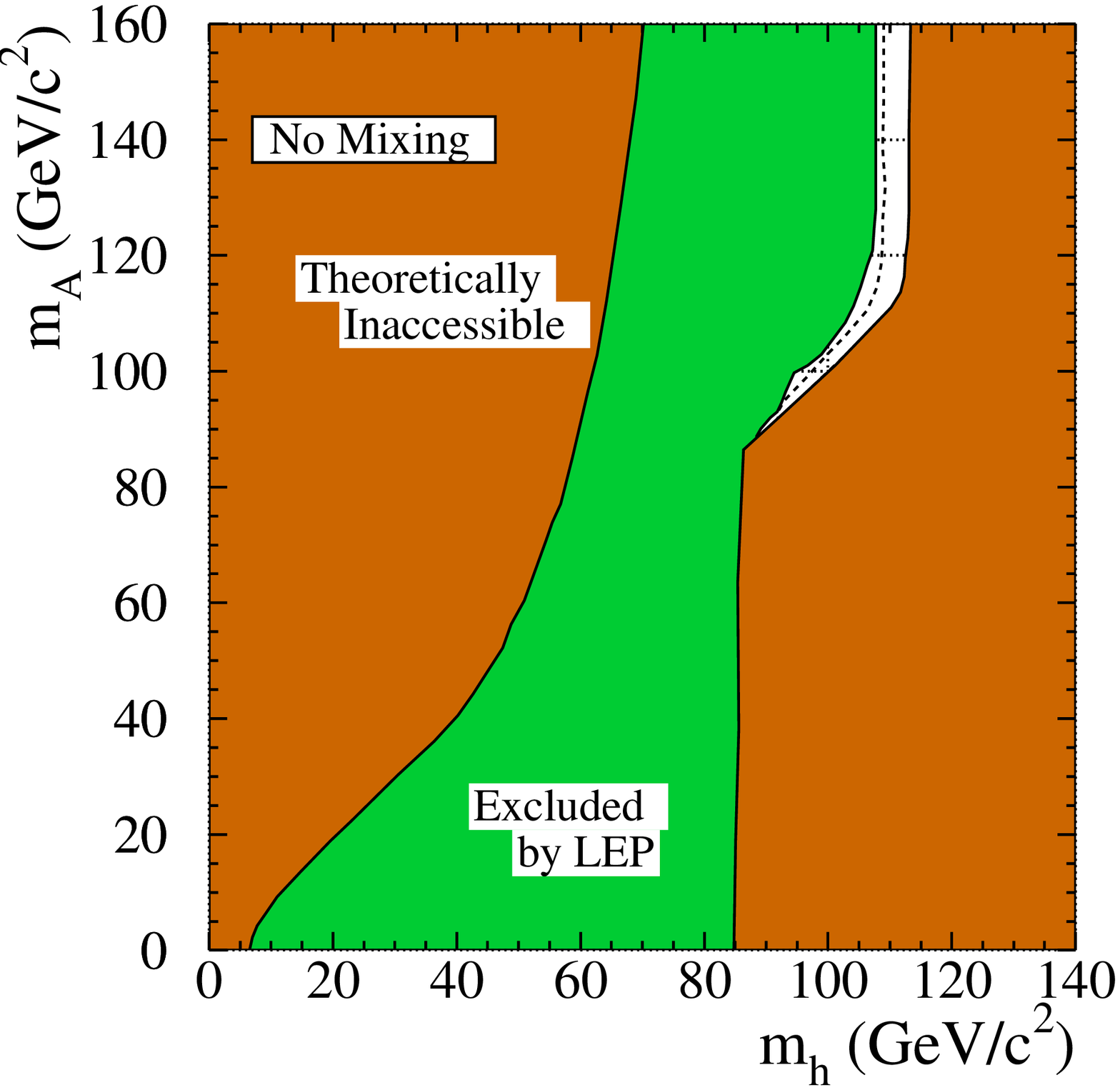}
\vspace{-8.5cm}

\hspace*{7.5cm}\epsfxsize=8.5cm
 \epsffile{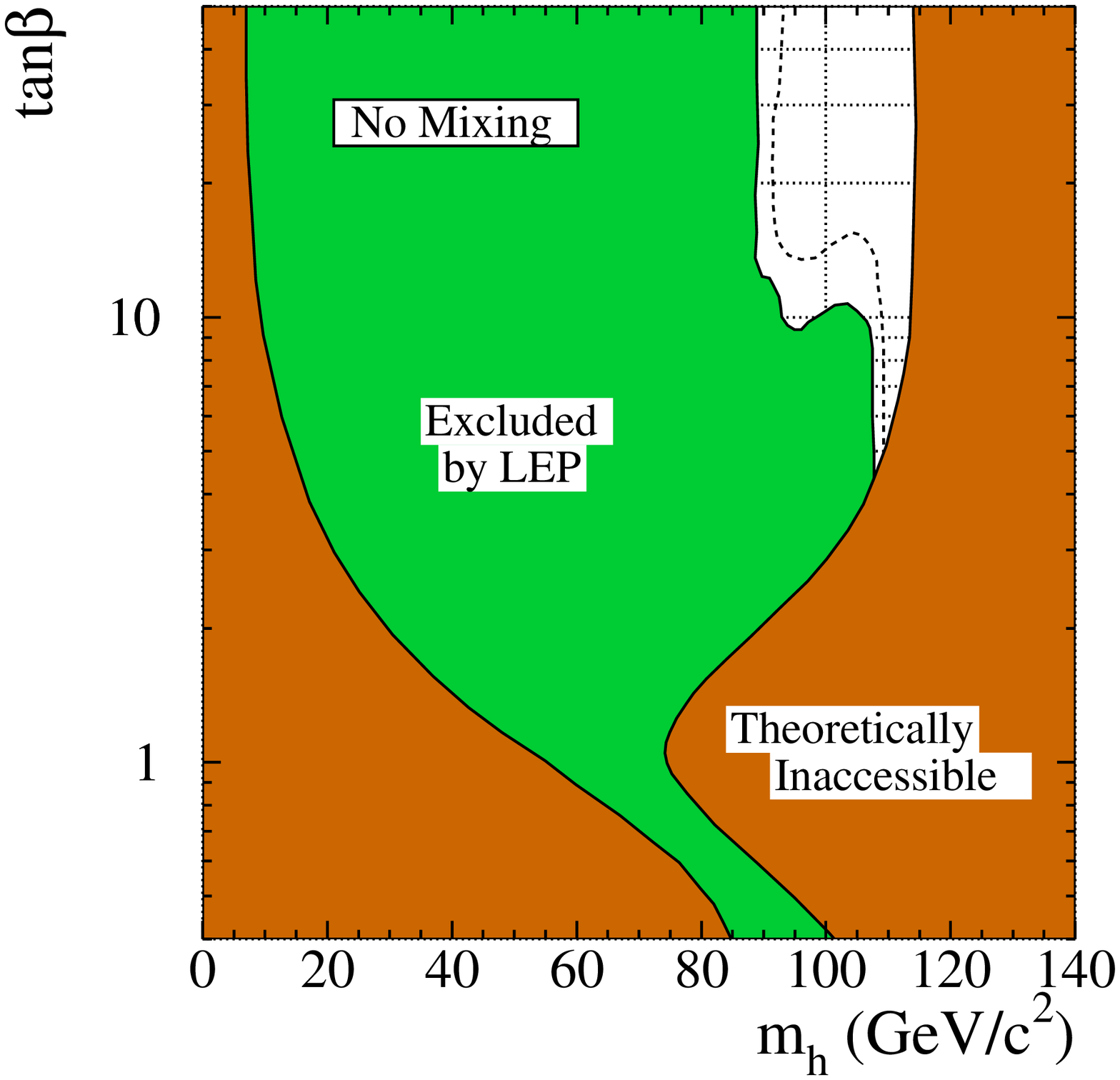}\vspace{1cm}

 \epsfxsize=9cm
\hspace*{3cm} \epsffile{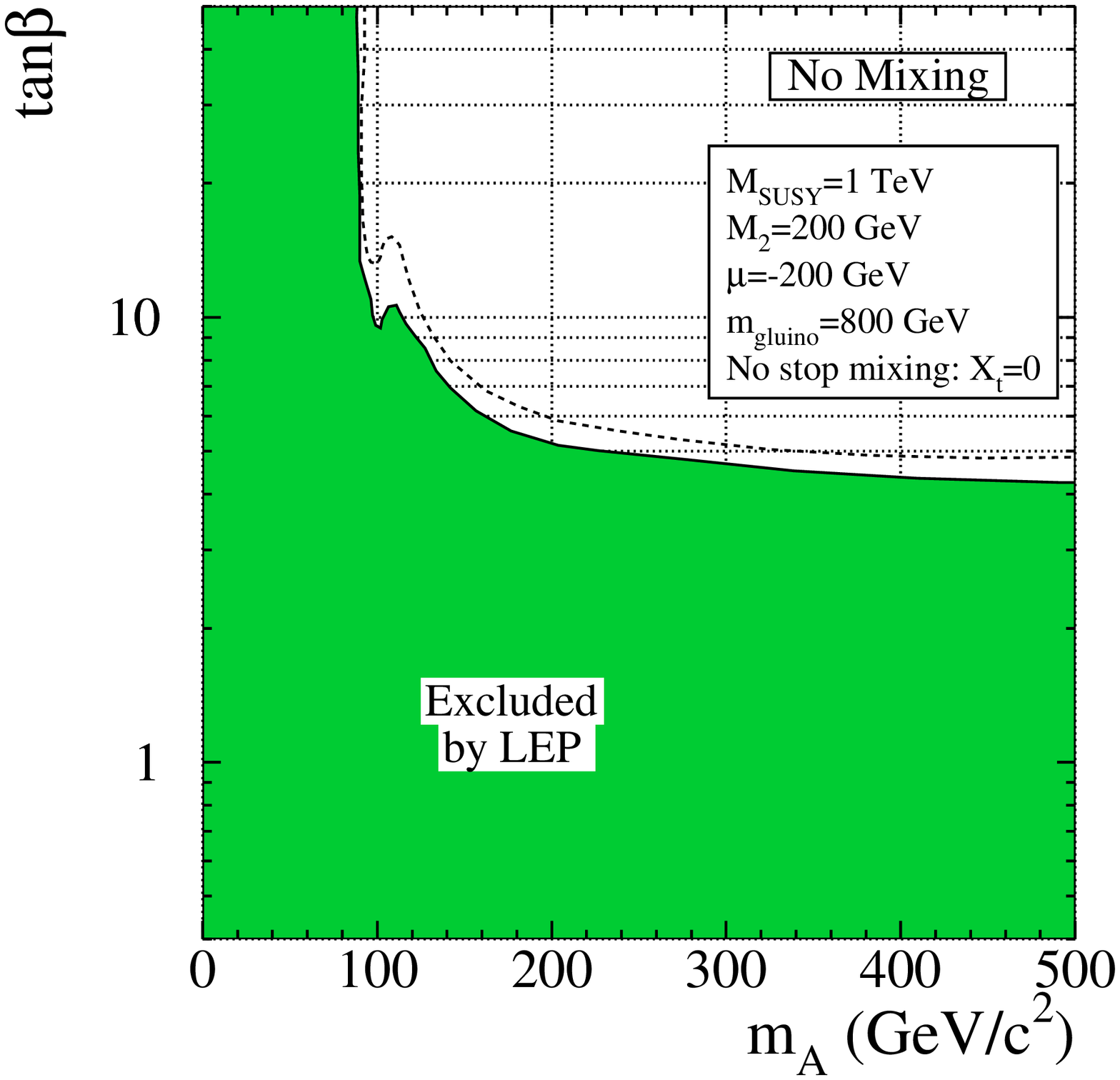}
\caption{Excluded regions for the neutral Higgs bosons search in
the MSSM in the no-mixing case} \label{fig:7}
\end{figure}

\begin{figure}[ht]
 \leavevmode \hspace*{-1cm}
  \epsfxsize=8.5cm
 \epsffile{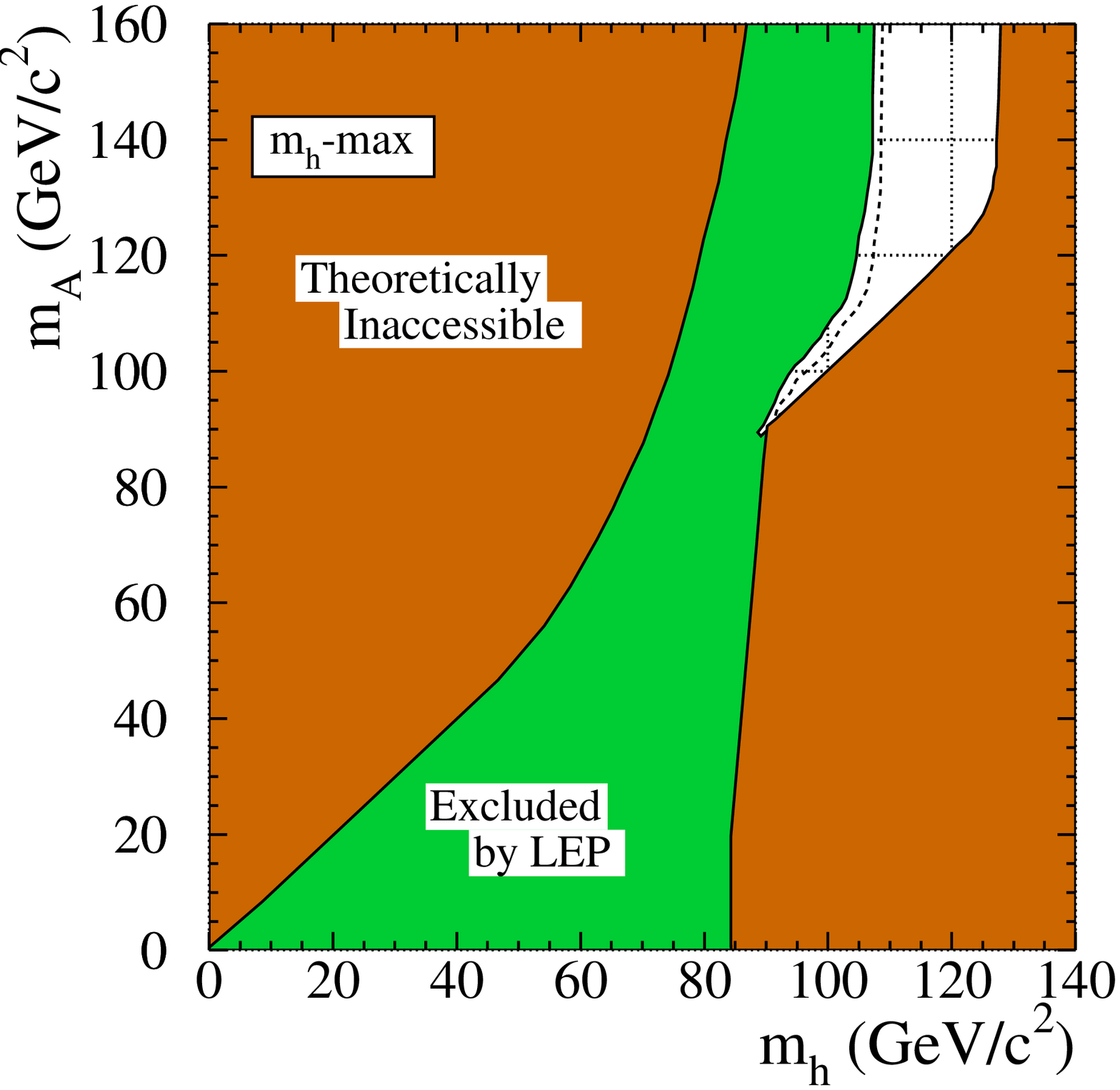}
\vspace{-8.5cm}

\hspace*{7.5cm}\epsfxsize=8.5cm
 \epsffile{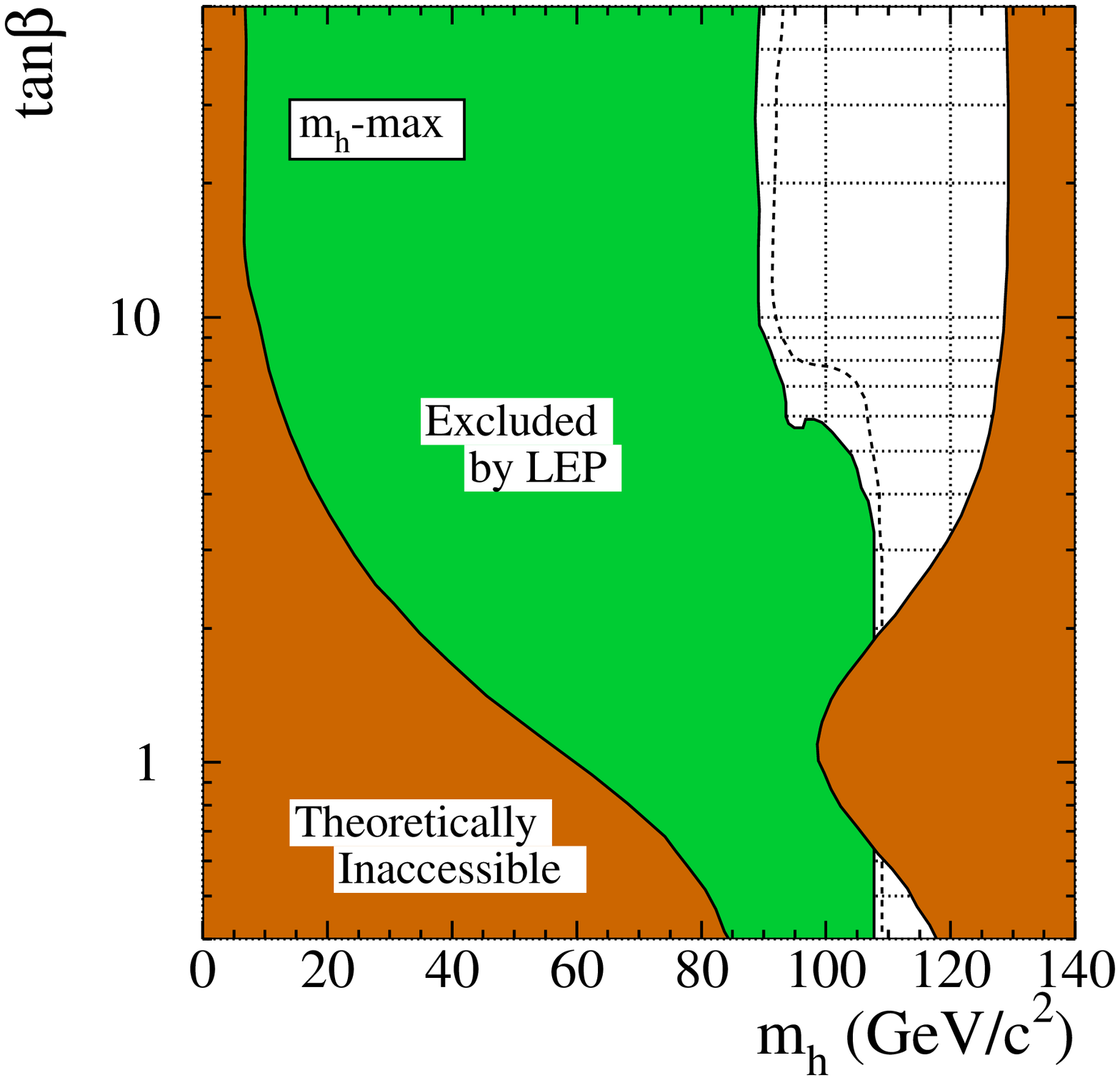}\vspace{1cm}

 \epsfxsize=9cm
\hspace*{3cm} \epsffile{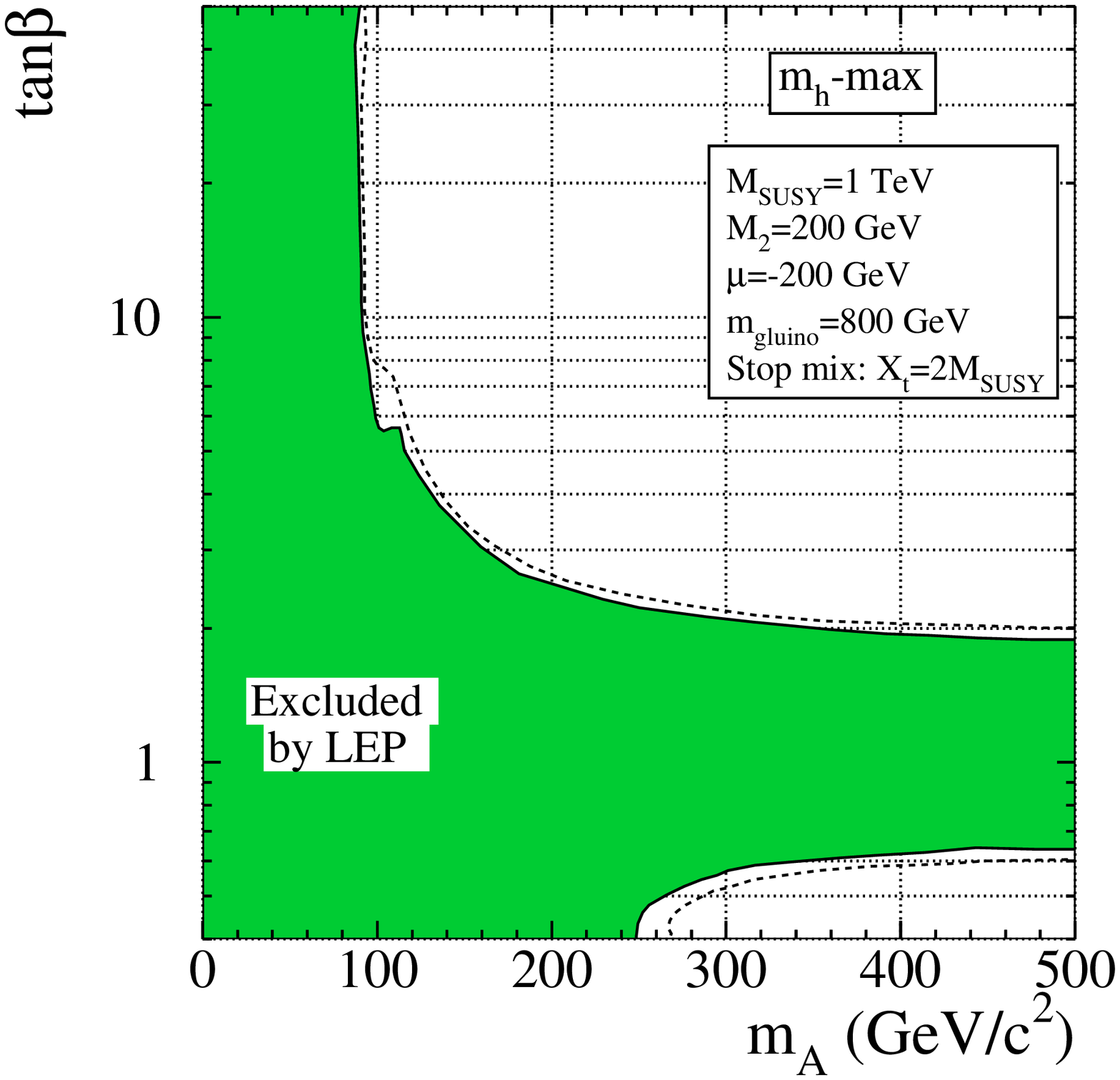}
\caption{Excluded regions for the neutral Higgs bosons search in
the MSSM in the maximal mixing case} \label{fig:8}
\end{figure}

 However, for a heavy pseudoscalar boson $A$ the second
process is decoupled and one basically has the same production
rate as in the SM. Therefore, in this case the SM experimental
limit is applicable also to the MSSM.

To present the result for the Higgs search in the MSSM, various
variables can be used. The most popular ones are $(m_h, m_A)$,
$(m_h,\tan\beta)$ and $(m_A,\tan\beta)$ planes. They are shown
below in Figs.\ref{fig:7}-\ref{fig:8} for  two particular cases:
no-mixing and maximal mixing in the stop sector~\cite{EWWG}. For
comparison the theoretically allowed regions are shown. One can
see that

 a) low $\tan\beta$ solution ($0.5 < \tan\beta < 3.3$) is already excluded;

 b) very small region for the lightest neutral Higgs boson mass
  is left (specially for the no-mixing case).

\begin{figure}[ht]
  \begin{center}
 \leavevmode
  \epsfxsize=8cm \epsfysize=7cm
  \epsffile{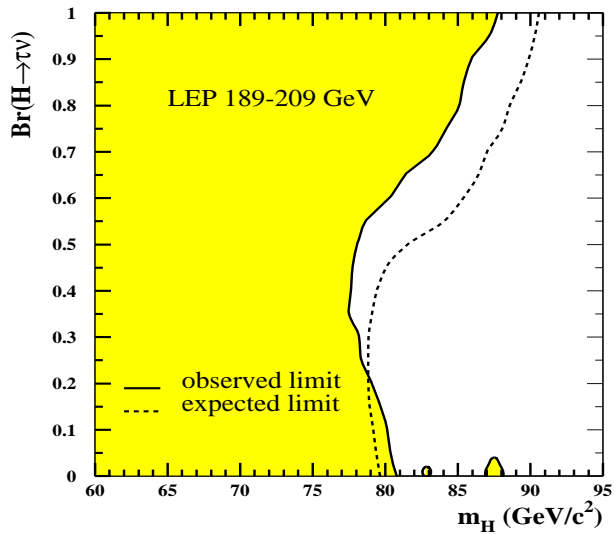}
\end{center}
 \caption{Combined exclusion plot for the charged Higgs boson}
 \label{fig:a35}
 \end{figure}

 As it has been explained, in the MSSM one has also the
charged Higgs bosons. The searches for the charged Higgs bosons
are the attempts to look beyond the Standard Model. It is
basically the same in the MSSM and in any two Higgs doublet model.
The charged Higgs bosons are produced in pairs in an annihilation
process like any charged particles. The couplings are the standard
EW couplings and the only unknown quantity is the charged Higgs
mass. However, the branching ratios for the decay channels depend
on the mass and the model. A large background comes from the
$W$-pair production. Nonobservation of charged Higgs bosons at LEP
gives the lower limit on their masses. The combined exclusion plot
for various channels is shown in Fig.\ref{fig:a35}. This imposes
the absolute lower limit on the charged Higgs boson
mass~\cite{nielsen}
 \begin{equation}
 m_{H^\pm} > 77.5 \ {\rm GeV/c}^2 \ \ \ \ {\mbox @}\ \ 95\%\ \
 C.L.
 \end{equation}

 \vspace{0.5cm}
\underline{Tevatron and LHC}
 \vspace{0.5cm}

 With the LEP shut down, further attempts to discover the Higgs boson
are connected with the Tevatron and LHC hadron colliders.

\begin{figure}[ht]\vspace{-0.8cm}
\begin{center}
 \leavevmode
 \epsfxsize=14cm
 \epsffile{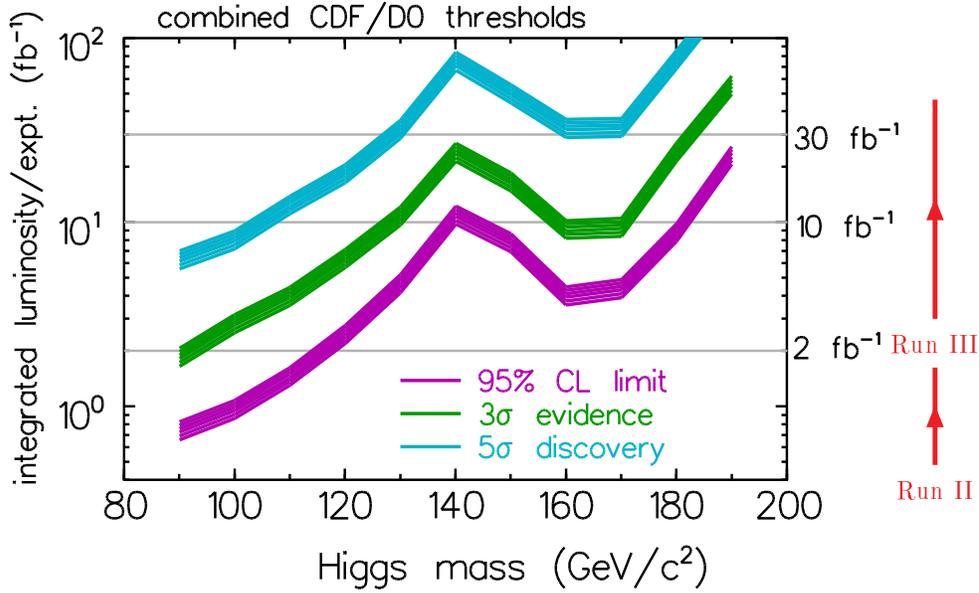}
\end{center}\vspace{-1.3cm}
\caption{Integrated luminosity needed for exclusion ($(2\sigma)$,
evidence $(3\sigma)$ and discovery $(5\sigma$) of the Higgs boson
at Tevatron} \label{tev}
\end{figure}

Tevatron will start the Run II next year and will reach the c.m.
energy of 2 TeV with almost 10 times greater luminosity than in
RUN I. However, since it is a hadron collider, not the full energy
goes into collision taken away by those quarks in a proton that do
not take part in the interaction. Having a very severe background,
this collider needs a long time of running to reach the integrated
luminosity required for the Higgs discovery.  A combined CDF/D0
plot~\cite{CDF/D0} shows the integrated luminosity at Tevatron as
a function of the Higgs mass (see Fig.\ref{tev}).  The three
curves correspond to $2\sigma$ (95\% confidence level), $3\sigma$
and $5\sigma$ signal necessary for exclusion, evidence and
discovery of the Higgs boson, respectively. One can see that the
integrated luminosity of 2 fb$^{-1}$, which is planned to be
achieved at the end of 2001, will allow one to exclude the Higgs
boson with the mass of an order of 115 GeV, i.e., just the limit
reached by LEP. One will need RUN III to reach  10 fb$^{-1}$ to
cover the most interesting interval, even at the level of
exclusion $(2\sigma)$.

\vspace{1.5cm}

\begin{figure}[h]
\begin{center}
 \leavevmode
  \epsfxsize=9.5cm
 \epsffile{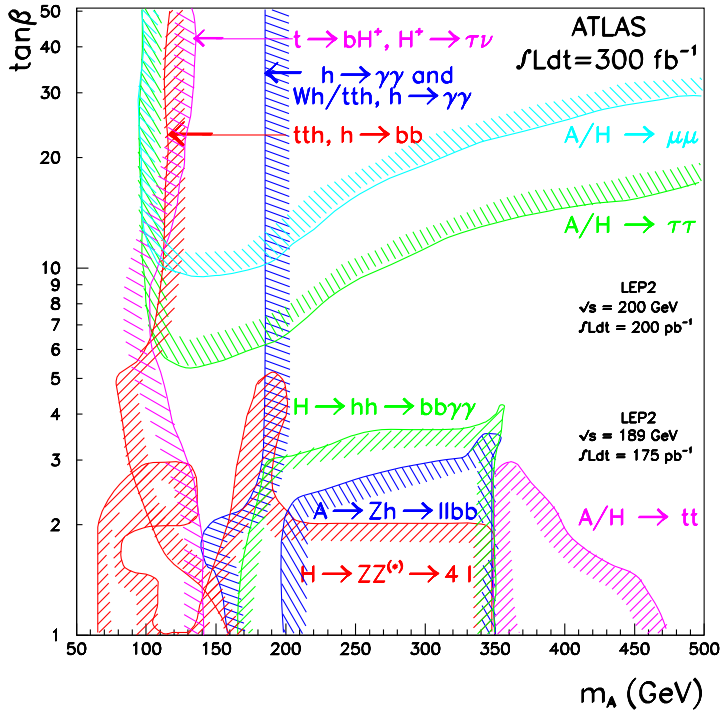}
\end{center}\vspace{-0.5cm}
\caption{Exclusion plots for LHC hadron collider for different
Higgs decay modes} \label{lhc}
\end{figure}

To find the Higgs boson, one will need still greater integrated
luminosity. The signatures of the Higgs boson are related to the
dominant decay modes which depend on the mass of the Higgs boson.
In the Tevatron region they are
 \begin{equation}
 \begin{array}{ll}
 H \to b\bar b, &\ \ \ \ 100 < m_H < 140\ {\rm GeV},\\
 H \to W W^*, &\ \ \ \ 140 < m_H < 175\ {\rm GeV}, \\
  H \to Z Z^*, &\ \ \ \ 175 < m_H < 190\ {\rm GeV}.
  \end{array}
 \end{equation}

The LHC hadron collider is the ultimate machine for a new physics
at the TeV scale. Its c.m. energy is planned to be 14 TeV with
very high luminosity up to a few hundred fb$^{-1}$. It is supposed
to start operating in 2006. In principle, LHC will be able to
cover the whole interval of SUSY and Higgs masses up to a few TeV.
It will either discover the SM or the MSSM Higgs boson, or prove
their absence. In terms of exclusion plots shown in
Figs.\ref{fig:7}, \ref{fig:8} the LHC collider will cover the
whole region~\cite{LHC}. Various decay modes allow one to probe
different areas, as shown in Fig.\ref{lhc}, though the background
will be very essential.

\section{Conclusion}

\begin{figure}[ht]\vspace*{-3cm}\hspace*{-1.5cm}
 \leavevmode
 \epsfxsize=18cm \epsfysize=26cm
 \epsffile{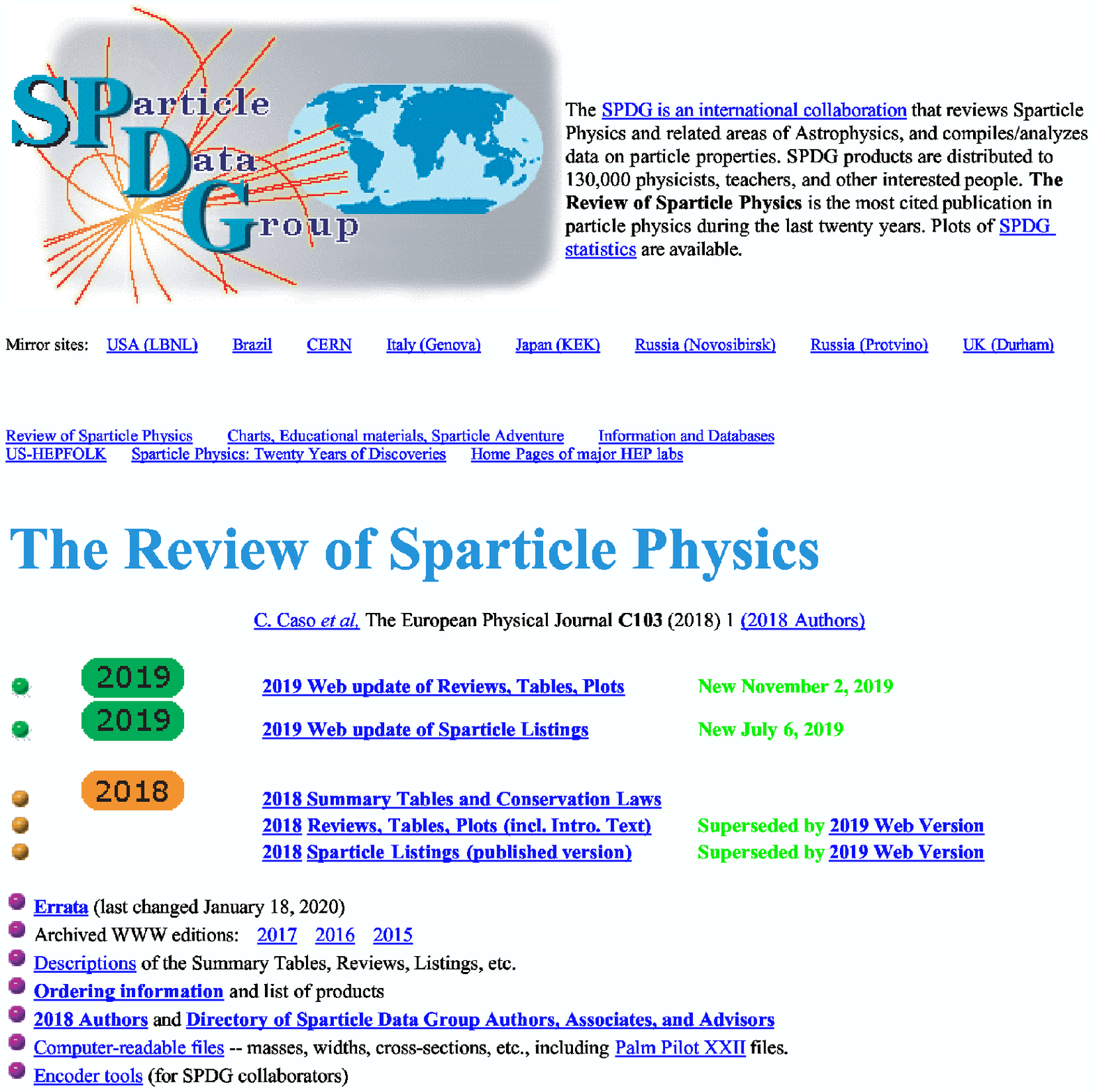}
\vspace{-5cm} \caption{Foreseeable future: SParticle Data Group}
\label{fig:pdg} \end{figure}

LEP II has neither discovered the new physics, nor has proven the
existence of the Higgs boson. However, it  gave us some indication
that both of them exist. Supersymmetry is now the most popular
extension of the Standard Model. It promises us that new physics
is round the corner at a TeV scale to be exploited at new machines
of this decade. If our expectations are correct, very soon we will
face new discoveries, the whole world of supersymmetric particles
will show up and the table of fundamental particles will be
enlarged in increasing rate. If we are lucky, probably we will
soon have the table of sparticles in new addition of Sparticle
Data Group (see Fig.\ref{fig:pdg})~\cite{JB}. This would be a
great step in understanding the microworld. If not, still new
discoveries are in agenda.%

\vspace{0.5cm} {\large \bf Acknowledgements} \vspace{0.5cm}

 I am grateful to the organizers of the School at Caramulo for
providing very nice atmosphere during the school and to the
students for their interest, patience and attention. I would like
to thank V.Velizhanin for his help in preparation of this
manuscript.  Financial support from RFBR grants \# 99-02-16650, \#
98-02-17453 and \# 00-15-96691 and Heienberg-Landau Programme  is
kindly acknowledged.


\clearpage \newpage
 \begin{thebibliography}{99}

\bibitem{SM}
D.~E.~Groom {\it et al.}, ``Review of Particle Physics'', Eur.\
Phys.\ J.\ {\bf C15} (2000) 1.

\bibitem{test-sm} LEP EWWG,
http://lepewwg.web.cern.ch/LEPEWWG/plots/summer2000/

\bibitem{super} Y. A. Golfand and E. P. Likhtman, {\em JETP Letters}
{\bf 13}  (1971) 452; D. V. Volkov and V. P. Akulov, {\em JETP
Letters} {\bf 16} (1972) 621; J. Wess and B. Zumino, {\em Phys.
Lett.} {\bf B49} (1974) 52.

\bibitem{Rev} P. Fayet and S. Ferrara, {\em Phys. Rep.} {\bf 32} (1977)
249; M. F. Sohnius, {\em Phys. Rep.} {\bf 128} (1985) 41;
 H. P. Nilles, {\em Phys. Rep.} {\bf 110} (1984) 1;
 H. E. Haber and G. L. Kane, {\em Phys. Rep.} {\bf 117} (1985) 75;
A. B. Lahanas and D. V. Nanopoulos,  {\em Phys. Rep.} {\bf 145}
(1987) 1.

\bibitem{WessB} J. Wess and J. Bagger, {\em "Supersymmetry and Supergravity"},
Princeton Univ. Press, 1983.

 \bibitem{West}  P. West, {\em "Introduction
to Supersymmetry and Supergravity"}, World Scientific, 1986.

\bibitem{sspace} S. J. Gates, M. Grisaru, M. Ro\v{c}ek and W. Siegel,
{\em "Superspace or One Thousand and One Lessons in
Supersymmetry"}, Benjamin \& Cummings, 1983.

\bibitem{Weinberg} S.~Weinberg, {\em "The quantum theory of fields.  Vol. 3:
Supersymmetry"}, Cambridge, UK: Univ. Press, 2000.

\bibitem{theorem} S. Coleman and J .Mandula, {\em Phys.Rev.} {\bf
 159} (1967) 1251.

\bibitem{SUGRA} P. Nath and R. Arnowitt, {\em Phys. Lett.} {\bf B56} (1975) 177;
D. Z. Freedman, P. van Nieuwenhuizen and S. Ferrara, {\em Phys.
Rev.} {\bf D13} (1976) 3214; S. Deser and B. Zumino, {\em Phys.
Lett.} {\bf B62} (1976) 335; see also "Supersymmetry", S.Ferrara,
ed. (North Holland/World Scientific, Amsterdam/Singapore, 1987)
and Refs.\cite{WessB}-\cite{Weinberg}.

\bibitem{GUT} G. G. Ross, {\em "Grand Unified Theories"},
Benjamin \& Cummings, 1985.

\bibitem{bethke} S.~Bethke, {\em J. Phys.} {\bf G26} (2000) R27; hep-ex/0004021.

\bibitem{msbar} G.~'t Hooft, {\em Nucl. Phys}. {\bf B61}, (1973) 455; \\
  W.~A.~Bardeen, A.~Buras, D.~Duke and T.~Muta, {\em Phys.~Rev}.~{\bf D 18},
(1978) 3998.

\bibitem{fine} A.~D.~Martin, J.~Outhwaite and M.~G.~Ryskin,  {\em Phys.
Lett.} {\bf B492} (2000) 69; hep-ph/0008078.

\bibitem{ABF} U. Amaldi, W. de Boer and H. F\"urstenau, {\em Phys. Lett.}
{\bf B260} (1991) 447.

\bibitem{akt} I.~Antoniadis, C.~Kounnas, and K.~Tamvakis, {\em Phys.
Lett.} {\bf 119B} (1982) 377.

\bibitem{string} M.~B.~Green, J.~H.~Schwarz and E.~Witten, {\em
"Superstring Theory"}, Cambridge, UK: Univ. Press,  1987. {\it
Cambridge Monographs On Mathematical Physics}.

\bibitem{Polyakov} A.~M.~Polyakov, {\em "Gauge Fields And Strings"},
 CHUR, Switzerland: Harwood, 1987, {\it Contemporary Concepts in
Physics, 3}.

\bibitem{WZgauge} J. Wess and B. Zumino, {\em Nucl. Phys.} {\bf B78} (1974)
1.

\bibitem{ber} F. A. Berezin, {\em "The Method of Second Quantization"},
Moscow, Nauka, 1965.

\bibitem{theta} K.~Huang, {\em "Quarks, Leptons And Gauge Fields"},
Singapore,  World Scientific, 1982.

\bibitem{Fayet} P. Fayet and J. Illiopoulos,  {\em Phys. Lett.} {\bf B51}
(1974) 461.

\bibitem{O'R} L. O'Raifeartaigh,  {\em Nucl. Phys.} {\bf B96} (1975) 331.

\bibitem{MSSM} R. Barbieri,  {\em Riv. Nuo. Cim.} {\bf 11} (1988)
1;\\ H. E. Haber, {\em "Introductory Low-Energy Supersymmetry"},
Lectures given at TASI 1992, (SCIPP 92/33, 1993),
hep-ph/9306207.\\ W. de Boer, {\em "Grand Unified Theories and
Supersymmetry in Particle Physics and Cosmology"}, {\em Progr. in
 Nucl. and Particle Phys.,} {\bf 33} (1994) 201,
hep-ph/9402266;\\ D. I. Kazakov, {\em "Mimimal Supersymmetric
Extension of the Standard Model"}, {\em Surveys in High Energy
Physics,} {\bf 10} (1997) 153, wwwtheor.itep.ru/school96/ \\ D. I.
Kazakov, {\em "Supersymmetry in Particle Physics: Renormalization
Group Viewpoint"}, Review talk at the Conf. RG-2000,
hep-ph/0001257.

\bibitem{haber} see e.g. H. Haber in \cite{MSSM}.

\bibitem{shadow}
http://atlasinfo.cern.ch/Atlas/documentation/EDUC/physics14.html

\bibitem{r-parity}
P. Fayet, {\em Phys. Lett}. {\bf B69} (1977) 489; G. Farrar and P.
Fayet, {\em Phys. Lett.} {\bf B76} (1978) 575.

\bibitem{r-symmetry} P.Fayet, {\em Nucl. Phys.} {\bf B90}(1975) 104;
 A.Salam and J.Srathdee, {\em Nucl. Phys.} {\bf B87}(1975) 85.

\bibitem{r-con} H. Dreiner and G. G. Ross, {\em Nucl. Phys.} {\bf B365}
(1991) 597,
 K. Enqvist, A. Masiero and A. Riotto, {\em Nucl. Phys.} {\bf
B373} (1992) 95, \\ H. Dreiner and P. Morawitz, {\em Nucl. Phys.}
{\bf B428} (1994) 31;  H. Dreiner and H. Pois, preprint
NSF-ITP-95-155; hep-ph/9511444, \\ V. Barger, M. S.  Berger , R.
J. N. Philips and T. W\"ohrmann, {\em Phys. Rev.} {\bf D53} (1996)
6407.

\bibitem{hidden}
L. Hall, J. Lykken and S. Weinberg, {\em Phys. Rev.} {\bf D27}
(1983) 2359;  S. K. Soni and H. A. Weldon, {\em Phys. Lett.} {\bf
B126} (1983) 215; I. Affleck, M. Dine and N. Seiberg, {\em Nucl.
Phys.} {\bf B256} (1985) 557.

\bibitem{gravmed} H. P. Nilles, {\em Phys. Lett.} {\bf
B115} (1982) 193; A. H. Chamseddine, R. Arnowitt and P. Nath, {\em
Phys. Rev. Lett.} {\bf 49} (1982) 970; {\em Nucl. Phys.} {\bf
B227} (1983) 121; R. Barbieri, S. Ferrara and C. A. Savoy, {\em
Phys. Lett.} {\bf B119} (1982) 343; E. Cremmer, P. Fayet and L.
Girardello, {\em Phys. Lett.} {\bf B122} (1983) 41; L. Ib\'a\~nez,
{\em Phys. Lett.} {\bf B118} (1982) 73; H. P. Nilles, M. Srednicki
and D. Wyler, {\em Phys. Lett.} {\bf B120} (1983) 346.

\bibitem{gaugemed} M. Dine and A. E. Nelson, {\em Phys. Rev.} {\bf D48} (1993)
1277,  M. Dine, A. E. Nelson and Y. Shirman, {\em Phys. Rev.} {\bf
D51} (1995) 1362, hep-ph/9408384; M. Dine, A. E. Nelson, Y. Nir
and Y. Shirman, {\em Phys. Rev.} {\bf D53} (1996) 2658,
hep-ph/9507378.

\bibitem{anommed} L. Randall and R. Sundrum, {\em Nucl. Phys}. {\bf B557} (1999) 79,
hep-th/9810155; G.~F.~Giudice, M.~A.~Luty, H.~Murayama and
R.~Rattazzi, {\em JHEP}, {\bf 9812} (1998) 027, hep-ph/9810442.

\bibitem{gauginomed}D.~E.~Kaplan, G.~D.~Kribs and M.~Schmaltz, {\em Phys.
Rev.}  {\bf D62} (2000) 035010, hep-ph/9911293; Z.~Chacko,
M.~A.~Luty, A.~E.~Nelson and E.~Ponton, {\em JHEP}, {\bf 0001}
(2000) 003, hep-ph/9911323.

\bibitem{GG} L. Girardello and M. Grisaru, {\em Nucl. Phys.} {\bf B194} (1982)
65.

\bibitem{Peskin}  M.~E.~Peskin, {\em "Theoretical summary lecture for EPS
HEP99"},  hep-ph/0002041, see also Ref.\cite{JB}.

\bibitem{polonsky} N. Polonsky and A. Pomarol, {\em Phys. Rev. Lett.} {\bf
73} (1994) 2292; {\em Phys. Rev}. {\bf D51} (1995) 6532.

\bibitem{spectrum}  G. G. Ross and R. G. Roberts, {\em Nucl. Phys.} {\bf
B377} (1992) 571.  \\
 V. Barger, M. S. Berger and P. Ohmann, {\em Phys. Rev.} {\bf
D47} (1993) 1093.

\bibitem{BEK} W. de Boer, R. Ehret and D. Kazakov, {\em Z. Phys.}
 {\bf C67} (1995) 647;\\ W. de Boer et al., {\em Z. Phys.} {\bf C71} (1996)
415.

\bibitem{bound1} N. Cabibbo, L. Maiani, G. Parisi,
and R. Petronzio, {\em Nucl. Phys.} {\bf B158} (1979) 295;\\
 M. Lindner, {\em Z. Phys}. {\bf C31} (1986) 295; M. Sher, {\em Phys. Rev.}
 {\bf D179} (1989) 273;\\ M. Lindner,  M. Sher and H. W. Zaglauer, {\em
Phys. Lett.} {\bf B228} (1989) 139.

\bibitem{Ibanez}  L. E. Ib\'a\~nez, C. Lop\'ez and C. Mu\~noz, {\em
Nucl. Phys.} {\bf B256} (1985) 218.

\bibitem{Barger} W. Barger, M. Berger, P. Ohman, {\em Phys. Rev.} {\bf D49} (1994)
4908.

\bibitem{mtop} G.~Brooijmans  [CDF and D0
Collaborations], hep-ex/0005030.\\ %
http://www-d0.fnal.gov/public/top/top\\ 
http://www-cdf.fnal.gov/physics/new/top/results/mass/combine

\bibitem{bbog} V.~Barger, M.~S. Berger, P.~Ohmann and R.~J.~N. Phillips,
  {\em Phys.~Lett.}~{\bf B314} (1993) 351.

\bibitem{lanpol} P.~Langacker and N.~Polonsky, {\em Phys. Rev.} {\bf D49} (1994) 1454.

\bibitem{bmaskln} S.~Kelley, J.~L. Lopez and D.V. Nanopoulos, {\em
Phys.~Lett.}~{\bf B274} (1992) 387.

\bibitem{cpr} P.~H. Chankowski, S.~Pokorski and J.~Rosiek, {\em Nucl.
Phys.} {\bf B423} (1994) 437;

\bibitem{copw1} M.~Carena, M.~Olechowski, S.~Pokorski and C.E.M.
Wagner, {\em Nucl.~Phys.}~{\bf B426} (1994) 269.

\bibitem{ara91} H.~Arason et~al., {\em Phys.~Rev.~Lett.}~{\bf 67} (1991)
2933.

\bibitem{CLEO} R. Poling (University of Minnesota), Talk at the Lepton - Photon '99
 Conference, Stanford University, USA, 9-14 August 1999;
         see also CLEO Collaboration, CLEO CONF 98-17, ICHEP98 1011.

\bibitem{ALBSG} R. Barate et al. (ALEPH Collaboration), {\em Phys. Lett.} {\bf B429}
                (1998) 169.

\bibitem{borz} F. Borzumati, {\em Z. Phys.} {\bf C63} (1994) 291.

\bibitem{bsgamm3} S. Bertolini, F. Borzumati, A. Masiero, and G. Ridolfi,
{\em Nucl. Phys.} {\bf B353} (1991) 591 and references therein; N.
Oshimo, {\em Nucl. Phys.} {\bf B404} (1993) 20.

\bibitem{LEPSUSY} ALEPH Collaboration, {\em
"Search for supersymmetric particles in e+ e- collisions at $\sqrt
s$ up to 202-GeV and mass limit for the lightest neutralino"},
hep-ex/0011047.

\bibitem{TEVSUSY}S.~Abel {\it et al.}  [SUGRA Working Group
Collaboration], {\em Report of the SUGRA working group for run II
of the Tevatron}, hep-ph/0003154.\\ %
D0 Coll., {\em "Search for Squarks and Gluinos in Events
Containing Jets and a Large Imbalance in Transverse Energy"}, {\em
Phys. Rev. Lett.} {\bf 83} (1999) 4937.

\bibitem{borner} G.~B\"orner, {\em "The early Universe"}, Springer Verlag,
1991.

\bibitem{kolb} E. W. Kolb and M. S. Turner, {\em "The early Universe",}
Addison-Wesley, 1990.

\bibitem{relic} G. Steigman, K. A. Olive, D. N. Schramm, M. S. Turner, {\em
Phys. Lett.}  {\bf B176} (1986) 33; \\  J. Ellis, K. Enquist, D.V.
Nanopoulos, S. Sarkar,  {\em Phys. Lett.} {\bf B167}
 (1986) 457;\\ G. Gelmini and P. Gondolo, {\em Nucl.  Phys.} {\bf
B360} (1991) 145.

\bibitem{relictst} M. Drees and M. M. Nojiri, {\em Phys. Rev.} {\bf D47}
(1993) 376;\\
  J. L. Lopez, D. V. Nanopoulos and H. Pois, {\em Phys. Rev.} {\bf D47} (1993)
  2468;\\ P. Nath and R. Arnowitt, {\em Phys. Rev. Lett.} {\bf 70} (1993)
  3696;\\ J. L. Lopez, D. V. Nanopoulos and K. Yuan, {\em Phys. Rev.} {\bf
D48} (1993) 2766.

\bibitem{roskane} G.~L. Kane, C.~Kolda, L.~Roszkowski and J.~D. Wells,
  {\em Phys.~Rev.}~{\bf D49} (1994) 6173.

\bibitem{rosdm} L. Roszkowski, {\em Univ. of Michigan Preprint},
UM-TH-93-06; {\em Phys. Rev.} {\bf D50} (1994) 4842.

\bibitem{bop} F. M.~Borzumati, M. Olechowski and S. Pokorski, {\em Phys.
Lett.} {\bf B349} (1995) 311.

\bibitem{BGGK} W.de Boer, H.-J.Grimm, A.V.Gladyshev and D.I.Kazakov, {\em
Phys. Lett.} {\bf B438} (1998) 281.

\bibitem{BHGK} W. de Boer, M. Huber, A. V. Gladyshev and D. I. Kazakov,
hep-ph/0007078.

\bibitem{higgs}
J. Gunion, H. Haber, G. Kane and S. Dawson, {\em "Higgs Hunter's
Guide"}, Addison-Wesley, New York, 1990.

\bibitem{EWWG} The LEP Higgs Working group, R. Bock et al.,
CERN-EP-2000-055 and LEP experiments, ALEPH 2000-28, DELPHI
2000-050, L3-Note 2525, OPAL TN646.

\bibitem{bound2}  M. Sher, {\em Phys. Lett.} {\bf B317} (1993) 159; C. Ford,
D. R. T. Jones, P. W. Stephenson and M.~B.~Einhorn, {\em Nucl.
Phys.} {\bf B395} (1993) 17; G. Altarelli and I. Isidori, {\em
Phys. Lett.} {\bf B337} (1994) 141; J. A. Casas, J. R. Espinosa
and M. Quiros, {\em Phys. Lett.} {\bf B342} (1995) 171.

\bibitem{HR} T. Hambye, K. Reisselmann, {\em Phys. Rev.} {\bf D55} (1997)
7255; H. Dreiner, hep-ph/9902347.

\bibitem{nielsen}
P.~Bock {\it et al.}  [ALEPH, DELPHI, L3 and OPAL Collaborations],
CERN-EP-2000-055;
http://lephiggs.web.cern.ch/LEPHIGGS/papers/osaka\_note.ps

\bibitem{ALEPH}
R.~Barate {\it et al.}  [ALEPH Collaboration], {\em "Observation
of an excess in the search for the standard model Higgs  boson at
ALEPH,}" hep-ex/0011045.

\bibitem{L3}
M.~Acciarri {\it et al.}  [L3 Collaboration], {\em "Higgs
candidates in e+ e- interactions at $\sqrt s$ = 206.6\ GeV"},
hep-ex/0011043.

\bibitem{radcorr} J. Ellis, G. Ridolfi, F. Zwirner, {\em Phys. Lett.} {\bf
B262} (1991) 477; A. Brignole,
 J. Ellis, G.~Ridolfi, F. Zwirner, {\em Phys. Lett.} {\bf B271} (1991) 123.

\bibitem{carena}  M. Carena, J. R. Espinosa, M. Quiros and C. E. M. Wagner, {\em
Phys. Lett.} {\bf B355} (1995) 209; \\ J. Ellis, G. L. Fogli and
E. Lisi, {\em Phys. Lett.} {\bf B333} (1994) 118.

\bibitem{G} A.V.Gladyshev, D.I.Kazakov, W.de Boer, G.Burkart,
R.Ehret, {\em Nucl. Phys.} {\bf B498} (1997) 3.

\bibitem{CW}  M. Carena, M. Quiros and C. E. M. Wagner, {\em
Nucl. Phys.} {\bf B461} (1996) 407.

\bibitem{feynhiggs} S. Heinemeyer, W. Hollik and G. Weiglein,
{\em Phys. Lett.} {\bf B455} (1999) 179, hep-ph/9903404; {\em Eur.
Phys. J.} {\bf C9} (1999) 343, hep-ph/9812472.

\bibitem{pomarol} M. Masip, R. Mu\~noz-Tapia and A. Pomarol, {\em Phys. Rev.} {\bf
D57} (1998) 5340.

\bibitem{CDF/D0} M.~Carena {\em et al.}, {\em "Report of the Tevatron Higgs
working group"}, hep-ph/0010338.

\bibitem{LHC} ATLAS TDR,
http://www.cern.ch/ATLAS/GROUPS/PHYSICS/TDR.

\bibitem{JB} J. Bagger, summary talk at SUSY2K\\
http://wwwth.cern.ch/susy2k/susy2ktalks/bagger\_plenary.pdf.gz

\end{thebibliography}
\end{document}